\documentclass[%
  twocolumn,
  amsfonts,
  amsmath,
  amssymb,
  superscriptaddress,
  floatfix,
  nobibnotes,
  nofootinbib,
  aps,
  prd
]
{revtex4-2}

\usepackage[utf8]{inputenc}
\usepackage{xargs}
\usepackage[english]{babel}
\usepackage{graphicx}
\usepackage{color}
\usepackage{lineno}
\usepackage{xspace}
\usepackage{dcolumn}
\usepackage{url}
\usepackage{bbm}
\usepackage{bm}
\usepackage{bbold}
\usepackage{float}
\usepackage{rotating}
\usepackage{siunitx}
\usepackage{mathrsfs}
\usepackage{slashed}
\usepackage{tensor}
\usepackage{nicefrac}
\usepackage{enumitem}
\usepackage{wrapfig}
\usepackage{xcolor}
\usepackage{booktabs}
\usepackage{orcidlink}
\usepackage{multirow}
\usepackage[normalem]{ulem}
\usepackage{makecell}
\usepackage{autobreak}

\usepackage{hyperref}
\hypersetup{
  colorlinks=true,
  citecolor=blue,
  citebordercolor=red,
  linktoc=all,
  linkcolor=blue,
  urlcolor=blue
}
\usepackage{cleveref}

\usepackage[acronym]{glossaries-extra}
\glssetcategoryattribute{acronym}{nohyperfirst}{true}
\setabbreviationstyle[acronym]{long-short}

\newacronym{QCD}{QCD}{quantum chromodynamics}
\newacronym{RG}{RG}{renormalization group}
\newacronym{FRG}{FRG}{functional renormalization group}
\newacronym{MFA}{MFA}{meanfield approximation}
\newacronym{RPA}{RPA}{random-phase approximation}
\newacronym{DSE}{DSE}{Dyson-Schwinger equations}
\newacronym{STI}{STIs}{Slavnov-Taylor identities}
\newacronym{mSTI}{mSTI}{modified STI}
\newacronym{mSTIs}{mSTIs}{modified STIs}
\newacronym{UV}{UV}{ultraviolet}
\newacronym{IR}{IR}{infrared}
\newacronym{DCSB}{D$\chi$SB}{dynamical chiral symmetry breaking}
\newacronym{CSB}{$\chi$SB}{chiral symmetry breaking}
\newacronym{EoS}{EoS}{equation of state}
\newacronym{LEC}{LEC}{low-energy constant}
\newacronym{LECs}{LECs}{low-energy constants}

\newacronym{LEM}{LEM}{low-energy model}
\newacronym{QMD}{QMD}{quark--meson--diquark}
\newacronym{NJL}{NJL}{Nambu--Jona-Lasinio}
\newacronym{SPS}{SPS}{scalar-pseudoscalar}
\newacronym{CSC}{CSC}{color-superconducting}
\newacronym{2SC}{2SC}{two-flavor color-superconductivity}
\newacronym{CFL}{CFL}{color-flavor-locked}

\AddToHook{cmd/appendix/before}{
  
  \crefalias{section}{appendix}
  \crefalias{subsection}{appendix}
}

\crefformat{equation}{#2Eq.~(#1)#3}
\crefformat{appendix}{#2App.~#1#3}
\crefformat{section}{#2Sec.~#1#3}
\crefformat{table}{#2Tab.~#1#3}
\crefformat{figure}{#2Fig.~#1#3}

\crefmultiformat{section}{#2Secs.~#1#3}{ and~#2#1#3}{, #2#1#3}{
and~#2#1#3}
\crefmultiformat{equation}{#2Eqs.~(#1)#3}{ and~#2(#1)#3}{, #2(#1)#3}{
and~#2(#1)#3}

\crefrangeformat{equation}{#3Eqs.~(#1)#4 to #5(#2)#6}
\crefrangeformat{figure}{#3Figs.~#1#4 to #5#2#6}

\Crefformat{figure}{#2Fig.~#1#3}

\newcommand{\ccite}[1]{Ref.~\cite{#1}}
\newcommand{\ccites}[1]{Refs.~\cite{#1}}

\definecolor{umcol}{RGB}{247, 0, 37}
\definecolor{hgcol}{RGB}{143, 0, 255}
\definecolor{sycol}{RGB}{143, 0, 255}
\definecolor{chcol}{RGB}{143, 0, 255}
\definecolor{bjcol}{RGB}{143, 0, 137}
\definecolor{frcol}{RGB}{143, 0, 255}

\graphicspath{
  {plots/}
  {figures/}
}

\newcommand{\Darmstadt}{
  Technische Universität Darmstadt, %
  Fachbereich Physik, %
  Institut für Kernphysik, %
  Theoriezentrum, %
  Schlossgartenstr. 2, %
  D-64289 Darmstadt, %
  Germany %
}

\newcommand{\JLU}{%
  Institut f\"{u}r Theoretische Physik, %
  Justus-Liebig-Universit\"{a}t Gie\ss{}en, %
  35392 Gie\ss{}en, %
  Germany%
}
\newcommand{\HFHFGiessen}{%
  Helmholtz Forschungsakademie Hessen f\"{u}r FAIR (HFHF), %
  GSI Helmholtzzentrum f\"{u}r Schwerionenforschung, %
  Campus Gie\ss{}en, %
  35392 Gie\ss{}en, %
  Germany%
}

\newcommand{\onefig}{0.49\textwidth}
\newcommand{\twofigs}{0.49\textwidth}

\newcommand{\Tr}{\mathrm{Tr}}

\newcommand{\tr}{\mathrm{tr}}

\newcommand{\sqvec}[1]{\ensuremath{ \vec{#1}^{\, 2}}}
\newcommand*{\tp}{\mathsf{T}}

\DeclareMathOperator{\diag}{diag}

\renewcommand{\d}{\; .}
\renewcommand{\c}{\; ,}

\newcommand{\MeV}{\,\mathrm{MeV}}
\newcommand{\GeV}{\,\mathrm{GeV}}

\newcommand{\cfa}{\mathbf{a}}
\newcommand{\cfb}{\mathbf{b}}
\newcommand{\cfc}{\mathbf{c}}

\newcommand{\da}{\mathfrak{a}}
\newcommand{\db}{\mathfrak{b}}
\newcommand{\dc}{\mathfrak{c}}
\newcommand{\dd}{\mathfrak{d}}

\newcommand{\qbqbqq}{\bar{q}\bar{q}qq}

\newcommand{\mgapUV}{m_{\text{gap},\text{UV}}}
\newcommand{\mgapscal}{m_{\text{gap},\text{scaling}}}

\newcommand{\citeDiquarkMass}{\cite{
Rapp:1997zu,
Hess:1998sd,
Oettel:2000jj,
Maris:2002yu,
Nicmorus:2008vb,
Eichmann:2009zx,
Eichmann:2016hgl,
Watanabe:2021nwe,
Kelvin-Lee:2025vyf,
Kelvin-Lee:2026owm%
}}

\newcommand{\citeScalingExponent}{\cite{
vonSmekal:1997ohs,
Lerche:2002ep,
Pawlowski:2003hq%
}}

\makeatletter
\def\l@subsubsection#1#2{}
\makeatother

\begin{document}

\title{Scalar diquarks in the QCD vacuum}

\author{Hosein Gholami \orcidlink{0009-0003-3194-926X}}
\affiliation{\Darmstadt}

\author{Ugo Mire \orcidlink{0009-0009-2345-691X}}
\email[Corresponding author: ]{ugo.louis.tryphon.mire@physik.uni-giessen.de}
\affiliation{\JLU}

\author{Fabian Rennecke \orcidlink{0000-0003-1448-677X}}
\affiliation{\JLU}
\affiliation{\HFHFGiessen}

\author{Bernd-Jochen Schaefer \orcidlink{0000-0003-0659-2679}}
\affiliation{\JLU}
\affiliation{\HFHFGiessen}

\author{Shi Yin \orcidlink{0000-0001-5279-6926}}
\affiliation{\JLU}

\begin{abstract}

While QCD fundamentally only depends on the values of the strong coupling and the quark masses, it exhibits a rich nonperturbative structure at low energies, where composite fields emerge as the relevant degrees of freedom. In this work, we present a first-principles framework that captures the transition from fundamental QCD to its low-energy sector in vacuum. It builds on the dynamical hadronization technique within the functional renormalization group approach to two-flavor QCD. In this framework, the low-energy constants relevant for effective models, including effective masses and coupling strengths, naturally emerge from the underlying renormalization group flow without introducing free parameters beyond those of QCD itself. We investigate the dynamical emergence of the pion, the $\sigma$-meson and the scalar diquark in both imaginary and real time, and determine a set of QCD low-energy constants which can be used to fix the free parameters of models of dense quark matter with a two-flavor color superconducting phase. In particular, this includes previously unknown properties of the scalar diquark. Our results provide important microscopic input for constraining color superconducting phases, which are expected to play a key role in our understanding of dense neutron star matter.
    
\end{abstract}

\maketitle

\tableofcontents

\section{Introduction}
\label{sec:introduction}

The phase structure of dense, strongly interacting matter continues to
be an important open problem of modern physics. In particular, only
limited knowledge exists about cold \gls{QCD} matter at densities
above a few times nuclear saturation density. Yet, this region is of
particular relevance for the physics of neutron stars and their
mergers \cite{Baym:2017whm,  Annala:2019puf}. Owing to the rapid
progress in multi-messenger astronomy, a wealth of data sensitive to
the equation of state and transport properties of cold and dense
\gls{QCD} matter is being expected \cite{LIGOScientific:2017vwq,
  Miller:2019cac, Riley:2021pdl}.  It is hence imperative to
understand this region.

Matter in the core of neutron stars can potentially be compressed to a
point where hadrons break apart and exotic quark matter is
revealed. Since there is an attractive interaction between quarks in
the color--anti-triplet channel \cite{Bailin:1983bm}, a Cooper
instability towards the formation of diquark states is expected at
sufficiently high density. Indeed, at asymptotically large $\mu_B$
where perturbation theory is applicable, it is known that cold
\gls{QCD} matter is in a \gls{CFL} color-superconducting state
\cite{Alford:1998mk,Pisarski:1999tv,Alford:2007xm,Schmitt:2025cqi}.
At lower density, a rich structure
of homogeneous and inhomogeneous phases is expected based on model
calculations \cite{Fukushima:2010bq}. In particular, at scales where
the mass difference between light, i.e., up and down, and strange
quarks becomes relevant, \gls{2SC} seems to be likely. This is
supported by the observation that the formation of nucleons in nuclear
matter is well described by the emergence of an intermediate
scalar diquark state \cite{Eichmann:2016yit}, whose condensation
gives rise to the \gls{2SC} phase; see also \ccite{Barabanov:2020jvn} for a general review on diquark correlations. It is therefore conceivable
that, upon dissociation of nucleons at sufficiently high densities
in the cores of neutron stars, a \gls{2SC} phase may form.
Consequently, \gls{2SC} matter plays a central role in
theoretical studies of dense QCD matter and in the construction of
quark-matter equations of state for neutron stars
\cite{Alford:2002kj, Alford:2007xm, Schmitt:2025cqi}.

Ideally, cold and dense \gls{QCD} matter should be described from
first principles.  This is a formidable task. For all
phenomenologically relevant densities QCD is strongly coupled and may
exhibit
multiple competing phases. Consequently, neither perturbative methods
nor controlled effective field-theory descriptions are generally
applicable throughout the entire density range of interest. Moreover,
owing to the sign problem, importance-sampling methods such as lattice
\gls{QCD} are severely limited at finite baryon chemical potential,
$\mu_B$ \cite{Pasztor:2024dpv}. These challenges can be addressed with
functional approaches, in particular the \gls{FRG} and \gls{DSE}; for
\gls{QCD}-related reviews, see \ccites{Pawlowski:2005xe, Gies:2006wv,
  Rosten:2010vm, Braun:2011pp, Dupuis:2020fhh, Fu:2022gou} for the
\gls{FRG} and \ccites{Alkofer:2000wg, Roberts:2000aa, Fischer:2006ub,
  Eichmann:2016yit, Fischer:2018sdj, Huber:2018ned} for
\gls{DSE}. Rapid progress has been made in description of dense
\gls{QCD} matter from first principles in recent years
\cite{Rennecke:2025bcw, Fischer:2026uni, Fischer:2026vkc}. This
includes predictions for the \gls{QCD} critical point
\cite{Fischer:2012vc, Fu:2019hdw, Gao:2020fbl,
  Gunkel:2021oya, Fu:2026qnl}, spatially modulated moat regimes
\cite{Fu:2019hdw, Fu:2024rto,Fu:2026qnl}, the \gls{EoS} of hot and
dense matter \cite{Isserstedt:2020qll, Lu:2023mkn, Lu:2025cls}, and
first results on color-superconducting matter \cite{Muller:2013pya,
  Muller:2016fdr, Leonhardt:2019fua, Braun:2021uua}. A key result of
\gls{FRG} studies is that, in two-flavor \gls{QCD}, the \gls{2SC} phase is
indeed favored over the chirally broken phase
at intermediate and high baryon densities \cite{Braun:2019aow}.

Despite this progress, a first-principles determination of the
neutron-star equation of state is not yet available.  Consequently,
phenomenological models remain indispensable for predicting and
interpreting astrophysical observations.
Such models inevitably involve parameters that must be constrained
by external input, most commonly by experimental data.
Typical examples include particle masses, decay constants and other vacuum
observables \cite{Rehberg:1995kh,Buballa:2003qv}.  However, this
already entails a first subtlety that  is often ignored: while
experiments usually determine  pole masses, which manifest themselves as peaks in
scattering processes, most model calculations are often calibrated
using the curvature masses.
For a propagator $G_\phi(p_0,\vec{p}\,)$ of a field $\phi$ with
Euclidean frequency $p_0$, the pole mass, $m_{\phi,{\rm pole}}$, is
defined via
\begin{align}\label{eq:mpole}
G_\phi^{-1}(-i m_{\phi,{\rm pole}},0) = 0\,,
\end{align}
while the curvature mass is defined as
\begin{align}\label{eq:mcurv}
m_{\phi,{\rm curv}}^2 =
G_\phi^{-1}(0,0)/Z_\phi\,,
\end{align}
where $Z_\phi$ is an appropriate wave function renormalization factor to ensure RG invariance.
Given that the \gls{EoS} is most naturally computed in Euclidean
space, the determination of pole masses requires an additional
analytic continuation to Minkowski space, whereas curvature masses are
directly accessible. Consequently,
the latter are widely employed in model calculations. Fortunately, for
very light mesons, such as pions, curvature and pole masses are known
to be nearly identical \cite{Helmboldt:2014iya, Fu:2024rto}. Whether
a similarly close correspondence exists 
for other mesons remains largely unexplored.
A related issue arises in the determination of effective couplings. 
While experimental information is extracted from scattering processes
involving on-shell particles,
model calculations typically employ
off-shell
interactions, often approximated as pointlike.

However, even in models whose parameters can, in principle, be fixed
from pole masses and on-shell couplings, some of the required input
quantities are not experimentally accessible.
Of particular relevance in the present context are diquarks, whose
masses and interactions have never been measured directly.
Yet, they are essential input for effective models of cold and dense
\gls{QCD} matter.
Various theoretical approaches have predicted the vacuum pole mass of
the scalar diquark $\Delta$, consistently placing it at approximately
twice the constituent quark mass,
$m_{\Delta,{\rm pole}} \approx 2 m_q \approx 600-700 \MeV$
\citeDiquarkMass. This can be understood naturally within the
quark--diquark picture of nucleons mentioned above. If nucleons are
viewed as quark--diquark bound states, their mass is expected to be
approximately given by $m_{\Delta,{\rm pole}} + m_q$ minus some
binding energy, at least for physical quark masses \cite{Bender:1996bb}. Thus, while  no direct experimental determination of
the diquark mass exists, theoretical considerations provide at least a
reasonably well 
constrained estimate 
that can be used to fix one of the model parameters.

In many model studies, however, the diquark pole mass is often
identified with the corresponding curvature mass. As we will
demonstrate below, the relation between these quantities depends
sensitively on the quark--diquark interaction, which is considerably
less constrained.  Information on diquark self-interactions is even
more scarce. While one may attempt to constrain the corresponding
model parameters using astrophysical observations
\cite{Raaijmakers:2019qny, Raaijmakers:2021uju, Mroczek:2024mbo,
  Gholami:2024ety, Gao:2024lzu, Ayriyan:2024zfw, Christian:2025guq},
the large number of poorly known parameters substantially limits the
predictive power of such approaches.

In the present work, we address these shortcomings by determining
quantities relevant for the modeling of dense two-flavor quark matter
directly from \gls{QCD} within a \gls{FRG} approach. In particular, we
compute the diquark curvature and pole masses, the pointlike
quark--diquark coupling, and diquark self-interactions. To this end,
we build on the techniques for the self-consistent treatment of
\gls{QCD} developed in \ccites{Pawlowski:2003hq, Fischer:2004uk,
  Cyrol:2016tym}, in particular the efficient framework introduced in
\ccite{Goertz:2024dnz}, as well as on methods for describing emergent
bound states developed in \ccites{Gies:2001nw, Gies:2002hq,
  Pawlowski:2005xe, Braun:2009ewx, Mitter:2014wpa, Braun:2014ata,
  Rennecke:2015eba, Fu:2019hdw}. This framework enables the
determination of
\gls{LECs} associated with  emergent \gls{QCD} degrees of freedom  without
requiring  any phenomenological input beyond  the scale setting that
is inherent to  any
first-principles calculation. Consequently, we obtain genuine
first-principles  predictions for
a variety of phenomenologically  relevant vacuum \gls{LECs} based
solely  on the values of the
strong coupling and the current quark masses specified at a perturbatively
high energy scale.

This paper is organized as follows. In
\cref{sec:rg-invariant-formulation}, we introduce the \gls{FRG}
approach to \gls{QCD} and briefly review key aspects of infrared
\gls{QCD}. In \cref{sec:truncation}, we present our truncation for the
gluonic and low-energy sectors, while in
\cref{sec:correlation_function} is devoted to the \gls{RG} flows of
the corresponding  correlation functions. Readers primarily  interested in the
results  may wish to proceed directly to
\cref{sec:qcd_flow_results}.
There, we present and discuss our numerical results for the \gls{QCD}
flow in Euclidean space. In \cref{sec:composite-two-point}, we compute
the spectral functions of the pion, the $\sigma$-meson, and the scalar
diquark directly from \gls{QCD}, with particular emphasis on
clarifying the relation between pole and curvature masses.  Finally,
\cref{sec:conclusion}, contains our conclusions and an
outlook. Technical details are collected in the appendices, including
explicit expressions for all flow equations employed in this work.

\section{Functional Renormalization Group Approach to QCD}
\label{sec:rg-invariant-formulation}

To set the stage and fix our notation, we start with briefly introducing \gls{QCD}. The gauge-fixed \gls{QCD} action is given by 
\begin{equation}\label{eq:action}
  S[A,\bar{c},c,\bar{q},q] = S_{\text{QCD}}[A,\bar{q},q]
  + S_{\text{gf}}[A] + S_{\text{gh}}[A,\bar{c},c] \; ,
\end{equation}
with the \gls{QCD} action $S_{\text{QCD}}$, the gauge fixing term $S_{\text{gf}}$ and the ghost contribution $S_{\text{gh}}$ arising from the Faddeev-Popov procedure. It describes the classical dynamics of quarks $q$, gluons $A$ and ghosts $c$. As usual, the bar denotes the Dirac-conjugate.

While we will also consider analytic continuation to Minkowski space in \cref{sec:composite-two-point}, we are focusing on equilibrium properties in this work so that our basic setup is in Euclidean space. The \gls{QCD} action then is
%
\begin{align} \label{eq:qcd_action}
  S_{\text{QCD}}[A,\bar{q},q] =
  \int_x \Big\{
    \frac{1}{4} (F_{\mu\nu}^a)^2
    + \bar{q} \left(
      \slashed{D}
      + m_q
    \right) q
  \Big \}\,,
\end{align}
%
where we use the shorthand $\int_x = \int d^4x$. We only consider the physically relevant case of three-colors $N_c=3$. As alluded to in the introduction, we focus on the $N_f=2$ light quark sector, $q=\left(u,d\right)$, as the strange quark is known to be subleading for the quantities of interest here, \cite{Alford:2007xm, Fu:2019hdw}. In addition, we only consider isospin symmetric matter. We use Euclidean $\gamma$-matrices defined such that
$\{\gamma_\mu,\gamma_\nu\}=2\delta_{\mu\nu}$ and $\gamma_\mu^\dagger = \gamma_\mu$.

The field strength tensor is given by
%
\begin{align}
  F^a_{\mu\nu} =
  \partial_\mu A^a_\nu
  - \partial_\nu A^a_\mu
  + g_s f^{abc} A^b_\mu A^c_\nu \, ,
\end{align}
%
and the covariant derivative in the fundamental representation is
%
\begin{align} \label{eq:covariant_deff}
 (D_\mu)_{\da\db} =
  \partial_\mu \delta_{\da\db}
  - i g_s (A_\mu)_{\da\db} \, ,
\end{align}
%
where we use Latin letters $a,b,c,\dots \in \{1, \dots, N_c^2-1\}$ for adjoint color indices and Fraktur letters $\da,\db,\dc,\dots \in \{1, \dots, N_c\}$ for fundamental color indices. The strong quark-gluon coupling strength is given by $g_s=\sqrt{4\pi\alpha_s}$. The gluon field
and the generators of the gauge group $SU(N_c)$ are
%
\begin{align}
  A_\mu = A_\mu^a T^a \c 
  \qquad T^a = \lambda^a/2 \c
\end{align}
%
where for $N_c=3$ the generators can be given in terms of the Gell-Mann matrices $\lambda^a$. They obey
%
\begin{align}
  [T^a,\,T^b]=i\,f^{abc}\,T^c\,,\quad
  \mathrm{Tr}(T^a\,T^b)=\frac{1}{2}\delta^{ab}\,.
\end{align}
%
The gauge is fixed by
%
\begin{equation}
  S_{\text{gf}}[A] = \frac{1}{2\xi} \int_x (\partial_\mu A^a_\mu)^2.
\end{equation}
%
Landau gauge, $\xi=0$, is used throughout this work. To ensure consistent gauge fixing in functional \gls{QCD}, ghosts are required,
%
\begin{align}
  S_{\text{gh}}[A,\bar{c},c] = - \int_x \bar{c}^a \partial_\mu D_\mu^{ab} c^b\,.
\end{align}
%
The covariant derivative in the adjoint representation is
%
\begin{align}
  D_\mu^{ab} =
  \partial_\mu \delta^{ab}
  - g_s f^{abc} A_\mu^c  \,,
\end{align}
%
with the $SU(3)$ structure constant $f^{abc}$.

\subsection{FRG with scale-dependent fields}
\label{sec:frg_general}

The \gls{FRG} implements Wilson's \gls{RG} idea to gradually include
quantum fluctuations from the \gls{UV} to the \gls{IR}
\cite{Wetterich:1992yh}. \gls{QCD} is particularly well suited for
such an approach, as asymptotic freedom provides a well-controlled
starting point at high momentum scales. The \gls{FRG} describes a
\gls{RG} flow of the scale-dependent effective action $\Gamma_k$,
which is given by a modified Legendre transformation,
\begin{align}\label{eq:Gakdef}
    \Gamma_k[\Phi] = \sup_J\bigg(J_\cfa\Phi^\cfa - \ln Z_k[J]\bigg) - \Delta S_k[\Phi] \c
\end{align}
of the scale-dependent generating functional
\begin{align}\label{eq:Zjkdef}
    Z_k[J] = \int\!\mathcal{D}\hat\Phi\, \exp\!\left(-S[\hat\Phi] - \Delta S_k[\hat\Phi] + J_\cfa \hat\Phi^\cfa_k[\hat\Phi] \right) \d
\end{align}
$\hat\Phi$ are the quantum fields, i.e., quarks, gluons and ghosts for \gls{QCD}, which are sourced by $J$. The \gls{RG} scale $k$-dependence enters through the cutoff term,
\begin{align} \label{eq:DeltaSk_def}
    \Delta S_k[\Phi] = \frac{1}{2} R_{\cfa\cfb,k} \Phi^\cfb \Phi^\cfa \d
\end{align}
Summation over bold latin letters assumes a non-trivial metric in field space including appropriate signs and Dirac conjugations for fermions, see \ccites{Pawlowski:2005xe, Rennecke:2015lur} for details. Furthermore, it entails momentum integration and summation over all internal indices.
The field space matrix $R_k$ is a regulator that suppresses all field modes with $p^2 \lesssim k^2$ through a mass-like contribution $\sim k$, while modes with $p^2 \gtrsim k^2$ are left unaltered. Our explicit choices for this regulator are specified below.

$Z_k[J]$ describes the generating functional where all ``hard modes" with $p^2 \gtrsim k^2$ are taken into account, while the ``soft modes" are decoupled. Hence, all quantum fluctuations from the deep \gls{UV} down to the \gls{IR} scale $k$ are included. So at $k=0$ the full quantum theory is resolved. In practice, it is advantageous to express this in terms of the scale-dependent effective action $\Gamma_k$ in \cref{eq:Gakdef}, where the sources $J$ are replaced by mean-fields
\begin{align}
    \Phi(x) \equiv \frac{\delta \ln Z_k[J]}{\delta J(x)}\bigg|_{J = J_{\rm sup}} = \big\langle \hat\Phi(x)\big\rangle\,.
\end{align}
$\Gamma_k$ is hence the quantum effective action that contains all quantum fluctuations down to the scale $k$. $\Gamma_0 = \Gamma$ is the full effective action.

The basic idea of applying \gls{FRG} to \gls{QCD} is to start from a large initial scale $k = \Lambda \gg 1 \GeV$, where perturbation theory is valid and the initial effective action is, essentially (this is specified below), the classical \gls{QCD} action in \cref{eq:action}. The fluctuations of increasingly soft modes are then successively included in order to finally arrive at the full \gls{QCD} effective action. In the course of this procedure, one faces all the nontrivial, nonperturbative features characteristic for \gls{QCD}, i.e., chiral symmetry breaking, confinement and the formation of bound states like hadrons. We will go though these features in more detail below, but will already point out that, while this in principle can be done solely in terms of the fundamental fields of \gls{QCD}, it is most efficient to account for the formation of bound states explicitly. This is done by also introducing sources for composite operators into the path integral. Here, we consider the lightest scalar and pseudoscalar mesons in terms of the $O(4)$ field $\phi = (\sigma, \vec{\pi})$, and the complex diquark field $\Delta$. Note that this does not imply that these are the only composite operators in the system. It simply facilitates the direct extraction of information about these operators. All the details are given in \cref{sec:qmd-truncation} and \cref{sec:four-quark-int}. 

Of course, composite operators like mesons and diquarks can only exist as dynamical degrees of freedom at low energies. Hence, the corresponding quantum fields must be considered to be scale dependent, $\hat \phi = \hat\phi_k$ and $\hat \Delta = \hat\Delta_k$. The scale dependence is determined by the quark scattering channels where these composite states emerge as resonances. The dynamical hadronization technique ensures that the transition from quark scattering to composite operators is done in an exact manner within the \gls{FRG} \cite{Gies:2001nw, Gies:2002hq, Pawlowski:2005xe,
Floerchinger:2009uf, Mitter:2014wpa, Braun:2014ata, Rennecke:2015eba, Fu:2019hdw}, see \cref{sec:four-quark-int}.

%
\begin{figure}[t]
  \centering
  \includegraphics[width=0.42\textwidth]{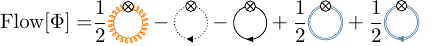}
  \caption{Diagrammatic representation of the flow contribution for the different fields considered in this work: gluons (orange curly line), ghosts (dotted
    line), quarks (solid line), mesons (double line) and diquarks (double
  line with arrow). The circle with a cross denotes the regulator insertions $\partial_t R_k$.}
  \label{fig:potential_flow_diagrams}
\end{figure}
%

By taking a $k$-derivative, we turn the path integral in
\cref{eq:Zjkdef} into a differential equation
for $\Gamma_k$. It is given by the Wetterich flow equation in the
presence of scale-dependent quantum fields \cite{Pawlowski:2005xe,
  Fu:2019hdw},

\begin{align} \label{eq:flow_eq_general}
    \partial_t \Gamma_k[\Phi] 
    + \dot\Phi^\cfa_k[\Phi] \left(\frac{\delta \Gamma_k[\Phi]}{\delta\Phi^\cfa} + h \, \delta_{\cfa\sigma}\right)
    & = \mathrm{Flow}_k[\Phi] \c
\end{align}
with the flow contribution
\begin{equation}
    \mathrm{Flow}_k[\Phi] =
    \frac{1}{2} G^{\cfa\cfb}_k[\Phi]
    \left( 
        \delta^\cfc_\cfa 
        \partial_t 
        + 2\frac{\delta \dot\Phi^\cfc_k[\Phi]}{\delta\Phi^\cfa} 
    \right)
    R_{\cfc\cfb,k}
\end{equation}
and where
\begin{align} \label{eq:propagator_def}
     G_k[\Phi] =
  \left(\Gamma^{(2)}_k[\Phi] + R_k\right)^{-1} \c
\end{align}
is the scale-dependent generalized (field-dependent) propagator. The \gls{RG}-time is defined as $t=\ln k / \Lambda$.
The $n$-point functions in momentum space are defined as
\begin{equation} \label{eq:corfunc_def}
  \Gamma^{(n)}_{\Phi_{i_1} \cdots \Phi_{i_n},k}(p_1,\dots,p_n) =
  \frac{\delta^n \Gamma_k[\Phi]}{\delta \Phi_{i_1}(p_1) \cdots \delta
  \Phi_{i_n}(p_n)} \d
\end{equation}
If only $n-1$ momentum indices are given, the momentum conserving $\delta$-function $\delta(p_1 + \cdots + p_n)$ is excluded,
\begin{equation} \label{eq:corfunc_def2}
  \Gamma^{(n)}_{\Phi_{i_1} \cdots \Phi_{i_n},k}(p_1,\dots,p_{n-1}) =
  \int_{p_n} \Gamma^{(n)}_{\Phi_{i_1} \cdots \Phi_{i_n},k}(p_1,\dots,p_{n}) \, ,
\end{equation}
with the shorthand $\int_{p} = \int \frac{d^4p}{(2\pi)^4}$.
For scale-independent fields $\dot\Phi_k=0$, \cref{eq:flow_eq_general} reduces to the conventional Wetterich equation \cite{Wetterich:1992yh}. 
The superfield $\Phi$ includes all the degrees of freedom we consider
in our truncation. As we want to provide prediction for the (pseudo-)scalar meson and diquark
correlations, we use
%
\begin{align}
  \Phi^\cfa =
  \big(A_\mu,\bar{c},c,\bar{q},q,\phi,\Delta^\dagger,\Delta\big)^\cfa \d
\end{align}
%
In \cref{eq:flow_eq_general}, the 
term $\sim h$ reflects the explicit chiral symmetry breaking from nonzero current quark masses. It ensures that chiral symmetry is manifest in the flow equation, with a linear term $\hat\sigma_k \sim \hat{\bar q} \hat q$ as the only source of chiral symmetry breaking \cite{Fu:2019hdw}. 
A diagrammatic representation of the flow equation for \gls{QCD} is shown in \cref{fig:potential_flow_diagrams}.

In order to make the \gls{RG}-invariance of the flow equation manifest, it is customary to rescale all fields and correlation function with appropriate powers of the wave function renormalizations. Note that we mean \gls{RG}-invariance in the conventional sense, i.e.\ in terms of a renormalization scale $\mu$, not in terms of the \gls{RG} scale $k$, see also \cite{Pawlowski:2005xe, Braun:2022mgx, Ihssen:2023xlp}. For the fields, this entails the we are working with renormalized fields,
\begin{align}\label{eq:phiZ}
    \Phi \rightarrow Z_{\Phi,k}^{1/2}\, \Phi\,.
\end{align}
While the renormalized fields are \gls{RG}-invariant, $\mu\frac{d \Phi}{d\mu} = 0$, they are scale dependent,
\begin{align}
    (\partial_t \Phi_k)[\Phi] \equiv - \frac{1}{2} \eta_{\Phi,k} \Phi\,,
\end{align}
with the anomalous dimension
\begin{align}\label{eq:ano_def}
  \eta_{\Phi,k} = - \partial_t \ln Z_{\Phi,k}\,.
\end{align}
The running of the fields in \cref{eq:flow_eq_general} therefore has two contributions,
\begin{align} \label{eq:field_transfo_general}
    \dot\Phi_k[\Phi] = \big\langle \partial_t \hat\Phi_k \big\rangle - \frac{1}{2} \eta_{\Phi,k} \Phi\,.
\end{align}
The first contribution stems from dynamical hadronization, and is specified in \cref{sec:four-quark-int}. It is important to note that this scale dependence arises from the fluctuating fields, not the mean fields \cite{Pawlowski:2005xe}. The last term follows from \gls{RG} invariance. 

As usual, \gls{RG}-invariant correlation functions are simply determined by the external fields,
\begin{align}\label{eq:RGinvGammas}
    \Gamma^{(n)}_{\Phi_{i_1} \cdots \Phi_{i_n},k} \rightarrow
  \frac{\Gamma^{(n)}_{\Phi_{i_1} \cdots
  \Phi_{i_n},k}}{Z_{\Phi_{i_1},k}^{1/2} \ldots Z_{\Phi_{i_n},k}^{1/2}}\,,
\end{align}
so that the effective action remains unchanged under the transformation in \cref{eq:phiZ}.
This way, the wave function renormalization factors
$Z_{\Phi,k}$ do not appear explicitly, and only the
anomalous dimensions enter the flow equations. From
the anomalous dimension $\eta_{\Phi,k}$, one can directly integrate
over $k$ to recover the wave function renormalization through
\begin{equation} \label{eq:wave_func_from_ano}
  Z_{\Phi,k} = Z_{\Phi,\Lambda}
  \exp \bigg\{ - \int_\Lambda^k \frac{dk}{k} \eta_{\Phi,k} \bigg\} \c
\end{equation}
where $Z_{\Phi,\Lambda}$ is the value of the wave function
renormalization at the ultraviolet scale $\Lambda$. Note that the
choice of $Z_{\Phi,\Lambda}$ is arbitrary and corresponds to a
given choice of global normalization for the field $\Phi$. Hence,
the physical content of the wave function renormalization is given by
its $k$-derivative, or, equivalently, by the anomalous dimension.

For later convenience, we introduce the (generalized) regulator scale derivative,
\begin{align} \label{eq:tilde_derivative_def}
    \tilde\partial_t = 
    \left(
    \delta^{\cfc}_{\cfa}
    \partial_t
    + 2 \frac{\delta \dot\Phi^\cfc_k[\Phi]}{\delta\Phi^\cfa}
    \right) R_{\cfc\cfb,k} 
    \; \frac{\delta}{\delta R_{\cfb\cfa,k}} \c
\end{align}
so that the the flow contribution may be expressed as
\begin{align}
    \mathrm{Flow}_k[\Phi] =
    \frac{1}{2}
  \tilde{\partial}_t\,\Tr \Big( \ln G_k[\Phi]^{-1} \Big) \d
\end{align}
For the field running \cref{eq:field_transfo_general}, the regulator scale derivative ensures proper insertion of the regulator, including the appropriate anomalous dimension contribution $\sim \partial_t R - \eta R$ in the one-loop diagrams.

To simplify notation, the dependence on the \gls{RG}-scale $k$ of most quantities will be left implicit in what follows.

\subsection{Key aspects of infrared QCD}
\label{sec:key_aspect_qcd}

In this section we briefly review some key aspects of low-energy \gls{QCD}, with a special attention to the properties of quark and gluon correlation functions related to confinement and chiral symmetry breaking from first principles. While the main focus of
this work is on a semi-quantitative description of the scalar
diquark, these properties play a crucial role in the construction of
our self-consistent approximation and the interpretation of our results.

At \gls{UV} energy scales, where perturbative \gls{QCD} is applicable,
the system contains free quarks and gluons. In this regime the strong
coupling is relatively weak. As the renormalization scale is lowered
toward the \gls{IR}, the coupling becomes stronger and nonperturbative effects emerge. Crucially, these include color confinement and chiral symmetry breaking. Confinement in \gls{QCD} means the absence of colored asymptotic states. Instead of individual massless gluons, \gls{QCD} features massive glueballs in the \gls{IR} \cite{Greensite:2011zz}. In gauge-fixed functional approaches, this is directly related to the behavior of the gluon and ghost propagators in the \gls{IR}. In particular, in linear covariant gauges, including the Landau gauge used here, the gluon itself has a mass gap, $m_{\text{gap}}$, while ghosts remain ungapped in the confined phase, see, e.g., \ccites{Gribov:1977wm, Kugo:1979gm, Zwanziger:1991gz, vonSmekal:1997ohs, Lerche:2002ep, Fischer:2002hna, Zwanziger:2003cf, Pawlowski:2003hq, Fischer:2006ub, Cucchieri:2007rg, Alkofer:2008jy, Aguilar:2008xm, Fischer:2008uz, Cyrol:2017ewj, Huber:2018ned, Eichmann:2021zuv, Goertz:2024dnz, Li:2026ayh}.

There are two prominent scenarios regarding the behavior of ghost and gluon propagators in the \gls{IR}. One is the Kugo-Ojima confinement criterion, which relies on canonical quantization and hence the existence of a BRST charge \cite{Kugo:1979gm}. Defining the ghost and transverse gluon dressing functions $\bar Z_{A/c}(p^2)$ via the propagators,
\begin{align}
\label{eq:Zbarphi}
    \Big[ G_{AA}(p^2) \Big]_{\mu\nu}^{ab} & =  
    \bigg(
    \frac{ \Pi^\text{T}_{\mu\nu} }{\bar Z_{A}(p^2)\, p^2}
    + \xi \frac{ \Pi^\text{L}_{\mu\nu} }{\bar Z_A^\text{L}(p^2)\,p^2}
    \bigg) \delta^{ab} 
    \c \\
    \Big[ G_{c\bar{c}}(p^2) \Big]^{ab} & = 
    \frac{1}{\bar Z_{c}(p^2)\, p^2} \delta^{ab} \c
\end{align}
with the transverse and longitudinal projection operators $\Pi^\text{T/L}_{\mu\nu}$,
the Kugo-Ojima criterion is consistent with the scaling solution,
%
\begin{align}
  \label{eq:scaling_ghost}
  \lim_{p\to0} \bar Z_c(p^2) & \propto (p^2)^\kappa \, , \\
  \label{eq:scaling_gluon}
  \lim_{p\to0} \bar Z_A(p^2) & \propto (p^2)^{-2\kappa} \,,
\end{align}
%
with scaling exponent $1/2 < \kappa < 1$ 
\citeScalingExponent. Hence, the gluon propagator vanishes in the \gls{IR}, while the ghost propagator is enhanced over the free $1/p^2$ behavior. This may be interpreted as a divergent gluon mass gap.

The other, so-called decoupling scenario involves a constant ghost dressing function and a transverse gluon dressing function that diverges exactly like $1/p^2$
\cite{Aguilar:2008xm, Boucaud:2008ky, Dudal:2008sp, Fischer:2008uz},
%
\begin{align}
  \lim_{p\to0} \bar Z_c(p^2) & \propto 1 \, , \\
  \lim_{p\to0} \bar Z_A(p^2) & \propto (p^2)^{-1} \, .
\end{align}
%
This implies that in the deep \gls{IR} the ghost propagator diverges like the tree-level propagator, while the gluon propagator tends to a constant. Global BRST symmetry is broken in this case \cite{Kugo:1979gm, Dudal:2008sp}.

It has been argued that the difference in these scenarios is related to nonperturbative gauge fixing \cite{Fischer:2008uz,Maas:2009se}: The Fadeev-Popov procedure only fixes the gauge locally, globally there are Gribov copies. Dealing with these copies involves global constraints from nonperturbative gauge fixing, which modify the boundary conditions of $n$-point functions at vanishing momenta. One may hence view these scenarios as manifestations of different resolutions of the Gribov problem.  

For our purposes, two features of these scenarios are relevant. First, they only manifest in the deep \gls{IR} and involve gapped gluons. Hence, as we will also verify explicitly below, the physics is the same regardless of which scenario is realized. Second, while the scaling solution is unique \cite{Fischer:2006vf, Fischer:2009tn}, there is a whole class of decoupling solutions. In the present work, we use a bootstrap approach within the \gls{FRG} to arrive at these solutions \cite{Cyrol:2016tym, Goertz:2024dnz, Li:2026ayh}: we supplement the longitudinal part of the gluon propagator with a running mass gap parameter, $m_{\text{gap}}$, and tune its initial value in the \gls{UV} so that we find either scaling or decoupling. This is described in detail in the next section. We emphasize that this does not imply explicit gauge symmetry breaking. On the contrary, the modification of the conventional \gls{STI} due to the gluon regulator in the gauge-fixed \gls{FRG} framework requires such a mass term \cite{Ellwanger:1994iz, Ellwanger:1996wy, Fischer:2004uk, Cyrol:2016tym}. Instead of using these \gls{mSTIs} to determine $m_{\text{gap}}$, we can simply tune its initial value $m_{\text{gap,UV}}$ such that the system flows into the desired confining solution. Unless stated otherwise, we choose the scaling solution, as its uniqueness makes it most appealing.

Chiral symmetry breaking means the formation of a chiral condensate
$\langle \bar q q\rangle$. In \gls{QCD}, this formation proceeds
through an effective four-quark interaction in the scalar channel,
which is generated by gluon exchange between quark pairs. Spontaneous
chiral symmetry breaking is signaled by a peak in this channel at zero
momentum exchange, corresponding to a soft sigma mode at zero
momentum. In case of a second-order transition, the peak is a
divergence and the sigma mode is massless.

Within the \gls{FRG}, a transparent picture emerges: at small gauge coupling $\alpha_s$ the flow of the four-quark interaction has an \gls{IR} attractive, weakly interacting fixed point. This is expected in the perturbative regime, where no large coupling should be generated dynamically. There exists, however, a critical $\alpha_s$ where this fixed point vanishes, allowing for the four-quark interaction to diverge \cite{Gies:2002hq, Braun:2011pp}. An analogous picture holds for any spontaneous symmetry breaking with a quark bilinear order parameter, and can directly be generalized also to more exotic order parameters. The possibility to construct a complete set of such four-quark interaction channels facilitates an unbiased determination of the phase structure \cite{Braun:2011pp, Braun:2019aow}.

All this entails that both confinement and chiral symmetry breaking emerge within a self-consistent setup here, without any free parameters besides the strong coupling and the the current quark masses, which we fix in the perturbative regime. The \gls{FRG} flow for \gls{QCD} describes the scale evolution from the perturbative \gls{UV} regime to the nonperturbative \gls{IR} regime. Owing to the mass gap, gluons will decouple at low energies while quarks form hadronic bound states due to strong correlations. The soft pseudo-Goldstone bosons from chiral symmetry breaking will eventually take over as dominant degrees of freedom in the \gls{IR}. All features of \gls{QCD} at low energies emerge dynamically, without any free parameters. For the low-energy sector, this is intimately related to the intermediate fixed point mentioned above \cite{Gies:2002hq, Braun:2009ewx, Mitter:2014wpa, Braun:2014ata, Rennecke:2015eba}. This also includes the parameters of intermediate, non-asymptotic states like diquarks, which we extract here. In the following, we discuss the details of this setup and the approximations we use.

\section{Truncation Scheme}\label{sec:truncation}

An exact solution to Wetterich equation requires evaluating infinite
tower of coupled flow equations for all possible correlation functions
of the system. As such, this tower must be truncated, ensuring that
the most relevant physical processes for the quantities we are
interested in are captured for reliable results. It is convenient to
express this in terms of an effective action $\Gamma_k$, where we only
write down explicitly the terms related to the correlations that we
keep in our truncation. It is important to emphasize that this does
not mean that it is assumed that other correlations are zero, only
that their feedback into the quantities of interest is small, so that
we can neglect it.

Building on previous \gls{FRG} studies \cite{Gies:2002hq,
  Braun:2009ewx, Mitter:2014wpa, Braun:2014ata, Rennecke:2015eba,
  Cyrol:2016tym, Fu:2019hdw, Goertz:2024dnz}, we decompose the
effective action into three distinct parts,
%
\begin{align} \label{eq:full-truncation}
  \nonumber
  \Gamma_k[\Phi] = &
  \Gamma_{\text{glue},k}[A,\bar{c},c]
  + \Gamma_{\text{quark-gluon},k}[A,\bar{q},q] \\[2ex] &
  + \Gamma_{\text{qmd},k}[A,\bar{q},q,\phi,\Delta^\dagger,\Delta] \; ,
\end{align}
%
which describe the gluonic, quark-gluon, and composite low-energy
sectors of \gls{QCD}, respectively.  Here, $\Gamma_{\text{glue},k}$
denotes the pure glue contribution involving only ghosts and gluons,
$\Gamma_{\text{quark-gluon},k}$ contains the quark kinetic term
together with quark-gluon and four-quark interactions, and
$\Gamma_{\text{qmd},k}$ encodes our truncation of the composite
low-energy sector of \gls{QCD}.

\subsection{Glue sector}
\label{sec:glue-truncation}

In the glue sector, we follow \ccite{Goertz:2024dnz} and employ a
zero-momentum expansion around the gauge-fixed classical tensor
structure, providing a simple semi-quantitative truncation of gluonic
fluctuations. To this end, we define the classical tensor structure
derived from the classical action, \cref{eq:qcd_action}, as
\begin{equation} \label{eq:cl_tensor_struc_def}
  \mathcal{T}_{\Phi_{i_1}\dots\Phi_{i_n}}^{(1)}(p_1, \dots, p_n) =
  \left.
  \frac{\delta^{n} S[\Phi]}{\delta\Phi_{i_1}(p_1) \dots \delta\Phi_{i_n}(p_n)}
  \right|_{g_s=1} \; .
\end{equation}
The explicit expressions are given in \cref{sec:diagramatic-vertices}.
Here, the superscript $(1)$ indicates that the corresponding tensor
structure is the first element of a complete tensor basis in the
respective interaction channel. We reserve the number one for the
classical tensor structure defined through the classical action.  Both
\gls{FRG} and \gls{DSE} studies have investigated the impact of
additional tensor structures and higher-order tensor contributions,
see, e.g., \ccites{Mitter:2014wpa, Eichmann:2014xya,Cyrol:2014kca, Cyrol:2016tym, Corell:2018yil,
  Huber:2018ned, Huber:2020keu}. The resulting
truncation of the effective action is then given by
%
\begin{widetext}
  \begin{equation} \label{eq:glue-truncation}
    \begin{aligned}
      & \Gamma_{\text{glue},k}[A, \bar{c}, c] =
      \frac{1}{2} \int_p A_\mu^a(p) \left[
        \left( p^2 + m^2_{\text{gap}} \right) \Pi_{\mu\nu}^\text{T}(p) +
        \frac{1}{\xi} (p^2 + m^2_{\text{mSTI}}) \Pi_{\mu\nu}^\text{L}(p)
      \right] A_\nu^a(-p)                        \\[2ex] & \quad
      + \lambda_{A^3} \frac{1}{3!} \int_{p_1p_2} 
      \Big[\mathcal{T}_{A^3}^{(1)}(p_1,p_2)\Big]_{\mu_1\mu_2\mu_3}^{a_1a_2a_3}
      \prod_{i=1}^3 A_{\mu_i}^{a_i}(p_i)
      + \lambda_{A^4} \frac{1}{4!} \int_{p_1p_2p_3} 
      \Big[\mathcal{T}_{A^4}^{(1)}(p_1,p_2,p_3)\Big]_{\mu_1\mu_2\mu_3\mu_4}^{a_1a_2a_3a_4}
      \prod_{i=1}^4 A_{\mu_i}^{a_i}(p_i)         \\[2ex] & \quad
      + \int_p \bar{c}^a(p) p^2 \delta^{ab} c^b(-p)
      + \lambda_{A\bar{c}c} \int_{p_1p_2}
      \Big[\mathcal{T}_{c\bar{c}A}^{(1)}(p_1,p_2)\Big]_{\mu}^{a_1a_2a_3}
      \bar{c}^{a_2}(p_2) c^{a_1}(p_1) A_\mu^{a_3}(-p_1 -p_2) \; .
    \end{aligned}
  \end{equation}
\end{widetext}
%
The first line includes the transverse and longitudinal part of the
gluon kinetic term, as indicated by their associated projectors
\begin{align}
  \Pi_{\mu\nu}^\text{T}(p) & =
  \delta_{\mu\nu} - \frac{p_\mu p_\nu}{p^2} \; , \\[2ex]
  \Pi_{\mu\nu}^\text{L}(p) & =
  \delta_{\mu\nu} - \Pi_{\mu\nu}^\text{T}(p)
  = \frac{p_\mu p_\nu}{p^2} \; .
\end{align}
The second line of \cref{eq:glue-truncation} includes the three and
four gluon interactions, while the last line contains the ghost
kinetic term and the ghost-gluon coupling characteristic of a
gauge-fixed setup. It is crucial to distinguish between the different
strong-coupling tensor structures, as the degeneracy of the associated
couplings is lifted by non-perturbative effects. In the following, we
refer to these couplings as \textit{avatars} of the strong coupling
$\alpha_s$.  In the glue sector, they are directly related to
$\lambda_{A^3,k}$, $\lambda_{A^4,k}$ and $\lambda_{A\bar{c}c,k}$.

As discussed in \cref{sec:key_aspect_qcd}, a key part of the
truncation is the inclusion of a running transverse gluon mass
gap. The modifications of the \gls{STI} due to the explicit gauge
symmetry breaking of the regulator give rise to a nonzero longitudinal
gluon mass parameter $m_{\text{mSTI}}$ \cite{Ellwanger:1994iz}. At
$k=0$ the regulator vanishes and the \gls{mSTIs} reduce to the
conventional \gls{STI}. These state that the longitudinal gluon
propagator is fully determined by the gauge fixing term, with no
quantum corrections. Hence, $m_{\text{mSTI}}$ needs to vanish in the
\gls{IR}.

In the perturbative regime, the \gls{mSTIs} are identical for transverse and longitudinal couplings. Yet, as pointed out in \cref{sec:key_aspect_qcd}, confinement is related to a nonzero mass gap $m_{\text{gap}}$ at $k=0$ in the gluon propagator in Landau gauge. So while the appearance of a gluon mass parameter $m_{\text{gap,UV}} = m_{\text{mSTI,UV}}$ in the \gls{UV} is a direct consequence of gauge invariance in the presence of the momentum cutoff of the \gls{FRG}, the running of $m_{\text{gap}}$ and $m_{\text{mSTI}}$ need to become different below the confinement scale, $k<k_{\text{conf}}$, in order to generate confinement. This is why in the first line of \cref{eq:glue-truncation}, we included both mass parameters. The difference between these masses should be generated dynamically through irregular vertices in the nonperturbative regime \cite{Cyrol:2016tym}.

In the present bootstrap approach, we circumvent the need to generate the necessary irregularities by simply tuning the initial value of $m_{\text{gap}}$ so that the system runs into the confining scaling solution discussed in \cref{sec:key_aspect_qcd}. The flow of $m_{\text{gap}}$ is readily extracted from the flow of the gluon two-point function, see \cref{sec:strong_couplings}. Furthermore, we use Landau gauge, $\xi = 0$. In this case, the transverse system is closed, so information on the longitudinal contributions is not required. Hence, there is no need to explicitly deal with the \gls{mSTIs}, which in general involve both transverse and longitudinal contributions, here. Note that the agreement between the avatars of the strong couplings shown in \cref{fig:strong_coupling_avatars} suggests that these identities, and hence gauge symmetry, is respected at least at higher energies here.

Depending on the choice of $\mgapUV^2$, three distinct regimes are found relative to the initial value that leads to the scaling solution, $\mgapUV^2 = \mgapscal^2$ \cite{Cyrol:2016tym}:
\begin{enumerate}
    \item \textit{Higgs regime}: for $\mgapUV^2 \gg \mgapscal^2$ the high effective gluon mass suppresses non-perturbative effects in the glue sector, and the gluon propagator behaves like the one of an ordinary massive vector field. Furthermore, the theory displays no \gls{DCSB} or any of the non-trivial structure of \gls{IR} \gls{QCD}.
    \item \textit{confining regime}: for $\mgapUV^2 \gtrsim \mgapscal^2$ the non-perturbative effects in the glue sector are non-negligible, and lead to the appearance of a peak at a finite momentum in the gluon propagator \cite{Cyrol:2016tym}. This regime corresponds to physical \gls{QCD}, and depending on the exact value of $\mgapUV^2$, scaling or decoupling solutions are found, also featuring \gls{DCSB}.
    \item \textit{Landau pole regime}: for $\mgapUV^2 < \mgapscal^2$ a Landau pole appears for a finite \gls{RG}-scale $k>0$ at which the strong coupling avatars diverge and the ghost wave function renormalization $Z_{c}$ tends to 0. As such the \gls{IR} cannot be reached in this regime. 
\end{enumerate}
For further details we refer to \ccites{Cyrol:2016tym,Goertz:2024dnz} and \cref{sec:indep_params_uv}. 

Finally, we note that the gluon mass parameter $m_\text{gap}$ should not be identified with the physical gluon mass gap, which is clear in various ways. For instance, $m_\text{gap}$ is not an \gls{RG}-invariant quantity and depends on the infrared details of the gluon dynamics, see \cref{sec:indep_params_uv}, two properties that are not wanted in the definition of the gluon mass gap. As a proxy to the gluon mass gap, we follow \ccite{Goertz:2024dnz}, and use the peak location of the gluon dressing function. For a recent discussion on the gluon mass gap see \ccite{Ferreira:2025tzo}.

\subsection{Quark-gluon sector}
\label{sec:quark-gluon-truncation}

In this section, we introduce the quark fields and the interactions relevant for the emergence of the mesonic and diquark states of interest here. This part of the truncation hence forms the link between the high-energy
description in terms of quarks and gluons and the low-energy
description in terms of quarks and composites.

The contribution to the effective action is given by
\begin{widetext}
  \begin{align} \label{eq:quark-gluon-trunc}
    \nonumber
    \Gamma_{\text{quark-gluon},k}[A,\bar{q},q] = &
    \int_p \bar{q}(-p)\, i\slashed{p}\, q(p)
    - \sum_{i = 1,4}
    \lambda_{A\bar{q}q}^{(i)}
    \int_{p_1 p_2}
    \left[ \mathcal{T}_{A\bar{q}q}^{(i)}(p_1,p_2) \right]^{a}_{\mu\alpha\beta}
    A^a_\mu(p_1) \bar{q}_\alpha(p_2) q_\beta(-p_1-p_2)
    \\[2ex] &
    + \sum_{i=\text{sps},\text{csc}} 
    \lambda_{\qbqbqq}^{(i)}
    \frac{1}{4}
    \int_{p_1p_2p_3}
    \left[ \mathcal{T}_{\qbqbqq}^{(i)}(p_1,p_2,p_3)
    \right]_{\alpha\beta\gamma\delta}
    \bar{q}_\alpha(p_1) \bar{q}_\beta(p_2) q_\gamma(p_3) q_\delta(-p_1-p_2-p_3)
    \; ,
  \end{align}
\end{widetext}
where the first line consist of the quark kinetic term and the quark-gluon
interaction. A complete basis for the quark-gluon interaction is
given by eight tensor structures, three of which are relevant for
quantitative accuracy \cite{Cyrol:2017ewj,Gao:2021wun}. In this work
we retain  the  two dominant tensor structures
\begin{align} \label{eq:classical_tensor_struct_Aqq}
  \left[ \mathcal{T}_{A\bar{q}q}^{(1)}(p_1,p_2) \right]^{a}_{\mu}
  = & i \gamma_\mu T^a \c \\[2ex]
\label{eq:4_tensor_struct_Aqq}
  \left[ \mathcal{T}_{A\bar{q}q}^{(4)}(p_1,p_2) \right]^{a}_{\mu}
  = & \big( \slashed{p}_2 + \slashed{p}_3 \big) \gamma_\mu T^a \c
\end{align}
with the associated dressing functions $\lambda_{A\bar{q}q}^{(1)}$
  and $\lambda_{A\bar{q}q}^{(4)}$. The remaining quantitatively
  important tensor structure, conventionally denoted by
  $\mathcal{T}^{(7)}_{A\bar{q}q}$, is known to induce corrections of
  approximately $10\%$ to the quark mass
  \cite{Fu:2025hcm,Gao:2021wun}. Since this effect is within our
  estimated systematic uncertainty, we neglect it in the present
  study.  Through the quark–gluon interaction, the matter sector
  couples directly to the gluonic sector. Towards lower RG scales, the
  non-perturbative quark-gluon interaction generates strong effective
  four-quark interactions that drive \gls{DCSB} and the formation of
  bound-states, as mentioned in \cref{sec:key_aspect_qcd}.

  Accordingly, we include in the second line a sum of four-quark
  interaction channels. For two-flavor \gls{QCD}, and assuming a
  broken chiral symmetry, a Fierz complete basis invariant under
  $U(1)_V \times SU(2)_V$ for these channels consists of 10 elements
  \cite{Mitter:2014wpa}. In this work we focus on two four-quark
  channels.  First, we include the \gls{SPS} channel ($i=\text{sps}$)
  which is known to be the dominant channel at low densities, and is
  capital to properly capture \gls{DCSB}. Furthermore, an
  approximation using only this channels is known to give accurate
  results on the phase structure even at finite temperature and
  moderate baryon chemical potential \cite{Braun:2014ata,
    Rennecke:2015eba, Fu:2019hdw, Braun:2019aow}. In addition, we
  consider the \gls{CSC} channel ($i=\text{csc}$) required to access
  properties of the scalar diquark, which is the main focus of this
  work. This interaction channel is commonly used in \gls{LEM} such as
  in the \gls{NJL} model \cite{Buballa:2003qv} or, after bosonization,
  in the \gls{QMD} model
  \cite{Andersen:2024qus,Andersen:2025ezj,Gholami:2025afm,Andersen:2026xrf,Mire:2026auc}. It is also expected to be the dominant channel for
  two-flavor \gls{QCD} at high densities \cite{Braun:2019aow}.

The tensor structure of the \gls{SPS} and \gls{CSC} channels, $\mathcal{T}_{\qbqbqq}^{(\text{sps})}$ and $\mathcal{T}_{\qbqbqq}^{(\text{csc})}$, follow from the interaction channels
\begin{align}
  \label{eq:sps_lagrangian}
  \mathcal{L}^{(\text{sps})} & =
  (\bar{q}q)^2 - (\bar{q} \gamma_5 \vec{\tau} q)^2 \; , \\[2ex]
  \label{eq:csc_lagrangian}
  \mathcal{L}^{(\text{csc})} & = \frac{1}{2}
  \big( q^\tp C \gamma_5 i\epsilon_\da \tau_2 q \big)
  \big( \bar{q} \gamma_5 i\epsilon_\da \tau_2 C \bar{q}^\tp \big) \; ,
\end{align}
and are given explicitly in \cref{sec:diagramatic-vertices}. In the
\gls{SPS} channel we introduced the Pauli matrices
$\vec{\tau}$ in flavor space, while in the \gls{CSC}
channel we introduced the charge conjugation matrix
$C=\gamma_2\gamma_4$ and the epsilon tensor acting in color space
$(\epsilon_\da)_{\db\dc} = \epsilon_{\da\db\dc}$. The charge
conjugation matrix $C$ has the following properties
\begin{equation}
  C \gamma_\mu^\tp C = \gamma_\mu
  \quad \text{and} \quad
  C^\tp = C^{-1} = -C^\dagger = -C \; ,
\end{equation}
and ensures that the pairing interaction \cref{eq:csc_lagrangian} is
Lorentz invariant. Similarly, the epsilon tensor $\epsilon_\da$
ensures invariance under $SU(3)_c$ gauge transformations.

As explained in more detail in \cref{sec:four-quark-int}, the emergence of mesons and diquarks in the four-quark scattering described by \cref{eq:quark-gluon-trunc} is
made explicit through dynamical hadronisation. This procedure
then continuously generates the low-energy sector of \gls{QCD} in the
form of bosonic fields carrying the quantum numbers of the associated four-quark
interactions, in our case the sigma, pion and scalar diquark fields.

\subsection{Composite sector}
\label{sec:qmd-truncation}

The last part of our truncation consist of the composite sector of
\gls{QCD}, including here the sigma, $\sigma$, pions, $\vec{\pi}$ and
the scalar diquark, $\Delta$. The sigma and pion are understood as a
quark-antiquark bilinears of the form
\begin{align}
  \sigma \sim \bar{q} q
  \qquad \text{and} \qquad
  \vec\pi \sim \bar{q} \vec\tau \gamma_5 q \; ,
\end{align}
and are conveniently combined into the $O(4)$ chiral field
$\phi=(\sigma,\vec\pi)$, while the diquark field corresponds
to a quark-quark composite of the form
\begin{equation} \label{eq:diquark-comp}
  \Delta_\da \sim q^\tp C \gamma_5 i \epsilon_\da \tau_2 q \; .
\end{equation}
The effective action of this \gls{QMD} part is
\begin{widetext}
  \begin{align} \label{eq:qmd_truncation}
    & \nonumber
    \Gamma_{\text{qmd},k}[A,\bar{q},q,\phi,\Delta^\dagger,\Delta] =
    \int_x \bigg\{
      \frac{1}{2} (\partial_\mu \phi) (\partial_\mu \phi)
      +  (\partial_\mu \Delta)^\dagger (\partial_\mu \Delta)
      + U_k(\rho_\phi, \rho_\Delta) - h \sigma
      \\[2ex] & \quad
      + g_{\phi\bar{q}q} \bar{q} \left(
        \sigma + i \gamma_5 \vec{\pi} \cdot \vec{\tau}
      \right) q \nonumber
      + \frac{1}{2} g_{\Delta q q} \left(
        \Delta_\mathfrak{a} \bar{q} \gamma_5 \tau_2 i
        \epsilon_\mathfrak{a} C \bar{q}^\tp
        - \Delta_\mathfrak{a}^* q^\tp C \gamma_5 \tau_2 i
        \epsilon_\mathfrak{a} q
      \right)
      \\[2ex] & \quad
      - i \lambda_{A \Delta^\dagger \Delta}
      A_\mu^a \left[
        (\partial_\mu \Delta^\tp) T^a \Delta^*
        - \Delta^\tp T^a (\partial_\mu \Delta^*)
      \right]
      + \lambda_{A^2 \Delta^\dagger \Delta}
      A_\mu^a A_\mu^b \Delta^\tp T^a T^b \Delta^*
    \bigg\} \; ,
  \end{align}
\end{widetext}
In the first line, we introduced the meson and diquark kinetic terms, the
effective potential $U_k(\rho_\phi, \rho_\Delta)$ and an explicit
chiral symmetry breaking term $-h\sigma$, which can be directly
mapped onto the non-vanishing current quark mass of (degenerate) up and down quarks. The effective potential
is kept as an arbitrary function of the chiral invariant
\begin{equation}
  \rho_\phi = \frac{1}{2} \phi^2
  = \frac{1}{2} \left( \sigma^2 + \vec{\pi}^2 \right) \; ,
\end{equation}
and the diquark invariant
\begin{equation}
  \rho_\Delta = |\Delta|^2
  = \Delta^\dagger \Delta
  = \sum_{\da=1}^{N_c} \Delta_\da^* \Delta_\da \; .
\end{equation}
It includes all possible zero-momentum meson-meson, meson-diquark and
diquark-diquark interactions consistent with \gls{QCD} symmetries. A
selection of the most relevant interactions used in this work can be found in
\cref{sec:diagramatic-vertices}.

The second line includes the standard quark-meson and quark-diquark
interactions. The exact form of the interactions directly follows
from bosonizing the four quark interactions in
\cref{eq:quark-gluon-trunc}. By using the
dynamical hadronization procedure detailed in
\cref{sec:four-quark-int}, the duality between four-quark interactions in certain channels and fundamental meson or diquark fields is rigorously taken into account, especially avoiding  any double counting.

We see that the presence of a
non-vanishing condensate for the $\sigma$ meson directly leads to a nonzero finite quark mass
\begin{equation}
  m_q = g_\phi \langle \sigma \rangle \; .
\end{equation}
The presence of finite condensate for the diquark field, for example
in the $3$ direction, would lead to a gap around the
Fermi-surface $\Delta_{\text{gap}} = g_\Delta \langle\Delta_3\rangle$
\cite{Alford:2007xm}. While there is no such condensate in the \gls{QCD} vacuum, it can appear at large chemical potential \cite{Braun:2019aow}.

The last line on \cref{eq:qmd_truncation} corresponds to the
diquark-gluon interactions and follows from the kinetic term of the
diquark field expressed in terms of covariant derivatives. Under an
$SU(3)_c$ gauge transformation $q \to U q$, it follows from
\cref{eq:diquark-comp} and the relation
$U^\tp \epsilon_\da U = U^*_{\da\db} \epsilon_\db$ that the diquark
transforms as an anti-triplet and the anti-diquark as a triplet, see
also \cite{Mire:2026auc}, i.e.,
\begin{equation}
  \Delta^* \to U \Delta^*
  \quad \text{and} \quad
  \Delta^\tp \to \Delta^\tp U^\dagger \; .
\end{equation}
As such the covariant diquark kinetic term is
\begin{align}
  \nonumber
  & (D_\mu \Delta^*)^\dagger(D_\mu \Delta^*) =
  (\partial_\mu \Delta)^\dagger (\partial_\mu \Delta)
  + g_s^2 \Delta^\tp A_\mu A_\mu \Delta^*
  \\[2ex] & \quad
  - i g_s
  \left[
    (\partial_\mu \Delta^\tp) A_\mu \Delta^*
    - \Delta^\tp A_\mu (\partial_\mu \Delta^*)
  \right] \; ,
\end{align}
which directly leads to the tensor structure of the gluon-diquark
interactions in \cref{eq:qmd_truncation}, where we introduced two
additional strong coupling avatars $\lambda_{A\Delta^\dagger\Delta}$
and $\lambda_{A^2\Delta^\dagger\Delta}$.

Finally, we note that the interaction term proportional to
$\lambda_{A^2\Delta^\dagger\Delta}$ is responsible for the appearance
of a Higgs mass $m^2_{\text{Higgs}}$ for the gluon field in presence
of a finite diquark condensate.

\section{Correlation Functions}
\label{sec:correlation_function}

\begin{figure*}[!t]
  \centering
  \includegraphics[width=0.79\linewidth]{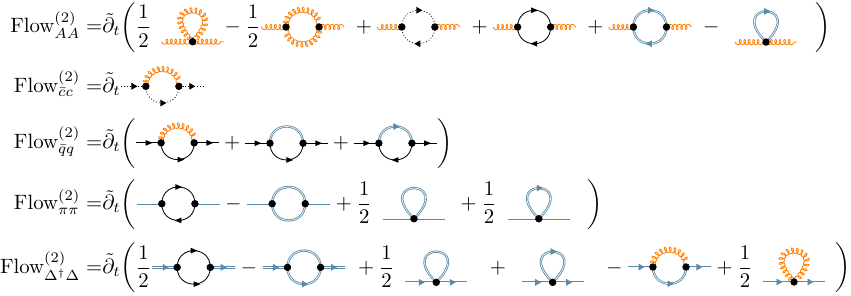}
  \caption{Diagrammatic representation of the flow contribution to
  the two-point function of the gluon (orange curly line), ghost (dotted line), quark (black solid line), meson (blue double line) and diquark fields (blue double line with an arrow). The black dots stand for the full vertex contributions originating from higher order flow equations. The tilde-derivative $\tilde{\partial}_t$ implies an \gls{RG}-time derivative performed only on the regulators and includes an anomalous dimension insertion, see \cref{eq:tilde_derivative_def}.}
  \label{fig:two_point_flows}
\end{figure*}

In this section we discuss the different correlation functions that
we will consider and specify the projections employed for the
different couplings in \cref{eq:full-truncation}. For this, it will
prove useful to introduce the flow contribution defined as
\begin{align} \label{eq:flow_def}
  & \nonumber \text{Flow}^{(n)}_{\Phi_{i_1}\cdots \Phi_{i_n}}(p_1,\cdots,p_n) =
  \\[2ex] & \qquad \qquad \qquad \qquad
  \left. \frac{\delta^n \text{Flow}[\Phi]}{\delta \Phi_{i_1}(p_1)
  \cdots \delta \Phi_{i_n}(p_n)}  \right|_{\Phi_c} \; ,
\end{align}
where $\Phi_c$ stands for a constant field configuration where every
field vanishes, expect for the sigma field, which assumes a spatially
homogeneous value $\sigma(x) = \sqrt{2\rho_\phi}$, with $\rho_\phi
\in \mathbb{R}$. This can be understood as the diagrammatic contributions
resulting from the flow equation \cref{eq:flow_eq_general}, see for
example \cref{fig:two_point_flows}.

Furthermore, we will formulate all flows in terms of dimensionless
couplings, rescaled by an appropriate factor of the \gls{RG}-scale $k$. The
dimensionless couplings will be denoted by a bar, for example
$\bar{m}_{\text{gap}}^2 = m^2_{\text{gap}} / k^2$ and, accordingly,
we define the dimensionless flow as
\begin{align} \label{eq:dimless_flow_def}
  & \nonumber
  \overline{\text{Flow}}^{(n)}_{\Phi_{i_1}\cdots
  \Phi_{i_n}}(p_1,\cdots,p_n) = \\[2ex] & \quad
  k^{-4} k^{d_{\Phi_{i_1}}} \cdots k^{d_{\Phi_{i_n}}}
  \text{Flow}^{(n)}_{\Phi_{i_1}\cdots \Phi_{i_n}}(p_1,\cdots,p_n) \; ,
\end{align}
with $d_{\Phi_i}$ the canonical dimension of the field $\Phi_i$.
Similarly to \cref{eq:corfunc_def} we sometimes omit one of the
momentum argument, see \cref{eq:corfunc_def2}. When no momentum
arguments are given, it is assumed that the flow is evaluated at
vanishing momenta, i.e,
\begin{align}
  & \text{Flow}^{(n)}_{\Phi_{i_1}\cdots \Phi_{i_n}} \equiv
  \text{Flow}^{(n)}_{\Phi_{i_1}\cdots \Phi_{i_n}}(0,\cdots,0) \d
\end{align}
Lastly, we note that the analytic expression of all flows discussed in the following can be found in \cref{app:zero-momentum-flows}.

\subsection{Anomalous dimensions}
\label{sec:anomalous_dimensions}

In the \gls{RG}-invariant formulation introduced in
\cref{sec:rg-invariant-formulation}, the anomalous dimensions are determined by imposing that all fields are canonically normalized at each \gls{RG}-scale $k$. This translates to
the following set of constraints on the scale-dependent two-point functions
\begin{align}
  \label{eq:rg_invariance_cond_A}
  \frac{1}{6}\, \lim_{p\to0}  \partial_p^2 \, \Pi_{\mu\nu}^\mathrm{T}(p)
  \left[\Gamma_{AA}^{(2)}(p)\right]_{\mu\nu}^{ab}
  & \overset{!}{=} \delta^{ab} \; , \\[2ex]
  - \frac{1}{2}\, \lim_{p\to0}  \partial_p^2 \left[ \Gamma_{\bar{c}c}^{(2)}(p) \right]^{ab}
  & \overset{!}{=} \delta^{ab} \; , \\[2ex]
  - \frac{1}{8N_cN_f}\, \lim_{p\to0} \partial_p^2 \, \tr\left[ i\slashed{p} \,
  \Gamma_{\bar{q}q}^{(2)}(p) \right]
  & \overset{!}{=} 1 \; , \\[2ex]
  \frac{1}{2}\, \lim_{p\to0} \partial_p^2 \left[\Gamma_{\phi\phi}^{(2)}(p)\right]_{ij}
  & \overset{!}{=} \delta_{ij} \; , \\[2ex]
  \label{eq:rg_invariance_cond_Delta}
  \frac{1}{2}\, \lim_{p\to0} \partial_p^2
  \left[\Gamma_{\Delta^*\Delta}^{(2)}(p)\right]_{\da\db}
  & \overset{!}{=} \delta_{\da\db} \; ,
\end{align}
where we only considered the transverse component of the gluon
two-point function as we work in Landau gauge.

Inserting the constraints
\crefrange{eq:rg_invariance_cond_A}{eq:rg_invariance_cond_Delta} into
the \gls{RG}-invariant formulation of the flow equation
\cref{eq:flow_eq_general} we obtain the following expression for
the anomalous dimensions of the different fields
\begin{align}
  \nonumber
  \eta_A & = - \frac{1}{6(N_c^2 - 1)}
  \\ & \quad \times \label{eq:etaA_proj}
  \lim_{p\to0} 
  \delta_{ab}
  \frac{\partial^2}{\partial p^2}
  \Pi_{\mu\nu}^{\mathrm{T}}(p)
  \left[ \text{Flow}_{AA}^{(2)}(p) \right]^{ab}_{\mu\nu} \; ,
\end{align}
\begin{align}
  \label{eq:etac_proj}
  \eta_c & = \frac{1}{2(N_c^2 - 1)} \lim_{p\to0}
  \delta_{ab} \frac{\partial^2}{\partial p^2}
  \left[ \text{Flow}^{(2)}_{\bar{c}c}(p) \right]^{ab} \; ,
\end{align}
\begin{align}
  \label{eq:etaq_proj}
  \eta_q & = -\frac{i}{8N_fN_c}
  \lim_{p\to0} \frac{\partial^2}{\partial p^2}
  \Tr \left[ \slashed{p} \; \text{Flow}_{\bar{q}q}^{(2)}(-p) \right] \; ,
\end{align}
\begin{align}
  \label{eq:etaphi_proj}
  \eta_\phi & = -\frac{1}{2(N_f^2 - 1)}
  \lim_{p\to0} \delta_{ij} \frac{\partial^2}{\partial p^2}
  \left[ \text{Flow}_{\pi\pi}^{(2)}(p) \right]_{ij} \; ,
\end{align}
\begin{align}
  \label{eq:etaDelta_proj}
  \eta_\Delta & = -\frac{1}{2 N_c}
  \lim_{p\to0} \delta_{\da\db} \frac{\partial^2}{\partial p^2}
  \left[ \text{Flow}_{\Delta^\dagger\Delta}^{(2)}(p) \right]_{\da\db} \; .
\end{align}
The diagrammatic representations of the different flow contributions for
the two-point functions are shown in \cref{fig:two_point_flows}.

\begin{figure*}[!t]
  \centering
  \includegraphics[width=1\linewidth]{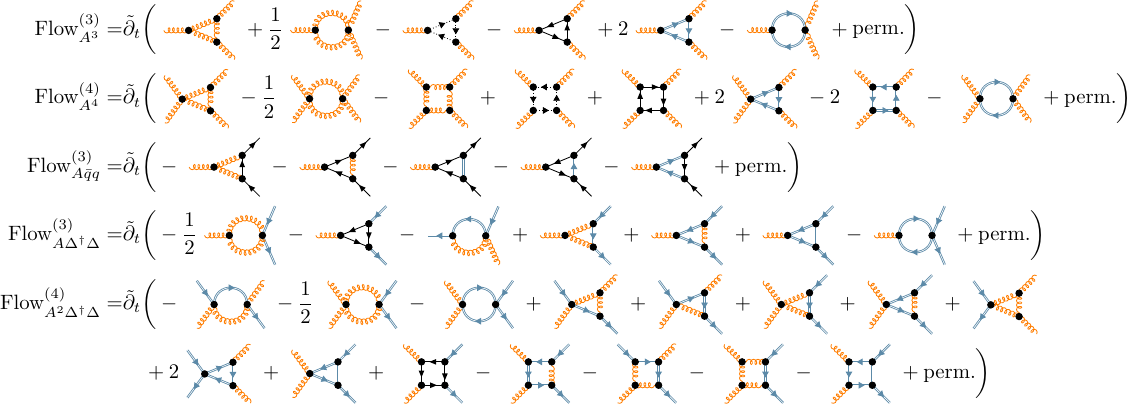}
  \caption{Diagrammatic representation of the flows of the different strong coupling avatars. Additional diagrams obtained
  by permuting external legs are implied.}
  \label{fig:strong_coupling_flows}
\end{figure*}

Although the anomalous dimension only takes into account the first correction in a small-momentum expansion, it is known to provide quantitatively accurate results for many systems. In particular, in \ccite{Goertz:2024dnz} it was shown to provide a good approximation to \gls{QCD}, for example providing a correct description of the gluon dressing function. Furthermore, in presence of a gluon mass parameter this approximation was shown to be sufficient in capturing both the scaling and decoupling solutions. We also note that the anomalous dimension of the composite fields plays an essential roles in the dynamical hadronization procedure as we will show in \cref{sec:numerical-results-composites}.

For the mesonic sector, we do not distinguish between the sigma and pion anomalous dimensions, and employ the pion two-point function to fix the common anomalous dimension $\eta_\phi \equiv \eta_\pi = \eta_\sigma$. This approximation has been repeatedly confirmed to be reliable in the low density regime, see, e.g., \ccites{Yin:2019ebz, Fu:2019hdw}. This is expected, because pions are the most relevant dynamical degrees of freedom at low energies.

In the infrared, the gluon anomalous dimension diverges
\cite{Cyrol:2016tym}. 
This is presumably an artifact of the present bootstrap approach. It is, however, inconsequential, as this only occurs in the deep \gls{IR}, where the gluon propagator of the scaling solution is dominated by the diverging gap parameter \cite{Goertz:2024dnz}.
To deal with this, we modify
the flow of the anomalous dimension as
\begin{equation} \label{eq:gluon_ano_freeze}
  \eta_A \to \eta_A \; k^{n-1} \big(
    k^n + k_{\text{conf}}^n
  \big)^{\frac{1-n}{n}} \c
\end{equation}
with $k_\text{conf}=1\GeV$ and $n=10$. This procedure ensures that for $k \lesssim
k_{\text{conf}}$ the gluon anomalous dimension remains constant. This does not affect the \gls{IR} observables, because the gluons are already decoupled. 
We checked explicitly that our results are independent of the exact choice of $n$ and $k_\text{conf}$. However, $k_\text{conf}$ must be chosen above the \gls{RG} scale where the gluon anomalous dimension diverges, i.e., $k \gtrsim 350 \MeV$, and below the gluon decoupling scale $k_\text{gap} \lesssim 1.2 \GeV$; see \cref{sec:qcd_flow_results}.

\subsection{Strong-coupling avatars and gluon mass gap}
\label{sec:strong_couplings}

In this section, we extend the previous discussion to the remaining
couplings related to the gauge sector. For the two-point functions, it
only remains to specify the flow of $m_{\text{gap}}$, which is found
by the $k$-dependence of the transverse part of the gluon two-point
function at vanishing momentum,
\begin{align}
  & \nonumber
  \partial_t \bar{m}_{\text{gap}}^2 = (-2 + \eta_A)
  \bar{m}^2_{\text{gap}} \\[2ex] & \quad
  + \frac{1}{3(N_c^2 - 1)} \lim_{p\to 0} \delta^{ab} \Pi_{\mu\nu}^\text{T}(p)
  \left[ \overline{\text{Flow}}^{(2)}_{A^2}(p) \right]^{ab}_{\mu\nu} \; .
\end{align}
This flow equation plays a crucial role in the infrared scaling behavior.

Next, we come to the strong couplings. As discussed earlier, the avatars of the strong coupling become degenerate in the perturbative regime, where they all take the same value. As they are evolved toward the infrared, nonperturbative quantum fluctuations drive them to exhibit different behaviors. The diagrammatic representations of the different contributions to the flow that we take into account can be found in \cref{fig:strong_coupling_flows}. 

For the three and four gluon couplings, we use their associated
classical tensor structure to perform the projection, which yields
\begin{align} \label{eq:three-gluon-proj}
  & \nonumber
  \partial_t \lambda_{A^3} = \frac{3}{2} \eta_A \lambda_{A^3}
  + \frac{i}{12 N_c(N_c^2 - 1)} \\[2ex] & \quad \times
  \lim_{p\to0}
  \frac{\partial^2}{\partial p^2}  f^{abc} \delta_{\mu\nu} p_\rho
  \left[ \text{Flow}^{(3)}_{A^3}(p,-p) \right]^{abc}_{\mu\nu\rho} \c
\end{align}
\begin{align}
  \partial_t \lambda_{A^4} = 2 \eta_A \lambda_{A^4}
  + \left[ \mathcal{P}_{A^4}^{(1)} \right]_{\mu\nu\rho\sigma}^{abcd}
  \left[ \text{Flow}_{A^4}^{(4)} \right]^{abcd}_{\mu\nu\rho\sigma} \c
\end{align}
where we introduced the projector 
\begin{align} \label{eq:four-gluon-proj}
  \left[ \mathcal{P}_{A^4}^{(1)} \right]_{\mu\nu\rho\sigma}^{abcd} =
  \frac{1}{
    108 N_c^2 (N_c^2 - 1)
  }
  \Big[\mathcal{T}_{A^4}^{(1)} \Big]^{abcd}_{\mu\nu\rho\sigma} \c
\end{align}
with the normalization
\begin{equation}
\Big[\mathcal{T}_{A^4}^{(1)} \Big]^{abcd}_{\mu\nu\rho\sigma} 
\Big[\mathcal{T}_{A^4}^{(1)} \Big]^{abcd}_{\mu\nu\rho\sigma} 
    = 108 N_c^2 (N_c^2 - 1) \d
\end{equation}
For the external momentum in the three gluon vertex, we have taken a simple configuration in which only two external gluon legs carry the momenta, similarly to \ccites{Braun:2014ata,Fu:2019hdw}. Note that all momentum configurations are expected to yield the same flow equation in the zero-momentum limit in our case\footnote{We checked this point explicitly by comparing the expressions of the purely gluonic contribution to  the flow of $\lambda_{A^3}$ obtained from the projection in \cref{eq:three-gluon-proj} and from a symmetric point configuration.}.

For the ghost-gluon vertex we employ the projection
\begin{align}
  & \nonumber
  \partial_t \lambda_{A\bar{c}c} =
  \Big( \frac{1}{2} \eta_A + \eta_c \Big) \lambda_{A\bar{c}c}
  - \frac{i}{2N_c(N_c^2 - 1)}  \\[2ex] & \quad \times
  \lim_{p\to0} \frac{\partial^2}{\partial p^2}
  f^{abc} p_\mu \left[ \text{Flow}^{(3)}_{A\bar{c}c}(0,p)
  \right]^{abc}_\mu \; .
\end{align}
However, in the limit of vanishing momentum, the flow contribution
vanishes because the ghost-gluon vertex is directly proportional to ghost momenta. Hence,
the flow of the ghost-gluon coupling only receives a canonical
contribution from the anomalous dimensions
\begin{equation}
  \partial_t \lambda_{A\bar{c}c} =
  \Big( \frac{1}{2} \eta_A + \eta_c \Big) \lambda_{A\bar{c}c} \; .
\end{equation}
We now turn to the strong coupling avatars involving matter fields. To obtain the flows of the two tensor structures of the quark-gluon coupling, we
choose the following projections
\begin{align}
  & \nonumber
  \partial_t \lambda_{A\bar{q}q}^{(1)} =
  \Big( \frac{1}{2} \eta_A + \eta_q \Big) \lambda_{A\bar{q}q}^{(1)} \\[2ex] & \quad
  - \frac{i}{8N_f(N_c^2 - 1)}
  \Tr\left\{ \gamma_\mu T^a \left[ \text{Flow}^{(3)}_{A\bar{q}q}
  \right]^a_\mu \right\} \; ,
\end{align}
\begin{align}
  & 
  \partial_t \bar{\lambda}_{A\bar{q}q}^{(4)} =
  \Big( 1 + \frac{1}{2} \eta_A + \eta_q \Big) \bar{\lambda}_{A\bar{q}q}^{(4)} 
  + \frac{k}{24N_f(N_c^2 - 1)} 
  \\[2ex] & \times \nonumber
  \lim_{p\to0} \frac{\partial^2}{\partial p^2}
  \Tr\left\{ \Pi_{\mu \nu}^{\mathrm{T}}(p) \gamma_\nu \slashed{p} T^a 
  \left[ \text{Flow}^{(3)}_{A\bar{q}q}(-2p,p) \right]^a_\mu \right\} \; ,
\end{align}
where the trace over quark indices is understood. 

Finally, we specify the flow of the two diquark-gluon strong coupling avatars. We choose the following projections
\begin{align}
  & \nonumber
  \partial_t \lambda_{A\Delta^\dagger\Delta} =
  \Big( \frac{1}{2}\eta_A + \eta_\Delta \Big)
  + \frac{1}{2(N_c^2-1)} \\[2ex] & \quad \times
  \lim_{p\to0} \frac{\partial^2}{\partial p^2} p_\mu T^a_{\da\db}
  \left[\mathrm{Flow}^{(3)}_{A\Delta^\dagger\Delta}(0,-p)\right]^{a}_{\mu
  \da\db} \; ,
\end{align}
\begin{align}
  & \nonumber
  \partial_t \lambda_{A^2\Delta^\dagger\Delta} =
  (\eta_A + \eta_\Delta) \\[2ex] & \quad
  + \frac{1}{4(N_c^2-1)} \delta^{ab} \delta_{\mu\nu} \delta_{\da\db}
  \left[\mathrm{Flow}^{(4)}_{A^2\Delta^\dagger\Delta}\right]^{ab}_{\mu\nu
  \da\db} \; ,
\end{align}
where the momentum configuration of the two-diquark-one-gluon avatar is similar to the one used for the three-gluon avatar.

\subsection{Four-quark interactions \& emergent composites}
\label{sec:four-quark-int}

We now move to the connection between the high- and low-energy degrees of freedom. It is established by exactly mapping emergent four-quark interactions onto effective meson or diquark exchange interactions. In the following we first discuss the projection onto the four-quark channels and then focus on the dynamical hadronization procedure.

\subsubsection{Four-quark interactions}

\begin{figure}[t]
  \centering
  \includegraphics[width=\linewidth]{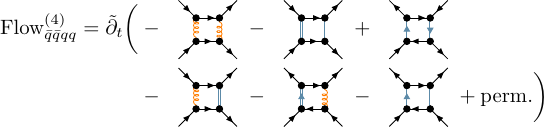}
  \caption{Diagrammatic representation of the flow of the four-quark interaction.}
  \label{fig:four_quark_flow}
\end{figure}

As one flows from the \gls{UV} to the \gls{IR}, effective four-quark interactions inevitably are generated through the quark-gluon interactions. For a given four-quark channel ($i=\text{sps},\text{csc}$) we
use the projection
\begin{equation} \label{eq:four-quark-flow}
  \partial_t \bar{\lambda}_{\qbqbqq}^{(i)} = (2 + \eta_q) \bar{\lambda}_{\qbqbqq}^{(i)}
  + \left[ \mathcal{P}^{(i)}_{\qbqbqq}
  \right]_{\alpha\beta\gamma\delta}
  \left[ \overline{\text{Flow}}_{\qbqbqq}^{(4)}
  \right]_{\alpha\beta\gamma\delta} \; ,
\end{equation}
with $\bar{\lambda}_{\qbqbqq}^{(i)} = k^2 \lambda_{\qbqbqq}^{(i)}$. The diagrammatic contribution to the flow of the four-quark channel are shown in \cref{fig:four_quark_flow}. As explained in the next section, we only need to take into account box diagrams, as the four-quark interactions themselves are zero owing to dynamical hadronization.

We define the projector for a given four-quark channel by
\begin{equation} \label{eq:projector_def}
  \left[ \mathcal{P}_{\qbqbqq}^{(i)}
  \right]_{\alpha\beta\gamma\delta} =
  \sum_{k} c_{ik} \left[ \mathcal{T}^{(k)}_{\qbqbqq}
  \right]_{\gamma\delta\alpha\beta} \; ,
\end{equation}
where the order of the fermionic indices is adjusted such that each
$\bar{q}$ index matches a $q$ index when applied on a tensor
structure. The coefficients
$c_{ik}$ are determined by imposing the condition
\begin{equation} \label{eq:tensor_applied_def}
  \left[ \mathcal{P}_{\qbqbqq}^{(i)} \right]_{\alpha\beta\gamma\delta}
  \left[ \mathcal{T}^{(j)}_{\qbqbqq} \right]_{\alpha\beta\gamma\delta}
  \stackrel{!}{=} \delta_{ij} \; ,
\end{equation}
from which we immediately conclude $c_{ij} = (a^{-1})_{ij}$
where the matrix $a$ has components
\begin{equation}
  a_{ij} = \left[ \mathcal{T}^{(i)}_{\qbqbqq}
  \right]_{\gamma\delta\alpha\beta}
  \left[ \mathcal{T}^{(j)}_{\qbqbqq}
  \right]_{\alpha\beta\gamma\delta} \; .
\end{equation}
In the case of a single tensor structure, this coefficient can be interpreted as a normalization factor similar to \cref{eq:four-gluon-proj}.

We find that the diquark sector is sensitive to the choice of projection onto the corresponding four-quark channel. In general, different choices of projections lead to different flow equations. Without considering a complete set of interaction channels, this can result in contamination from channels that are not taken into account. We found that the \gls{SPS} sector is mostly insensitive to the details of the projection, while the \gls{CSC} sector is more sensitive. This is clear in the \gls{QCD} vacuum, where the \gls{SPS} channel is by far the most dominant one due to spontaneous chiral symmetry breaking. In order to obtain a stable projection while keeping the calculation relatively simple, we choose to adopt a Fierz complete basis invariant under $U(1)_V\times SU(2)_L \times SU(2)_R$ for the construction of the projectors \cref{eq:projector_def}.
Such a basis consists of six elements, which we choose as
\begin{align}\label{eq:fierz}
    \begin{split}
  \mathcal{L}^{(\text{sps})} & =
  (\bar{q}q)^2 - (\bar{q} \gamma_5 \vec{\tau} q)^2 \; , \\[2ex]
  \mathcal{L}^{(\eta')} & =
  (\bar{q} \vec{\tau} q)^2 - (\bar{q} \gamma_5 q)^2 \; , \\[2ex]
  \mathcal{L}^{(\text{csc})} & = \frac{1}{2}
  \big( q^\tp C \gamma_5 i\epsilon_\da \tau_2 q \big)
  \big( \bar{q} \gamma_5 i\epsilon_\da \tau_2 C \bar{q}^\tp \big) \; ,    \\[2ex]
  \mathcal{L}^{(\text{V-A})} & =
  (\bar{q}\gamma_\mu q)^2 + (\bar{q} \gamma_\mu \gamma_5 q)^2 \; , \\[2ex]
  \mathcal{L}^{(\text{V+A})} & =
  (\bar{q}\gamma_\mu q)^2 - (\bar{q}\gamma_\mu \gamma_5 q)^2 \; , \\[2ex]
  \mathcal{L}^{(\text{V-A})^{\text{adj}}} & =
  (\bar{q}\gamma_\mu T^a q)^2 + (\bar{q}\gamma_\mu \gamma_5 T^a q)^2 \; .
  \end{split}
\end{align}
and where the associated tensor structures are derived similarly to \cref{eq:cl_tensor_struc_def}. 
This basis is taken from \ccite{Mitter:2014wpa} with the replacement of $\mathcal{L}^{(S+P)^{\text{adj}}_{-}}$ with $\mathcal{L}^{(\text{csc})}$, which both break $U(1)_A$ explicitly. Assuming invariance only under $U(1)_V \times SU(2)_V$, which is motivated by chiral symmetry breaking, leads to four additional tensor structures. However, we found only minor change upon including the last elements of the basis in \cref{eq:fierz} and hence chose to restrict ourselves to a basis invariant under $U(1)_V\times SU(2)_L \times SU(2)_R$.
For a recent discussion on the construction of tensor basis, see \ccite{Braun:2025gvq}. Based on this basis, we can extract the projections onto the \gls{SPS} and \gls{CSC} channels which are orthogonal to all other channels, hence avoiding contamination that will most drastically affect the subdominant \gls{CSC} channel.

In a truncation which includes only quark and gluons and do not resolve the full momentum dependency of the four-quark couplings, \gls{DCSB} is found through a resonance of the \gls{SPS} channel. In particular, in  such a truncation the \gls{SPS} channel evaluated at vanishing momenta diverges for a finite \gls{RG} scale $k>0$ \cite{Mitter:2014wpa, Braun:2014ata, Fu:2022uow}. The dynamical hadronization procedure provides a convenient way to capture this resonance by introducing an additional scale dependent field for the $\sigma$-meson and pions, and \gls{DCSB} then translate into a finite condensate for the $\sigma$-meson. Furthermore, this procedure can be directly generalized to capture additional resonances corresponding to other quark composite like the scalar diquark.

\subsubsection{Dynamical hadronization}
\label{sec:dynamical_hadronization}

\begin{figure}[!t]
  \centering
  \includegraphics[width=0.85\linewidth]{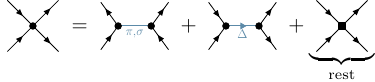}
  \caption{Diagrammatic illustration of the dynamical hadronization procedure. During the \gls{RG} flow, the sum of four quark interactions in \cref{eq:quark-gluon-trunc} is continuously mapped onto a sum of single-boson exchanges (here involving the $\sigma$-meson, pions and the scalar diquark) in a chosen momentum configuration and a ``rest". The rest contains all four-quark interaction channels which are not dynamically hadronized, and the momentum dependencies not taken into account in the dynamically hadronized channels.
  In this work, we focus purely on the dynamically hadronized \gls{SPS} and \gls{CSC} channels at zero momentum exchange and neglect the rest.}
  \label{fig:dynamical_hadro_illust}
\end{figure}

Dynamical hadronization is based on the generalized flow equation \cref{eq:flow_eq_general}, and is implemented through an appropriate choice for the flow $\dot\Phi$ of the composite fields. These fields serve a dual purpose: at high energies they simply encode quark correlations in certain channels, while at low energies they can become relevant degrees of freedom. Conventionally, four-quark interactions are bosonized into the corresponding meson and diquark fields through a Hubbard-Stratonovich transformation, replacing the four-quark interactions by Yukawa interactions and bosonic fields. This procedure is, however, not appropriate for \gls{QCD}. First, the scale where the bosonization is carried out must be chosen by hand. Second, four-quark interactions that are re-generated through boson exchanges between quarks at scales below the bosonization scale are not taken into account.

These problems are circumvented by bosonizing at each \gls{RG} scale $k$, hence neither introducing an arbitrary bosonization scale, nor losing any information on four-quark interactions generated at any scale. It has been shown in \ccite{Gies:2001nw} that this is can be done exactly by choosing the flows of the composite operators in \cref{eq:field_transfo_general} such that the flows of the associated four-quark interactions vanish. As illustrated in \cref{fig:dynamical_hadro_illust}, this amounts to an exact rewriting of parts of the fundamental four-quark interactions in terms of single-boson exchanges. Through the scale dependence, these bosons reflect their composite nature at high energies, while they can act as the relevant degrees of freedom at low energies. Crucially, as we will also explicitly demonstrate below, all features related to low-energy degrees of freedom and spontaneous symmetry breaking are emergent with this procedure, allowing for predictions from first principles.

The latter property is intimately tied to an \gls{IR} attractive fixed point in the flow of the four-quark couplings in the regime of small gauge coupling \cite{Gies:2002hq}. All information on the initial values of composite operators is lost as the system gets close to the fixed point. By the time the system enters the strong coupling regime, where the fixed point vanishes, all low-energy parameters have settled to values uniquely determined by \gls{QCD}. Crucially, this not only applies to composites build from the four-quark interactions themselves, but also higher-order operators \cite{Gies:2002hq, Rennecke:2015eba}; see also \cref{sec:indep_params_uv}.

In this work, we focus on six low-energy composite degrees of freedom, namely the $\sigma$-meson, the pions, the diquark $\Delta$, and the anti-diquark $\Delta^\dagger$. We hence parametrize the scale dependence of the fields with respect to these degrees of freedom,
\begin{equation}
  \dot\Phi_{\cfa}[\Phi] =
  \delta_{\cfa\phi} \dot\phi[\bar{q},q] +
  \delta_{\cfa\Delta} \dot\Delta[q] +
  \delta_{\cfa\Delta^\dagger} \dot\Delta^\dagger[\bar{q}] \d
\end{equation}
The scale dependence of the composite field is chosen directly proportional to their quark content. Hence for the mesonic fields $\phi=(\sigma,\vec\pi)$ we have
\begin{equation} \label{eq:phi_para}
  \dot \phi[\bar{q},q] = \dot A_{\phi}
  \begin{pmatrix}
      \bar{q} q \\
      i \bar{q} \vec{\tau}\gamma_5 q
  \end{pmatrix} \; ,
\end{equation}
while for the diquark and antidiquark we use
\begin{align} \label{eq:Delta_para}
  \dot\Delta_{\da}[q] & = \; \; \;
  \dot A_{\Delta} \; q^\tp C\gamma_5\tau_2i\epsilon_\da q \; , \\[2ex]
  \label{eq:Deltabar_para}
  \dot\Delta_{\da}^*[\bar{q}] & =
  - \dot A_{\Delta} \; \bar{q} \gamma_5\tau_2i\epsilon_\da C\bar{q}^\tp \c
\end{align}
where $\dot A_{\phi}$ and $\dot A_{\Delta}$ are two functions of the \gls{RG}-scale that must be determined.

Inserting these transformation on the left-hand side of
\cref{eq:flow_eq_general} and imposing that the flow of the \gls{SPS} and \gls{CSC} four-quark channel vanish at all scales, we obtain
\begin{equation} \label{eq:Adot_phi}
  \dot{{A}}_\phi = \frac{1}{g_{\phi\bar{q}q}}
  \left[ \mathcal{P}^{(\text{sps})}_{\qbqbqq}
  \right]_{\alpha\beta\gamma\delta}
  \left[ {\text{Flow}}_{\qbqbqq}^{(4)}
  \right]_{\alpha\beta\gamma\delta} \c
\end{equation}
and
\begin{equation} \label{eq:Adot_Delta}
  \dot{{A}}_\Delta =
  \frac{1}{2g_{\Delta q q}}
  \left[ \mathcal{P}^{(\text{csc})}_{\qbqbqq}
  \right]_{\alpha\beta\gamma\delta}
  \left[ {\text{Flow}}_{\qbqbqq}^{(4)}
  \right]_{\alpha\beta\gamma\delta} \d
\end{equation}
If in addition we assume that the four-quark
interactions vanish in the \gls{UV}, consistent with the \gls{QCD} classical action, we then ensure that the \gls{SPS} and \gls{CSC} four-quark interactions vanish for all $k$. Note that the criterion of vanishing four-quark flows can directly be used to motivate the parametrization of the composite fields \cref{eq:phi_para,eq:Delta_para,eq:Deltabar_para}. 

Through the generalized flow equation \cref{eq:flow_eq_general}, the dynamical hadronization functions $\dot A_{\phi}$ and $\dot A_{\Delta}$ now enter the flow of the quark-meson and quark-diquark Yukawa couplings, $g_{\phi\bar{q}q}$ and $g_{\Delta q q}$. Hence, as illustrated in \cref{fig:dynamical_hadro_illust}, the flow of the four-quark interactions is captured through those two couplings. In fact, in the present pointlike approximation, the four quark interactions are effectively given by the ratios
\begin{align}\label{eq:lambdaeff}
  \lambda_{\text{sps},\text{eff}}
  = \frac{g_{\phi\bar{q}q}^2}{2 (k^2+m^2_\phi)}\;,
\quad
  \lambda_{\text{csc},\text{eff}}
  = \frac{g_{\Delta\bar{q}\bar{q}}^2}{2 (k^2+m^2_\Delta)}\; .
\end{align}
These expressions are obtained by integrating out auxiliary meson and diquark fields.
As discussed in \cref{sec:key_aspect_qcd}, it is evident that peaks in the four-quark interaction directly translate into soft bosonic modes.

\subsubsection{Yukawa couplings}

\begin{figure}[!t]
  \centering
  \includegraphics[width=0.9\linewidth]{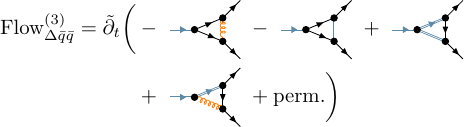}
  \caption{Diagrammatic representation of the flow of
  the quark-diquark coupling.}
  \label{fig:quark_diquark_flow}
\end{figure}

For the sake of simplicity, we follow \ccite{Pawlowski:2014zaa} and choose to project the flow of the quark-meson couplings using the
quark two-point function,
\begin{align} \label{eq:phiqbq_flow}
  \nonumber
  \partial_t g_{\phi \bar{q} q} = &
  \Big( \frac{1}{2} \eta_\phi + \eta_q \Big) g_{\phi\bar{q}q}
  - \dot{\bar{A}}_\phi \bar{m}^2_\pi
  \\[2ex] &
  - \frac{1}{4 N_f N_c} \frac{1}{\sigma}
  \Tr\left[ \text{Flow}_{\bar{q}q}^{(2)} \right] \; .
\end{align}
For the quark-diquark Yukawa coupling, we instead use the three point
function for the projection,
\begin{align}\label{eq:Deltaqbq_flow}
  \nonumber
  \partial_t g_{\Delta q q} = &
  \Big( \frac{1}{2} \eta_\Delta + \eta_q \Big) g_{\Delta q q}
  + 2\dot{\bar{A}}_\Delta \bar{m}^2_\Delta \\[2ex] &
  + \frac{1}{24 N_f}
  \Tr\left\{
    \Big( C\gamma_5 \tau_2 i\epsilon_\da \Big)
    \left[ \text{Flow}_{\Delta\bar{q}\bar{q}}^{(3)} \right]_{\da}
  \right\}
  \; .
\end{align}
This choice is motivated by the absence of a diquark condensate in the
derivation of the flows, which prevent us to use the quark-quark or
antiquark-antiquark two point function to project on the
quark-diquark Yukawa coupling. The diagrammatic contributions to the flow of the quark-diquark Yukawa coupling are shown in \cref{fig:quark_diquark_flow}.

\subsection{Effective potential}
\label{sec:eff_pot}

The effective potential corresponds to the vanishing momentum part of
the effective action, hence its flow is directly given by evaluating
the flow contribution \cref{eq:flow_eq_general} on a constant field
configuration $\Phi_c$. In order to facilitate the extraction of the flows of all zero-momentum meson-meson,
meson-diquark and diquark-diquark interactions, we express the effective potential as a general function of both, a non-vanishing sigma field
$\sigma(x)=\sqrt{2\rho_\phi}$ and a non-vanishing diquark condensate
\begin{equation}
  \Delta_\da(x) = \delta_{\da3} \sqrt{\rho_\Delta} \; ,
\end{equation}
with $\rho_\Delta\in\mathbb{R}$. For non-vanishing $\rho_\Delta$, this field configuration corresponds
to the so-called \gls{2SC} phase. In this phase, the diquark
condensate ``breaks"\footnote{This only holds in the present gauge-fixed description. Strictly speaking, since $SU(3)_c$ symmetry
is gauged, the diquark condensate signals that the system enters a Higgs phase.} the color symmetry $SU(3)_c$
to an $SU(2)_c$ subgroup involving only red ($\da=1$) and green
($\da=2$) quarks. This leads to important consequences, first the
emergence of a superconducting gap
\begin{equation}
  \Delta_{\text{gap}} = g_\Delta \sqrt{\rho_\Delta} \; ,
\end{equation}
in the dispersion relation of the red and green quarks, while the
blue ($\da=3$) quark dispersion is unchanged. Because we only
consider vanishing chemical potential in this work, the
superconducting gap only acts as a shift in the quark mass. 

Furthermore, the gluon acquires a gluon mass through the
Higgs mechanism,
\begin{equation}
  \left( m^2_{\text{Higgs}} \right)^{ab} =
  \lambda_{A^2\Delta^\dagger\Delta}\,\rho_\Delta
  \left( \left\{ T^a, T^b \right\} \right)_{33} \; ,
\end{equation}
see \cref{app:propagators} for the explicit expressions
of the quark and gluon propagators.

Lastly we note that, a
non-vanishing chiral and diquark condensate leads to a sigma-diquark
mixing in the two point function. This mixing can be resolved by
diagonalizing the two-point function, yielding modified dispersion
relation for the sigma meson and the blue ($\da=3$) diquark. For more
details in effective models, see, e.g.,
\ccites{Buballa:2003qv,Andersen:2024qus,Gholami:2025afm}, and for a
discussion in a \gls{FRG}-\gls{QCD} setup, see \ccite{Braun:2021uua}.

For arbitrary chiral and diquark condensates, the flow of the
dimensionless effective potential is given by
\begin{align} \label{eq:effective_potential_flow}
  \nonumber
   & \partial_t \bar{U}_k(\bar{\rho}_\phi,\bar{\rho}_\Delta) =  - 4 \bar{U}_k
  + (2 + \eta_\phi) \bar\rho_\phi \partial_{\bar\rho_\phi} \bar{U}_k \\[2ex] & \quad
  + (2 + \eta_\Delta) \bar\rho_\Delta \partial_{\bar\rho_\Delta} \bar{U}_k
  + \frac{1}{\mathcal{V}_4}
  \overline{\text{Flow}}^{(0)}(\rho_\phi,\rho_\Delta) \; ,
\end{align}
where $\mathcal{V}_4$ is the 4-dimensional Euclidean
spacetime volume. A diagrammatic representation of the flow
contribution is given in \cref{fig:potential_flow_diagrams}. For consistency with all other flow equations, we Taylor expand the effective potential around a
vanishing diquark condensate $\rho_\Delta=0$ and a flowing chiral
condensate $\rho_\phi=\rho_{\phi,0}$,
\begin{equation} \label{eq:Uk_definition}
  U_k(\rho_\phi, \rho_\Delta) =
  \sum_{n=0,m=0}^{m+n\leq N_\text{exp}}
  \frac{\lambda_{n,m}}{n! m!}
  (\rho_{\phi} - \rho_{\phi,0})^n
  \rho_\Delta^m \; .
\end{equation}
We choose $N_{\text{exp}}=5$, as the expansion is under control at this order. The running expansion coefficients $\lambda_{n,m}$ encode multi-meson and diquark interactions at zero momentum. In particular, the sigma and pion curvature masses are readily extracted from the coefficients,
\begin{align}
    \label{eq:curv_mass_pion}
  m^2_\pi & = \lambda_{1,0} \; , \\[2ex]
  \label{eq:curv_mass_sigma}
  m^2_\sigma & = \lambda_{1,0} + 2\rho_{\phi,0} \lambda_{2,0} \; ,
\end{align}
while the diquark curvature mass is given by
\begin{equation} \label{eq:curv_mass_Delta}
  m^2_\Delta = \lambda_{0,1} \; .
\end{equation}
The flow of the different couplings $\lambda_{n,m}$ is
found by taking appropriate derivative of \cref{eq:effective_potential_flow},
\begin{align} \label{eq:flow_eff_pot_couplings}
  \nonumber
  \partial_t  \bar\lambda_{n,m} & =
  \big( (-4+2n+2m) + n \, \eta_\phi + m \, \eta_\Delta \big) \bar\lambda_{n,m}
  \\[2ex] & \nonumber
  + \big(
    (2 + \eta_\phi) \bar\rho_{\phi,0}
    + \partial_t \bar\rho_{\phi,0}
  \big) \bar\lambda_{n+1,m} \\[2ex] &
  + \frac{1}{\mathcal{V}_4}  \left(
    \partial_{\bar\rho_\phi}^n \partial_{\bar\rho_\Delta}^m
    \overline{\text{Flow}}^{(0)}(\rho_\phi,\rho_\Delta)
  \right) \Big|_{\substack{\rho_\phi = \rho_{\phi,0} \hfill \\ \rho_\Delta = 0 \hfill}} \; .
\end{align}
It only remains to discuss the flow of $\rho_{\phi,0}$, which is derived by
imposing that the solution of the equation of motion resulting from the effective potential $\Omega_k=U_k -
h\sigma$ is located at $\rho_\phi=\rho_{\phi,0}$ for every \gls{RG}-scale $k$
\cite{Fu:2019hdw}. This can be expressed as the following condition
\begin{equation}
    \left[ \partial_{\rho_\phi}  \left(
      U_k - h \sigma
  \right) \Big] \right|_{\substack{\rho_\phi = \rho_{\phi,0} \hfill \\ \rho_\Delta = 0 \hfill}}
  = 0 \; ,
\end{equation}
from which we conclude that the running minimum is implicitly given by
\begin{equation} \label{eq:running_minimum}
  \rho_{\phi,0} = \frac{1}{2} \frac{h^2}{U_k^\prime(\rho_{\phi,0})^2} \; .
\end{equation}
Taking a $t$-derivative, we obtain the
flow of the running minimum
\begin{align}
  \nonumber
  & \partial_t \bar \rho_{\phi,0} =
  - \frac{\bar h^2}{\bar \lambda_{1,0}^3 + \bar \lambda_{2,0} \bar{h}^2}
  \bigg\{
    (1 + \frac{1}{2} \eta_\phi) \bar \lambda_{1,0}
    \\[2ex] & \quad
    + (2 + \eta_\phi) \bar \rho_{\phi,0} \bar \lambda_{2,0}
    + \frac{1}{\mathcal{V}_4} \left(
      \partial_{\bar\rho_\phi}
      \overline{\text{Flow}}^{(0)}
    \right) \Big|_{\substack{\rho_\phi = \rho_{\phi,0} \hfill \\ \rho_\Delta = 0 \hfill}}
  \bigg\} \; .
\end{align}
We note that even though the explicit symmetry breaking term
$h$ is scale independent, in an \gls{RG}-invariant, dimensionless formulation it receives a contribution from the anomalous dimension and its mass dimension, leading to the flow
\begin{equation}
  \partial_t \bar h =
  \Big( \! -3 + \frac{1}{2} \eta_\phi \Big) \bar h \; .
\end{equation}

We now briefly summarize this section. We have presented the projection procedure that defines the set of coupled flow equations we will solve to determine the relevant correlation functions for this work. Within the truncation employed here, all flow equations are evaluated at vanishing external momentum. We note, though, that owing to the combined \gls{UV} and \gls{IR} regularization of the flow equation, momenta around the \gls{RG} scale, $p^2 \!\approx\! k^2$, contribute to the flow of the correlation functions. Hence, information on the momentum dependence of the correlation functions is, to some extend and in a regulator-dependent way, still encoded and taken into account through their $k$-dependence; see, e.g., \cref{fig:gluon_dressing}. Furthermore, through the dynamical hadronization procedure, we get direct access to the correlations of composite operators. These include high-order correlations through the couplings $\lambda_{n,m}$, which are related to $(2n+2m)$-quark correlations in the hadronized channels. 
As we will show in the next section, the present truncation provides an accurate description of vacuum \gls{QCD}, allowing for simplified numerical computation while still providing reliable predictions.

\section{Low-Energy Constants from QCD Flows}
\label{sec:qcd_flow_results}

In this section, we present our numerical results and describe the
details of the parameter-fixing procedure. Unless stated
  otherwise, all results are shown as bands obtained from three
  different regulator shape functions: the flat shape function
  $r_{\text{flat}}$, \cref{eq:flat_shape_function}, and the
  exponential regulator $r_{\text{exp}}$,
  \cref{eq:exp_shape_function}, with $m=1$ and $m=2$. The
  corresponding central results are indicated by the faint dashed,
  dotted, and solid lines, respectively.

These regulators span a broad 
 range of momentum-shell widths (see
\cref{fig:shape-functions}), ranging from relatively sharp to
strongly  smeared
momentum shells  integrated out along  the \gls{RG}
flow. Consequently, the width of the
band provides a measure of  the residual  regulator dependence of our results. Since
this dependence vanishes for exact solutions by construction, the
bands serve as an estimate of the systematic uncertainty associated
with the present truncation.

\subsection{Parameter fixing}
\label{sec:parameter-fixing}

Within our self-consistent \gls{QCD} truncation, the only
parameters that require  fixing at the \gls{UV} scale $k=\Lambda$
are  the strong coupling and the current quark mass.
As we  demonstrate below, all other initial conditions for
the running couplings and masses  in our setup  including the gluon mass gap
parameter, the Yukawa couplings, and the parameters of the 
composite sector, are associated with  emergent low-energy degrees of
freedom.
Consequently, they are 
 either determined self-consistently  by the \gls{QCD} dynamics
or correspond to irrelevant directions whose  precise \gls{UV} values  leave no imprint
on physical observables.

Asymptotic freedom allows us to choose a \gls{UV} scale sufficiently deep in the
perturbative regime -- unless stated otherwise, we employ  $\Lambda = 20 \GeV$ 
 -- where the strong coupling is small and a perturbation theory
is applicable. In this regime, \gls{STI} ensure that the flows of
all strong-coupling avatars coincide  \cite{Schwartz:2014sze,
  Braun:2014ata}. Accordingly, at $k=\Lambda$ we identify them with a
common coupling 
$\alpha_{s,\text{UV}}$ such that
\begin{align}
  \nonumber
  \alpha_{A^3,\Lambda} &
  = \alpha_{A^4,\Lambda}
  = \alpha_{A\bar{c}c,\Lambda}
  = \alpha_{A\bar{q}q,\Lambda}
  \\[2ex] & 
  = \alpha_{A\Delta^\dagger\Delta,\Lambda}
  = \alpha_{A^2\Delta^\dagger\Delta,\Lambda}
  = \alpha_{s,\text{UV}} \; .
\end{align}
The value of  $\alpha_{s,\mathrm{UV}}$ is fixed by requiring  that the
\gls{IR} constituent quark mass satisfies $m_q=350\MeV$ in accordance
with \cite{Bowman:2005vx, Chang:2021vvx, Cyrol:2017ewj,Fu:2025hcm}.

This choice leads to good agreement between our computation and
lattice results for the gluon dressing function $\bar{Z}_A^{-1}$, see
\cref{fig:gluon_dressing} and the associated discussion.
We emphasize that the value 
of the strong coupling cannot be determined  directly from 
 more conventional renormalization schemes, such as
 $\overline{{\rm MS}}$, since  the perturbative running in our
  \gls{FRG} scheme differs beyond one-loop order
\cite{Gies:2006wv}.

\begin{table}[t]
  \centering
  \bgroup
  \def\arraystretch{1.5}
  \begin{tabular}{c c c c c}
    & $r_\text{flat}$
    & $r_\text{exp} \; (m=2)$
    & $r_\text{exp} \; (m=1)$
    \\ \hline\hline
    $\alpha_{s,\text{UV}}$
    & 0.20015
    & 0.19530
    & 0.19150
    \\
    $c$ [GeV$^3$]
    & 4.79
    & 4.76
    & 4.68
    \\
    $m_{\text{gap},\text{UV}}$ [MeV]
    & 751.67779
    & 1538.377019
    & 2283.226685
  \end{tabular}
  \egroup
  \caption{ \gls{UV} parameters for the three  regulator shape functions
    considered in this work.  The quoted precision of 
      $m_{\text{gap},\text{UV}}$ reflects the fine-tuning required to
      obtain  the scaling solution. Consequently, the least
      significant  digits
      may depend on  numerical details.  }
  \label{tab:uv_parameters}
\end{table}

As explained in \cref{sec:frg_general}, explicit chiral symmetry
breaking is encoded in the source term $h$. It is fixed through the
pion mass, $m_\pi=137\,\MeV$. Note that pole and curvature masses of
the pion are almost identical in the present setup; we return 
to this point in \cref{sec:composite-two-point}.

As discussed in \cref{sec:glue-truncation}, the initial value of
  the gluon mass-gap parameter, $m^2_{\text{gap},\text{UV}}$, is tuned
  such that the flow approaches the confining scaling solution. In
  practice, this amounts to adjusting $m^2_{\text{gap},\text{UV}}$
  such that it lies closest to the onset of the Landau-pole
  regime. This requires fine-tuning, and the resulting solution
  exhibits scaling behavior only down to the \gls{IR} scale
  $k_{\text{IR}}=10\MeV$ where we terminate the \gls{RG} flow. In
  \cref{sec:indep_params_uv}, we explicitly demonstrate that the
  choice between scaling and decoupling solutions affects only the
  deep-\gls{IR} behavior of the gapped glue sector and hence leaves
  physical low-energy observables unchanged.

\begin{table}[t]
  \centering
  \bgroup
  \def\arraystretch{1.5}
  \begin{tabular}{c @{\hspace{2em}} c c c c}
 [MeV]   & this work   & Fu et al. \cite{Fu:2019hdw}   & Ihssen et al.
    \cite{Ihssen:2024miv}   \\ \hline\hline
    $m_{\pi,\text{curv}}$            & 137       & 137
    & 138                                   \\
    $m_{q,\text{curv}}$              & 350       & 347
    & 350                                   \\
    $m_{\sigma,\text{curv}}$         & 435--441       & 531
    & 388.1                                 \\
    $\sigma_{0,l}$     & 71.6--80.3        & 68.6
    & 69.                                   \\
    $m_{\Delta,\text{curv}}$         & 1030--1187      & --
    & --
  \end{tabular}
  \egroup
  \caption{ \gls{IR} values of various quantities obtained in this
    work and compared with previous \gls{FRG} studies. The quoted
    uncertainties correspond to the variation under changes of
 the regulator. No uncertainty  is
shown for the pion and quark masses, since these quantities are used
as input in  the
    parameter-fixing procedure.}
  \label{tab:results_and_params}
\end{table}

\begin{figure*}[t]
  \centering
  \includegraphics[width=\onefig]{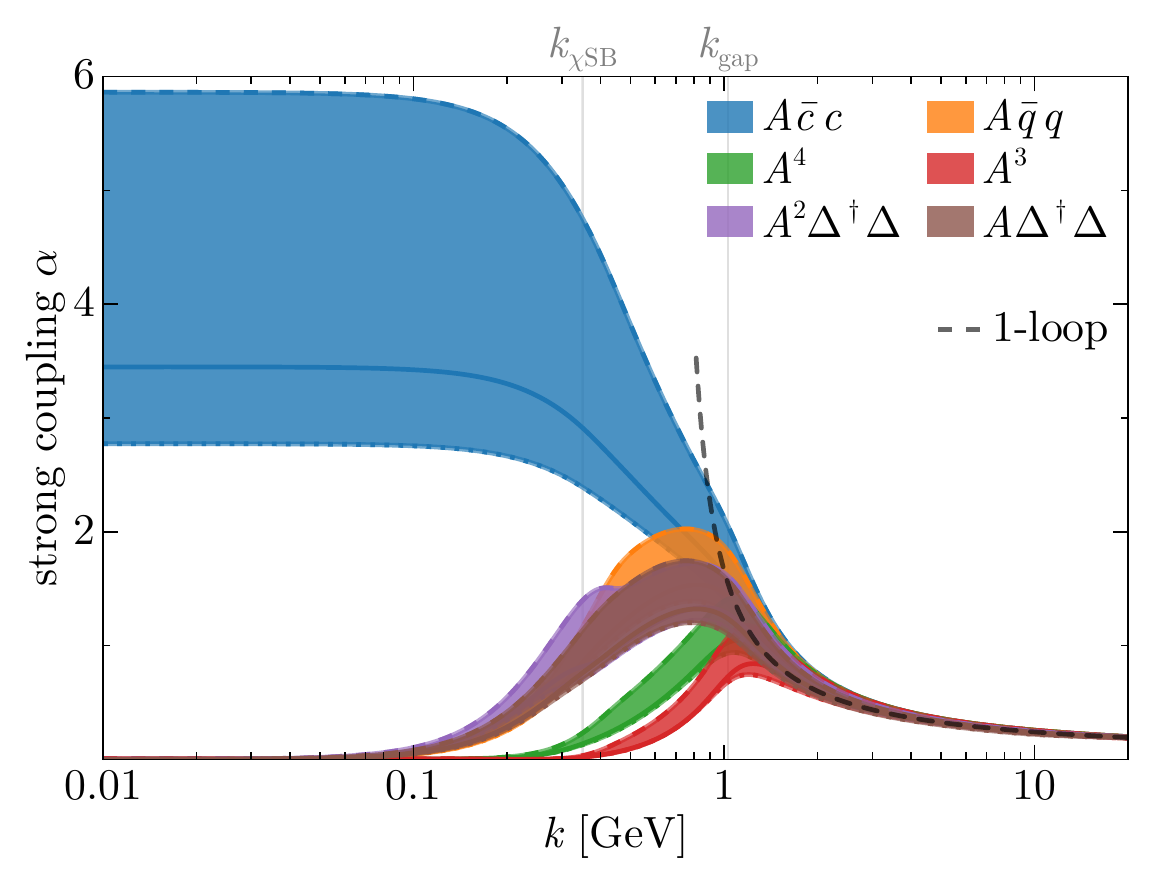}
  \includegraphics[width=\onefig]{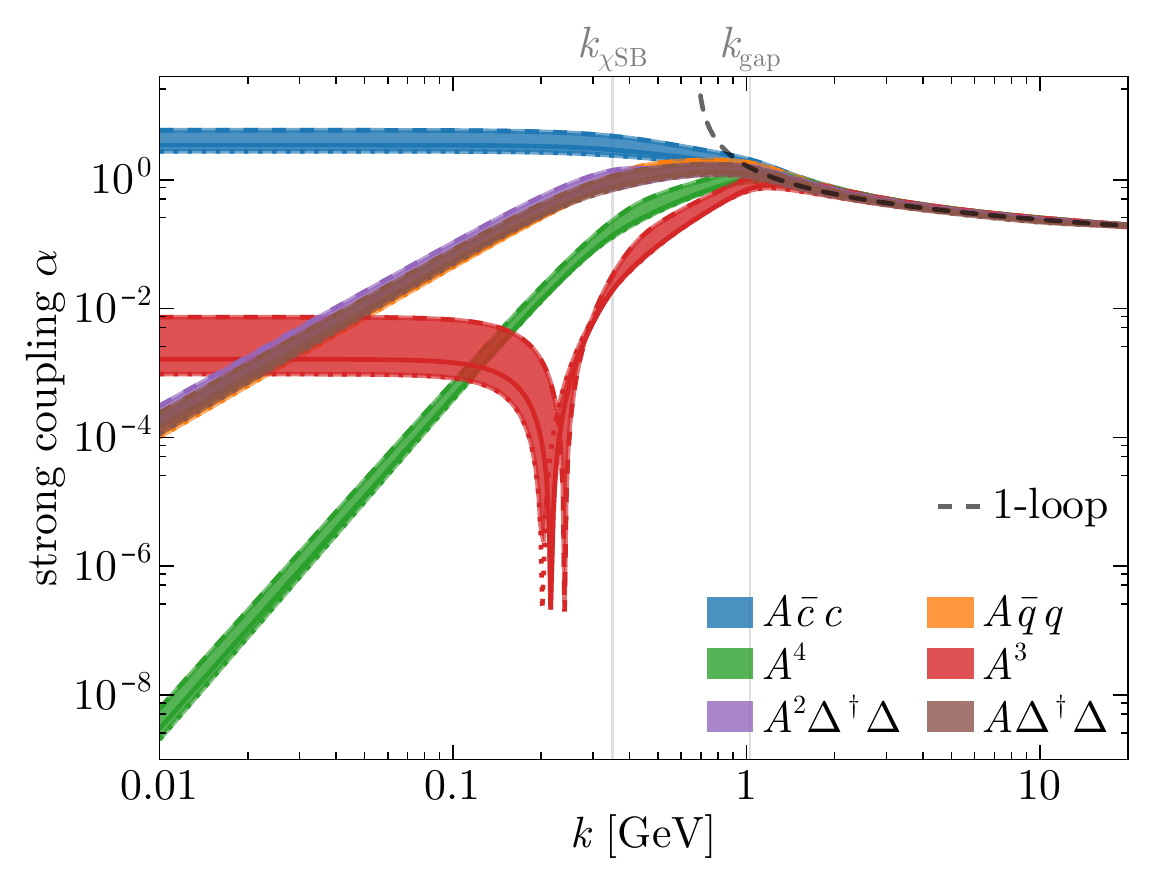}
  \caption{Running of the strong coupling avatars $\alpha$ 
  \crefrange{eq:alpha_A3}{eq:alpha_A2Delta2} on a linear
    (left) and a log (right) scale. The black dashed line indicates the
    perturbative running of the strong coupling $\alpha_s$ at one-loop.
    Note that the peculiar behavior of $\alpha_{A^3}$ is caused by a sign change below about $1\GeV$.}
  \label{fig:strong_coupling_avatars}
\end{figure*}

It only remains to specify the \gls{UV} value of the couplings in the composite sector. The \gls{UV} value of those couplings can be chosen mostly arbitrarily as an \gls{IR} attractive fixed point ensures independence of the \gls{IR} results on the initial conditions of the composite sector, see \cref{sec:dynamical_hadronization}.
One must only ensure that the initial conditions are on the critical surface of this fixed point. Otherwise, the system would describe a gauged \gls{NJL} model instead of \gls{QCD}. Fortunately, all initial conditions are on this critical surface as long as the four-quark interactions are sufficiently small in the \gls{UV}. With dynamical hadronization, this entails that the ratios of the corresponding Yukawa couplings and boson masses must be small, see \cref{eq:lambdaeff}. This is clear intuitively, as mesons and diquarks should be decoupled, auxiliary degrees of freedom in the perturbative regime.
In practice we make the choice
\begin{equation}
  g_{\phi\bar{q}q,\text{UV}} = g_{\Delta\bar{q}\bar{q},\text{UV}} = 1 \c
\end{equation}
and
\begin{equation}
  m^2_{\phi,\text{UV}} = m^2_{\Delta,\text{UV}} = \Lambda^2 \times
  10^4 \GeV^2 \c
\end{equation}
but note that any mass much larger than the initial scale $\Lambda$ suffices to ensure that these fields are auxiliary. The initial values for all other couplings in the composite sector can be chosen arbitrarily. For definiteness, we set them to zero.
In \cref{sec:indep_params_uv} we demonstrate explicitly the
independence of our results on the \gls{UV} behavior of the composite sector.

In \cref{tab:uv_parameters}, we show the \gls{UV} parameters used for the
different regulators we consider. For different regulators the initial values of the strong coupling, explicit chiral symmetry breaking strength and the \gls{IR}-enhancement exhibit only small differences. Compared to other studies, e.g., \ccites{Fu:2019hdw,Ihssen:2024miv,Goertz:2024dnz}, we need a slightly smaller strong coupling at the \gls{UV} scale $\Lambda=20\GeV$ to obtain a constituent quark mass $m_q=350\MeV$. This originates from the missing tensor structure $\mathcal{T}^{(7)}_{A\bar{q}q}$ in the quark-gluon vertex, which slightly overestimates the quark-gluon interaction strength. However, the initial condition for the mass gap parameter strongly depends on the regulator. This is expected, as $m_{\text{gap}}$ encodes the modifications of the \gls{STI} due to the presence of the regulator, see \cref{sec:glue-truncation}.

\cref{tab:results_and_params} shows a selection of results for
\gls{IR} observables in comparison to the \gls{FRG}-\gls{QCD} studies in \ccite{Fu:2019hdw, Ihssen:2024miv}. In this work, the parameters in \cref{tab:uv_parameters} are fixed using the pion mass $m_\pi=137\MeV$ and the constituent quark mass $m_q=350\MeV$, while the uncertainties in the sigma mass, light quark chiral condensate, and diquark mass originate from different regulator choices. The differences in the $\sigma$ mass can be related to different projections for the meson wave function renormalization \cite{Fu:2019hdw}. For the vacuum expectation value of the light scalar meson, our result for the flat regulator, $\sigma_{0,l}\sim71 \MeV$, is similar to the other two works, $\sigma_{0,l}\sim70 \MeV$. The reasonable agreement despite considerable differences in the truncations of these different works, especially regarding the gauge sector and momentum dependencies, underscores the predictive power of the \gls{FRG} approach to \gls{QCD}.

\subsection{Gauge sector}
\label{sec:numerical-results-qcd}

\begin{figure}[t]
  \centering
  \includegraphics[width=\columnwidth]{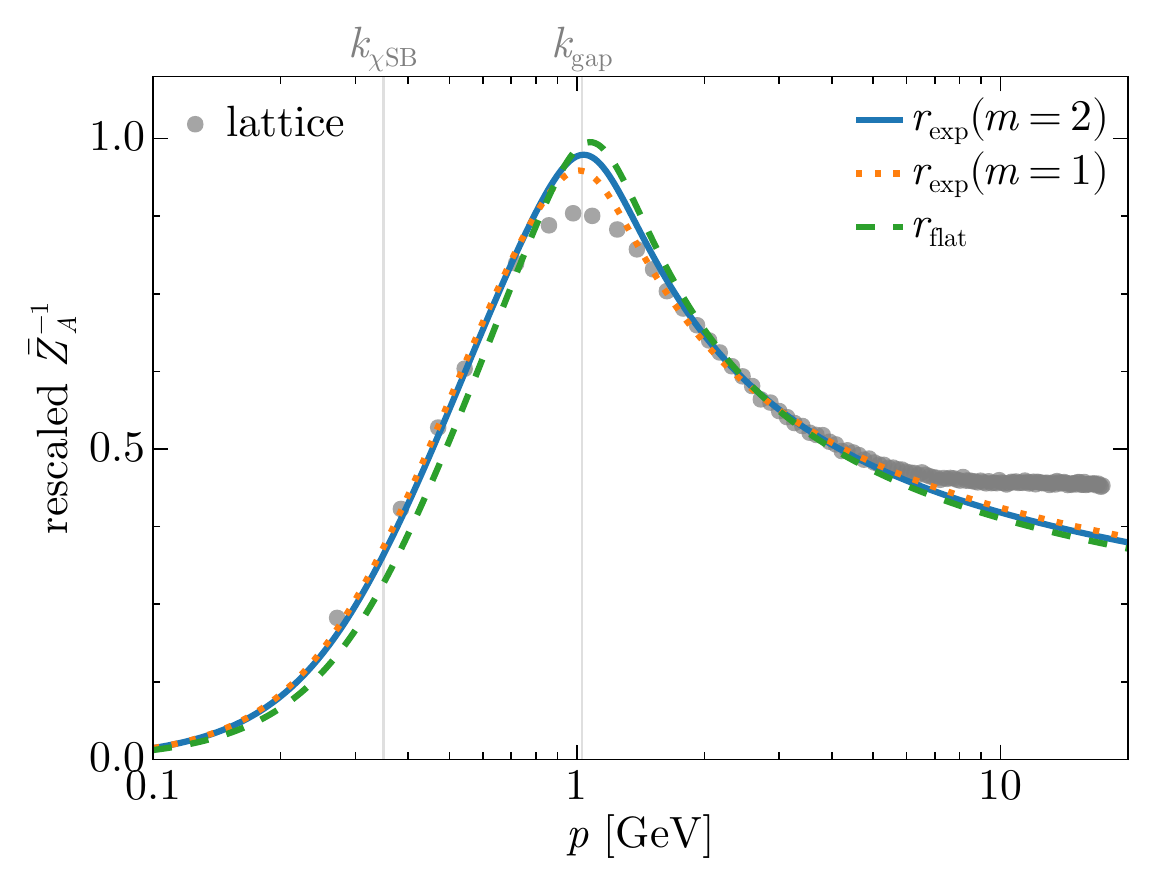}
  \caption{ Comparison between the transverse gluon dressing function
    $\bar{Z}_A^{-1}$ \cref{eq:gluon_dressing_flow} from this work
    (lines) and $N_f=2$ lattice results \cite{Sternbeck:2012qs}
    (dots). To facilitate
    a qualitative comparison, we approximate the momentum dependence
    with the $k$-dependence of our results. As the gluon propagator is
    an \gls{RG}-variant quantity, each \gls{FRG} result is rescaled to
    account for the different normalization of the gluon field in the
    lattice and \gls{FRG} computations.}
  \label{fig:gluon_dressing}
\end{figure}

We now move to a selection of numerical results in the gauge sector of \gls{QCD}. In most figures we show the chiral symmetry breaking
scale $k_{\chi\text{SB}}$, defined as the \gls{RG}-scale where the mass of
the pion and sigma are no longer degenerate, and the ``confinement''
scale $k_{\text{gap}}$, defined as the \gls{RG}-scale where the gluon
dressing function $\bar{Z}_A^{-1}$ peaks, see \cref{fig:gluon_dressing}.
At this scale gluonic fluctuations decouple from the flow, $m_{\text{gap}} \gtrsim k$ for $k\lesssim k_{\text{gap}}$.

We start with a discussion of the flow of the six strong coupling
avatars. To make a meaningful comparison between each coupling, we
express all couplings in a form similar to the strong coupling
constant defined as  $\alpha_s = g_s^2/4\pi$,
where $g_s$ is the strong coupling that appears in the covariant
derivative, see \cref{eq:covariant_deff}. Furthermore, to ensure \gls{RG} invariance, each strong
coupling avatar must be dressed by an appropriate power of the inverse gluon dressing function corresponding to the number of
external gluon legs of the coupling. This leads to the following definitions
{\allowdisplaybreaks
\begin{align}
    \label{eq:alpha_A3}
  \alpha_{A^3} & =
  \frac{\lambda_{A^3}^2}{4\pi} \frac{1}{(1 + \bar{m}^2_\text{gap})^3} \; , \\[2ex]
  \alpha_{A^4} & =
  \frac{\lambda_{A^4}}{4\pi} \frac{1}{(1 + \bar{m}^2_\text{gap})^2} \; , \\[2ex]
  \alpha_{A\bar{c}c} & =     \frac{\lambda_{A\bar{c}c}^2}{4\pi}
  \frac{1}{(1 + \bar{m}^2_\text{gap})} \; , \\[2ex]
  \alpha_{A\bar{q}q} & =
  \frac{\lambda_{A\bar{q}q}^2}{4\pi} \frac{1}{(1 + \bar{m}^2_\text{gap})} \; ,
\end{align}
}
while for the gluon-diquark avatars we have
\begin{align}
  \alpha_{A \Delta^\dagger \Delta} & =
  \frac{\lambda_{A \Delta^\dagger \Delta}^2}{4\pi} \frac{1}{(1 +
  \bar{m}^2_\text{gap})} \; , \\[2ex]
  \label{eq:alpha_A2Delta2}
  \alpha_{A^2 \Delta^\dagger \Delta} & =
  \frac{\lambda_{A^2 \Delta^\dagger \Delta}}{4\pi} \frac{1}{(1 +
  \bar{m}^2_\text{gap})} \; .
\end{align}
In \cref{fig:strong_coupling_avatars} we show the flows of these couplings. As explained in
\cref{sec:parameter-fixing}, all strong coupling avatars coincide in
the \gls{UV} as enforced by perturbative \gls{STI}.
Accordingly, at high \gls{RG} scale, all avatars follow the
perturbative running of the strong coupling
\begin{equation}
    \partial_t \alpha_s = - \frac{\alpha_s^2}{6\pi} \left(
    11 N_c - 2N_f
  \right)
  + \mathcal{O}(\alpha_s^3) \; ,
\end{equation}
as shown by the dashed black line.
We see that the perturbative running is recovered in the
\gls{UV} for all regulators, highlighting the \gls{RG}-scheme
independence of the one-loop perturbative running
\cite{Deur:2016tte}. We note that even though our truncation
includes an additional colored bosonic field, the scalar
diquark, it does not enter as an additional contribution to the perturbative running of the strong
coupling. This is a direct consequence of the dynamical hadronization
procedure, ensuring that the composite fields are fully decoupled from the \gls{UV} dynamic.

At intermediate \gls{RG} scale $k \lesssim 2 \GeV$, the flows of the
different strong coupling avatars start to differ as
non-perturbative effects are no longer negligible. At $k \sim
k_{\text{gap}} \sim 1 \GeV$, the flows of all avatars (except the
ghost-gluon vertex $\alpha_{A\bar{c}c}$) peak, signaling the suppression of gluonic
fluctuations due to the mass gap. As reflected in the width of the bands in \cref{fig:strong_coupling_avatars}, this coincides with the scale where different
regulators have the largest impact on the flow. At lower $k$,
the couplings decrease and reach an (almost) vanishing value,
again with the exception of the ghost-gluon vertex
$\alpha_{A\bar{c}c}$. 
In particular, as visible in the log-scale plot
in \cref{fig:strong_coupling_avatars} (right), the ghost-gluon and three-gluon couplings, $ \alpha_{A\bar c c}$ and $\alpha_{A^3}$, become constant in the \gls{IR}. This is a direct consequence of scaling 
\cite{Alkofer:2004it, Fischer:2008uz}. We would expect that at least the four-gluon vertex, but presumably also the other vertices shown in \cref{fig:strong_coupling_avatars} become constant in the \gls{IR} due to scaling. However, closed loops of the soft ghost modes do not contribute to the flows of these couplings within our approximations, see \cref{fig:strong_coupling_flows} and \cref{app:zero-momentum-flows}. As a result, the quark-gluon vertex, 
$\alpha_{A\bar{q}q}$, and the diquark-gluon vertices, 
$\alpha_{A\Delta^\dagger\Delta}$ and $\alpha_{A^2\Delta^\dagger\Delta}$, vanish with a slope proportional to $(k^2)^{2\kappa}$, while the four-gluon
$\alpha_{A^4}$ vertex vanishes with a slope
proportional to $(k^2)^{4\kappa}$, where $\kappa = 0.64$ is the
scaling exponent introduced in \cref{eq:scaling_ghost,eq:scaling_gluon}. These slopes stem from the gluon contributions in our definitions of the couplings in \crefrange{eq:alpha_A3}{eq:alpha_A2Delta2}. Similar observations have been made in Refs.\ \cite{Cyrol:2016tym, Goertz:2024dnz}. Note that for the decoupling solution, all of these vertices vanish in the deep \gls{IR} \cite{Cyrol:2016tym, Goertz:2024dnz}.

The ghost-gluon vertex displays a strong sensibility on the choice of regulator,
especially in the \gls{IR}. This has no effects on
observables as the ghost only couples to the gluons, which
decouple in the \gls{IR} due to the mass gap.  This sensibility is expected to some extent, as the ghosts encode the explicit gauge symmetry breaking from gauge fixing by construction, and the regulator acts as an additional source for this breaking. This is what gives rise to the modification of the \gls{STI} in the first place \cite{Ellwanger:1994iz}. The close connection between regularization and gauge fixing can be the basis for a gauge-invariant formulation of the \gls{FRG} see, i.e., \ccite{Asnafi:2018pre}.

\begin{figure}[t]
  \centering
  \includegraphics[width=\columnwidth]{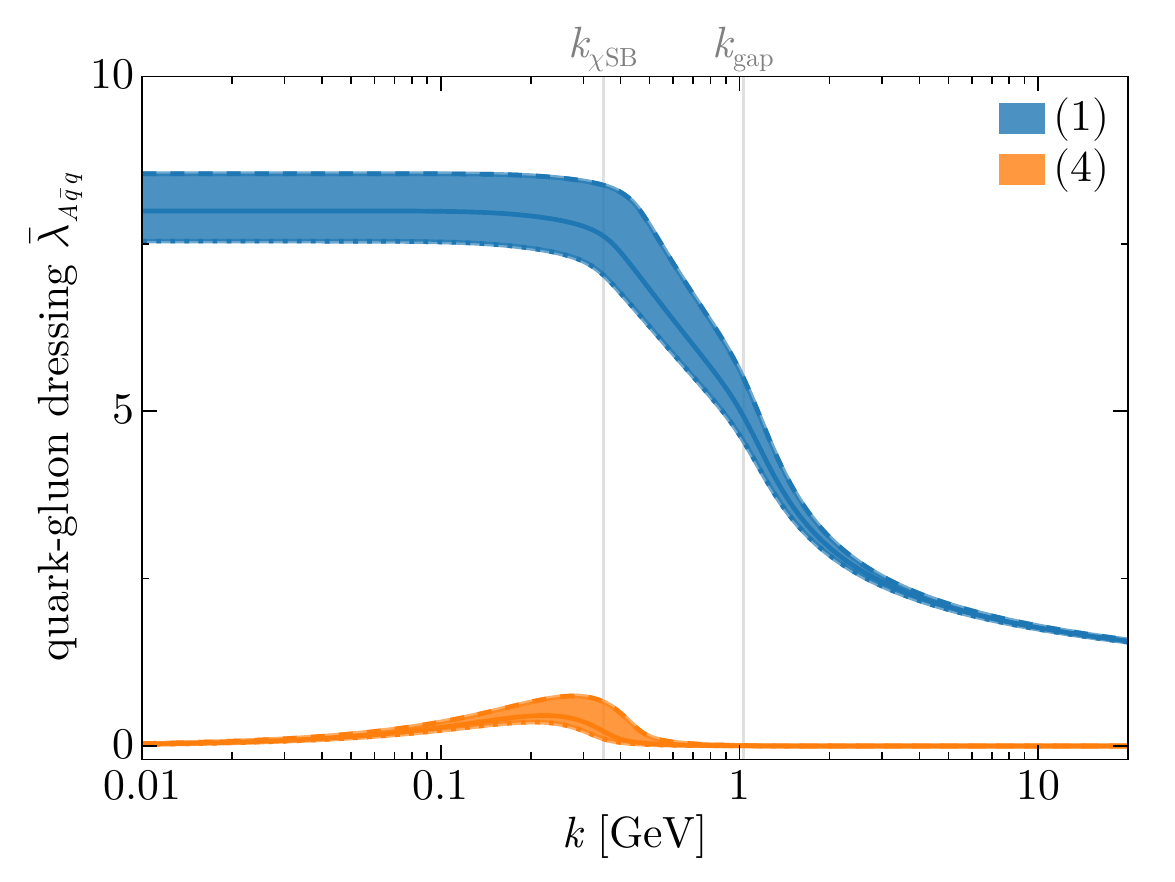}
  \caption{
  Dimensionless classical and non-classical dressings of the
  quark-gluon vertex, $\lambda^{(1)}_{A\bar{q}q}$ and
  $\lambda^{(4)}_{A\bar{q}q}$, as functions of the \gls{RG}-scale $k$. Since $\lambda^{(4)}_{A\bar{q}q}$ breaks chiral symmetry, it only becomes nonzero close to the chiral symmetry breaking scale $k_{\chi\mathrm{SB}}$.
  }
  \label{fig:quark_gluon_tensors}
\end{figure}

Furthermore, we checked explicitly that the two diquark-gluon vertices, $\alpha_{A\Delta^\dagger\Delta}$ and $\alpha_{A^2\Delta^\dagger\Delta}$, only
lead to minor modifications of around 1\% of the infrared properties of the system. This result is expected as the gluon
and diquark fields are relevant degrees of freedom in different regimes: the gluon is important in the \gls{UV} and
decouples in the \gls{IR},
while the diquark is a low-energy degree of freedom, and
hence only relevant at low \gls{RG}-scale; see \cref{fig:masses_flow}
and the associate discussion. However, the inclusion of the diquark-gluon
interactions becomes of high interest at finite densities, where
a diquark condensate is expected to form. In such a regime, these interactions play an important role, especially regarding the
generation of a gluon mass by means of the Higgs-mechanism. As such, this
work is a first step towards a non-perturbative treatment of the formation of a Higgs
mass in dense \gls{QCD} matter.

In \cref{fig:quark_gluon_tensors} we show the two dressings of the quark-gluon vertex $\lambda^{(1,4)}_{A\bar{q}q}$ against the \gls{RG}-scale $k$. Since $\lambda^{(1)}_{A\bar{q}q}$ is associated to the classical tensor-structure, it is nonzero in the \gls{UV}, while the non-classical dressing $\lambda^{(4)}_{A\bar{q}q}$ vanishes. $\lambda^{(4)}_{A\bar{q}q}$ breaks chiral symmetry explicitly, see \cref{eq:4_tensor_struct_Aqq}, and hence only becomes nonzero close to the chiral symmetry breaking scale $k_{\chi\mathrm{SB}}$. Consistent with previous results \cite{Mitter:2014wpa,Cyrol:2017ewj,Gao:2021wun,Fu:2025hcm}, we find that $\bar\lambda^{(4)}_{A\bar{q}q}$ is smaller than $\lambda^{(1)}_{A\bar{q}q}$ by roughly an order of magnitude. However, its inclusion leads to an increase of the constituent quark mass of more than $50\%$, highlighting the sensitivity of the chiral symmetry breaking strength to the strength of quark-gluon interactions \cite{Alkofer:2023lrl}.

In \cref{fig:gluon_dressing}, we show the transverse gluon dressing function $\bar{Z}_A^{-1}$ defined in \cref{eq:Zbarphi} as a function of momentum $p$. The momentum dependence is approximated by the \gls{RG}-scale dependence of the gluon propagator evaluated at vanishing momenta \cite{Braun:2014ata,Fu:2019hdw}, yielding
\begin{equation}
    \label{eq:gluon_dressing_flow}
    \bar{Z}_A(p) = Z_{A,k=p} \Big[ 1 + r_{A,k=p} + \bar{m}^2_{\text{gap},k=p} \Big] \c
\end{equation}
where $Z_{A}$ is the gluon wave function renormalization and $r_A$ the gluon regulator shape function; see \cref{app:regulators} for details.
As explained in \cref{sec:parameter-fixing}, the peak of $\bar{Z}_A^{-1}$ agrees for all regulator choice as a consequence of the parameter fixing procedure. Furthermore, we use the peak location to define the ``confinement' scale" $k_{\text{gap}}$.

Similar to \ccite{Goertz:2024dnz}, we find a good agreement for momenta above
the gluon peak, but some discrepancies around and below. This is a result of our zero-momentum approximation and neglected tensor structures, especially $\mathcal{T}^{(7)}_{A\bar{q}q}$ \cite{Cyrol:2017ewj}. Owing to the decoupling of the gluons around the scale $k_{\text{gap}}$, these discrepancies are expected to have a small quantitative effect. The deviations at large momentum are due to finite size effects on the lattice; our results correctly capture the expected logarithmic running in the \gls{UV}. Importantly, the identification $p=k$ is only an approximation which is sensitive to the regulator choice and not guaranteed to be quantitatively reliable, see for example \ccite{Ihssen:2024miv}.

\begin{figure}[t]
  \centering
  \includegraphics[width=\columnwidth]{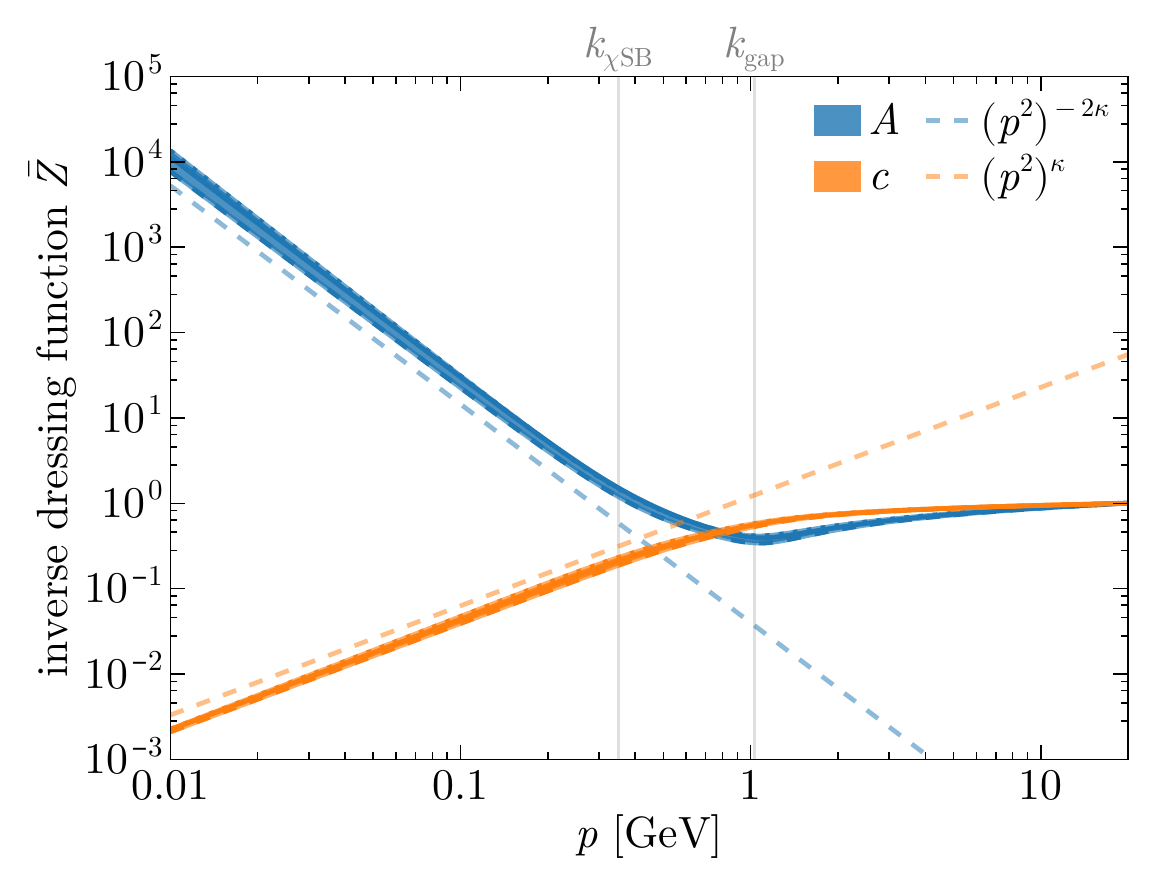}
  \caption{
  Inverse gluon \cref{eq:gluon_dressing_flow} and ghost \cref{eq:ghost_dressing_flow} scaling functions  
  against momentum $p$. The momentum dependence is again approximated by the \gls{RG}-scale dependence of the ghost and gluon propagators evaluated at vanishing momentum. The colored dashed lines indicate the
    behavior of the scaling solution, $\bar{Z}_A \propto (p^2)^{-2\kappa}$ and $\bar{Z}_c \propto (p^2)^{\kappa}$, with a scaling exponent $\kappa=0.64$.}
  \label{fig:ghost_gluon_dressing}
\end{figure}

In \cref{fig:ghost_gluon_dressing} we show the ghost
and gluon inverse dressing functions $\bar{Z}_c$ and $\bar{Z}_A$ as functions of momentum. Again, we identify $p = k$, which yields for the ghost inverse dressing function
\begin{equation}
    \label{eq:ghost_dressing_flow}
    \bar{Z}_c(p) = Z_{c,k=p} \Big[ 1 + r_{c,k=p} \Big] \; .
\end{equation}
Furthermore, we show the expected scaling
\cref{eq:scaling_ghost,eq:scaling_gluon} for the ghost and gluon
inverse dressing function for $\kappa=0.64$. As discussed in \cref{sec:parameter-fixing}, we find a solution of scaling type for both the gluon and ghost
propagator in the \gls{IR}. Furthermore, this \gls{IR} behavior can be used to extract the scaling exponent, the precise value of which we discuss in the next paragraph.

\begin{figure}[t]
  \centering
  \includegraphics[width=\columnwidth]{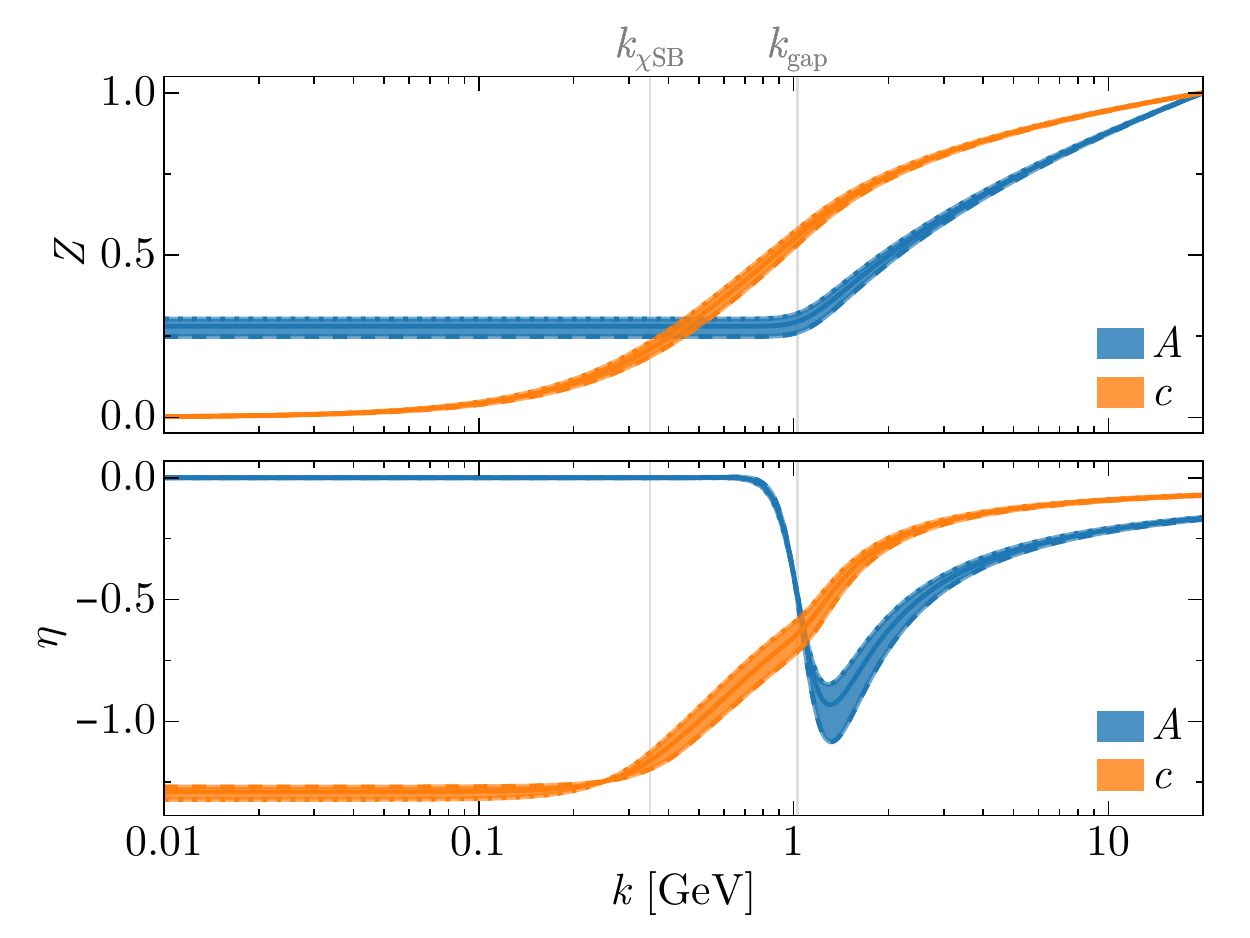}
  \caption{Wave function renormalizations, $Z_A$ and $Z_c$, and
    anomalous dimensions, $\eta_A$ and $\eta_c$, of the gluon and ghost
  fields as functions of the \gls{RG}-scale $k$.}
  \label{fig:ghost_gluon_anomalous}
\end{figure}

To conclude this section, we show in \cref{fig:ghost_gluon_anomalous}
the flow of the ghost and gluon anomalous dimensions and wave
function renormalizations. The latter are obtained from
\cref{eq:wave_func_from_ano} with the choice
$Z_{A,\Lambda}=Z_{c,\Lambda}=1$, but note that the choice of initial values is irrelevant in our \gls{RG}-invariant setup. As we tune the \gls{UV} mass gap to the scaling
solution, we find that the ghost anomalous dimension $\eta_c$
saturates to a negative value in the \gls{IR}, which in turn leads to
the vanishing of the \gls{IR} ghost wave function renormalization $Z_c$.
The scaling exponent $\kappa$ can be readily obtained
from the \gls{IR} value of the ghost anomalous dimension: using the
scaling anzatz \cref{eq:scaling_ghost} with the identification $p=k$, we find
\begin{equation}
  \lim_{k\to0} \eta_c = - \lim_{k\to0} \partial_t \log \left[ (k^2)^{\kappa}
  \right] = -2 \kappa \c
\end{equation}
yielding
\begin{equation}
    \kappa = 0.63\text{--}0.66
\end{equation}
consistent with we the expected value $1/2 < \kappa < 1$ \citeScalingExponent. We find only a slight
dependence on the choice of regulator for the value of the scaling
exponent $\kappa$. Finally, we note that the gluon anomalous dimension saturates to zero in the \gls{IR} as a consequence of the freezing procedure detailed in \cref{eq:gluon_ano_freeze}.

\subsection{Emergent composites}
\label{sec:numerical-results-composites}

\begin{figure}
  \centering
  \includegraphics[width=\onefig]{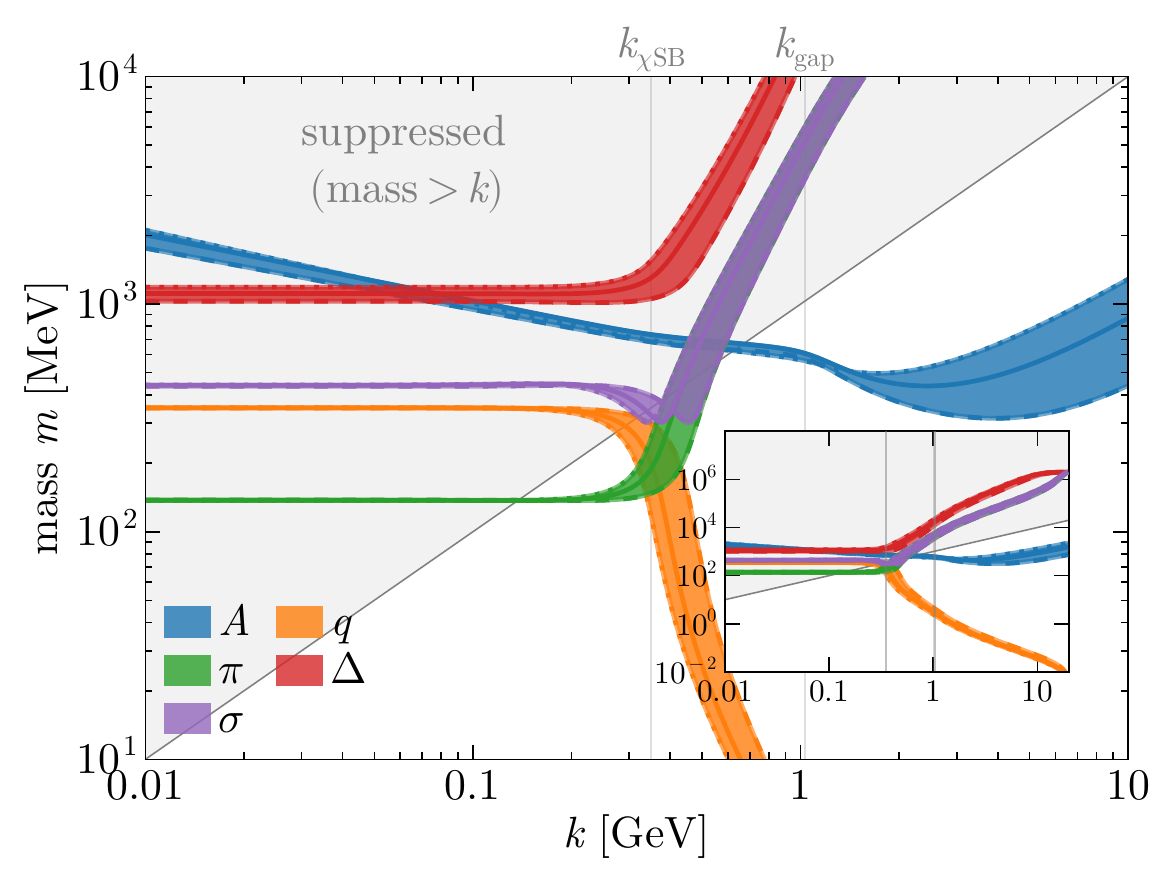}
  \caption{Curvature masses of the different fields
    considered in this work as functions of the \gls{RG}-scale $k$. The gluon mass is given by $m_{\text{gap}}$. In the
    gray region, the masses of the fields are greater than $k$, and thus decouple from the dynamics.}
  \label{fig:masses_flow}
\end{figure}

We now discuss the properties of the emergent composite in our
approach, i.e.\ the $\sigma$-meson, the pions, and the scalar diquarks. In
\cref{fig:masses_flow} we show the curvature masses of the different
fields that are considered in this work as functions of the \gls{RG}-scale. The gray region indicates where
the mass $m$ of a given field is higher than the \gls{RG}-scale $k$.
When this occurs, the associated field effectively decouples from the
dynamics of the flow equation. This can be inferred from the shape of the propagator at zero momentum,
$G_k=1/(k^2 + m^2)$, showing that for $m > k$ the propagator
becomes effectively constant and hence irrelevant for the flow.

As expected in \gls{QCD}, gluons and quarks dominate in the deep \gls{UV}. As explained in \cref{sec:parameter-fixing}, the
composite fields must be initialized with a mass well above the initial scale to ensure their decoupling in the \gls{UV}. Flowing down
to scales of a few GeV, the gluon mass gap parameter
$m_{\text{gap}}$ starts to decrease, signaling the increasing of the
strong coupling $\alpha_s$, while the quark mass $m_q$ starts rising
but remains negligibly small. The
mass of the composite fields drops sharply as a consequence of the
dynamical hadronisation procedure: quark-gluon interactions generates
effective four-quark interactions which are directly absorbed into
the composite sector.

Flowing to lower scales, the gluon mass gap $m_{\text{gap}}^2$ starts
to increase, correspondingly the gluon propagator develop a maximum
at the confinement scale $k_{\text{gap}}$, see
\cref{fig:gluon_dressing}. Rapidly after this point, the mass gap
parameter $m_{\text{gap}}^2$ reaches the decoupling region, and the
gluon field is the first to effectively decouple. In contrast to the other fields, the gluon mass gap parameter does not reach a plateau in the \gls{IR}. This is a direct consequence of the scaling solution. Because we freeze the anomalous dimension, see \cref{sec:anomalous_dimensions}, the gluon wave-function renormalization is constant in the \gls{IR} as shown in \cref{fig:ghost_gluon_anomalous}. Hence, the scaling condition of the gluon dressing function, \cref{eq:scaling_gluon}, can only be realized by a diverging $m_\text{gap}^2$. Imposing the scaling relation of the gluon dressing function $\bar Z_A(p^2)$ in \cref{eq:scaling_gluon} on the gluon propagator we find, again identifying $p=k$, that the \gls{IR} behavior of the mass gap parameters is given by 
\begin{equation}
    \lim_{k\to0} m^2_\text{gap} \propto (k^2)^{1-2\kappa} \d
\end{equation}
This scaling is explicitly realized in \cref{fig:masses_flow}. 

Close to the chiral
symmetry breaking scale $k_{\chi\text{SB}}$, the quark mass $m_q$ is
no longer negligible, signaling \gls{DCSB}, while pions and $\sigma$-mesons become light enough to be relevant. As the sigma meson is tightly connected to the soft mode associated with the chiral crossover \cite{Haensch:2023sig}, its mass dips at $k_{\chi\text{SB}}$. 
As a consequence of \gls{DCSB}, the degeneracy of the
masses of the $\sigma$-meson and pion is lifted around $k_{\chi\text{SB}}$.
Ultimately, the last fields to decouple from the dynamics
are the pions, as expected from pseudo-Goldstone bosons, the lightest \gls{QCD} excitations.

\begin{figure}[t]
  \centering
  \includegraphics[width=\onefig]{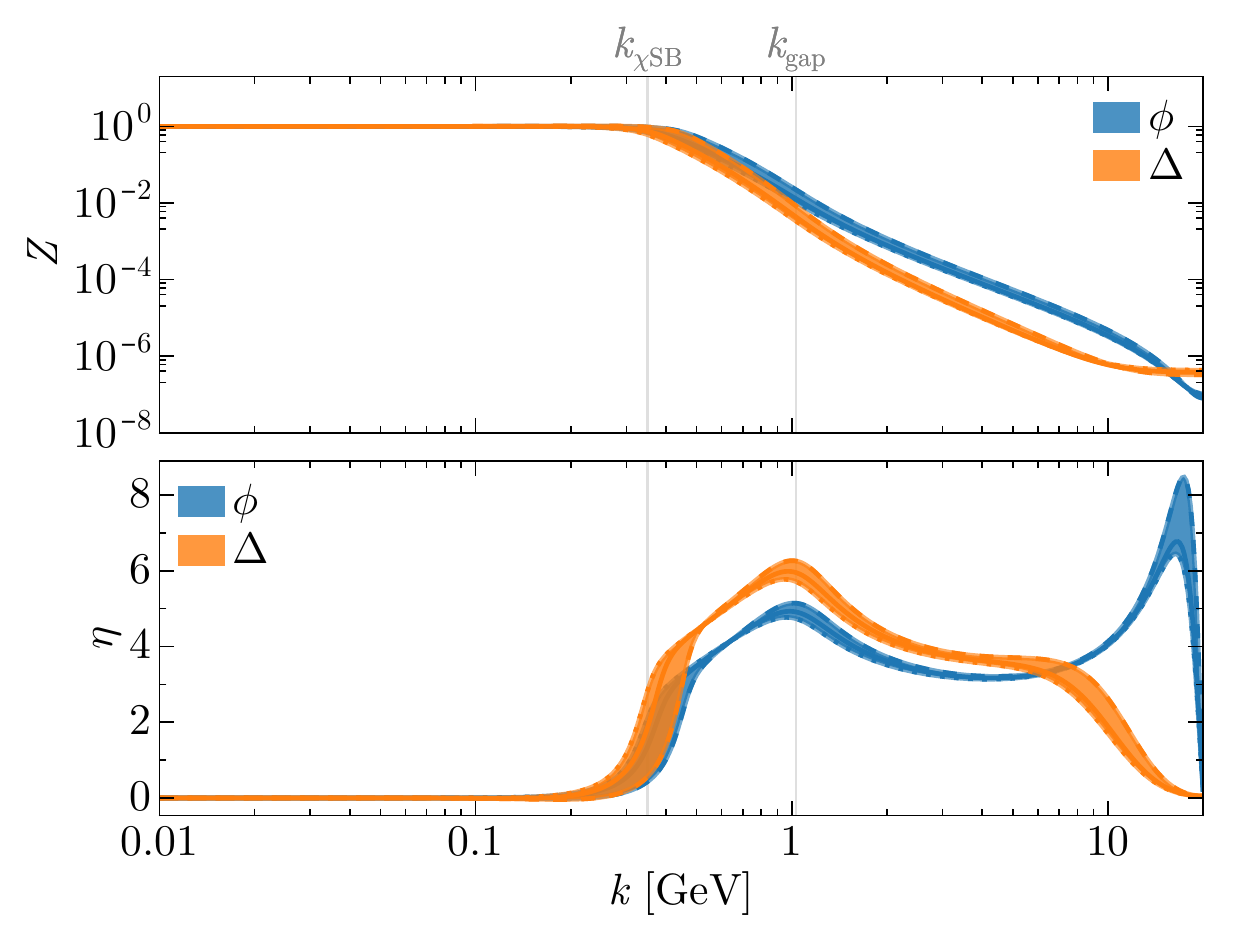}
  \caption{Wave function renormalization and anomalous dimension of
  the mesonic and diquark fields as a function of the \gls{RG}-scale $k$, where we choose $Z_{\phi,k_{\text{IR}}}=Z_{\Delta,k_{\text{IR}}}=1$.}
  \label{fig:boson_anomalous}
\end{figure}

We find that the scalar diquark is always decoupled. This implies that the
inclusion of diquark fluctuations only leads to sub-leading
correction in the vacuum compared to the other fields considered in this work. This
explains the observation that the diquark-gluon vertices only lead
to negligible modifications of the \gls{IR} observables, as mentioned
in the discussion of \cref{fig:strong_coupling_avatars}.
Strikingly, we find that scalar diquark curvature mass is around 1\,GeV. This is much heavier than the naive expectation of about twice the constituent quark mass. We will get back to this point in \cref{sec:composite-two-point}.

\begin{figure}[t]
  \centering
  \includegraphics[width=\onefig]{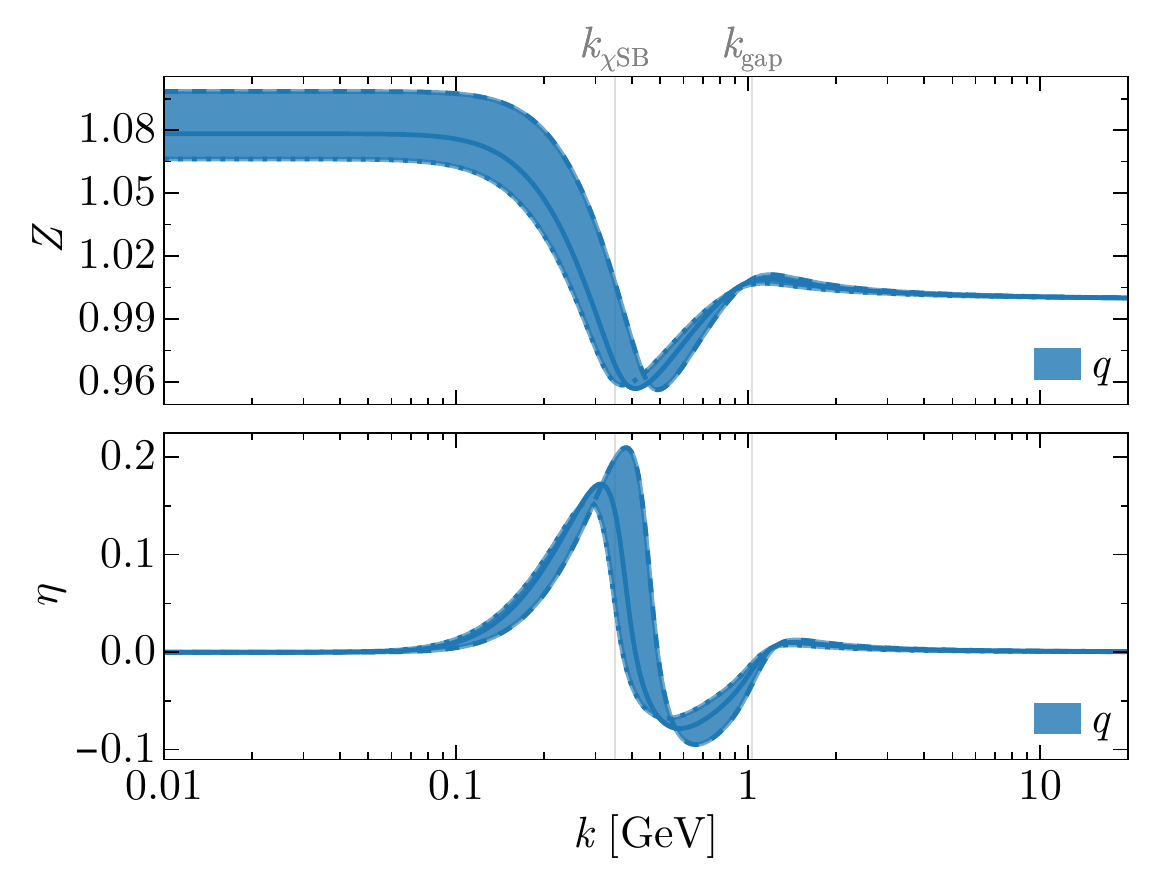}
  \caption{Wave function renormalization and anomalous dimension of
  the quark field as a function of the \gls{RG}-scale $k$, where we choose $Z_{\phi,\text{UV}}=Z_{\Delta,\text{UV}}=1$.}
  \label{fig:quark_anomalous}
\end{figure}

In \cref{fig:boson_anomalous}, we show the flow of the anomalous
dimension and wave function renormalization of the mesons $\phi$ and the diquark fields. In contradistinction to the ghost, gluon and quark fields, we choose here $Z_{\phi,k_{\text{IR}}}=Z_{\Delta,k_{\text{IR}}}=1$. We find
that the meson wave function renormalization $Z_\phi$ increases by
several orders of magnitude from the \gls{UV} to the \gls{IR},
consistent with previous studies \cite{Braun:2014ata,Rennecke:2015eba,Fu:2019hdw}.
This effects is a direct consequence of the dynamical hadronization
procedure and ensures that mesons become auxiliary fields in the \gls{UV}. This effect drives the sharp drop of the renormalized masses towards the \gls{IR} seen in \cref{fig:masses_flow}. The plateau in the anomalous dimension in the region of a few GeV reflects the intermediate-scale fixed point of the four-quark interaction discussed in \cref{sec:key_aspect_qcd,sec:dynamical_hadronization}.

For the scalar diquark, we find a similar behavior, however the
diquark wave function renormalization is smaller by roughly an order
of magnitude. This is consistent with the mass
difference between the $\sigma$-meson and scalar diquark. Notably, this
difference in the value of the meson and diquark wave function renormalization is not reflected in a generally smaller value of the
diquark anomalous dimension $\eta_\Delta$, but is caused by the
absence of a sharp rise of $\eta_\Delta$ for scales near the \gls{UV}
scale $\Lambda$ present in the meson anomalous dimension $\eta_\phi$. Furthermore, we checked
explicitly that this difference is caused by the behavior of the
\gls{UV} flow of the \gls{SPS} and the \gls{CSC} four-quark channels. While the former is nonzero, the latter almost vanishes.
We will see below that leaves an imprint in the \gls{UV} flow of all
observables related to the composite fields.

In \cref{fig:quark_anomalous} we show the quark anomalous dimension and wave
function renormalization. In contrast to the composite and gauge fields, $\eta_q$ remains small for all scales,
$|\eta_q|\lesssim0.15$. Again, this is consistent with other studies \gls{FRG} studies \cite{Mitter:2014wpa,Cyrol:2017ewj,Gies:2002hq,Braun:2014ata, Rennecke:2015eba,Fu:2019hdw}. It is related to the fact that, unlike the other fields, the decoupling of quarks in the \gls{IR} (in the sense of \cref{fig:masses_flow}) is due to  \gls{DCSB}. Furthermore, we find an \gls{IR} quark wave function renormalization larger than one, consistent with lattice results \cite{Oliveira:2016muq}.

\begin{figure}[t]
  \centering
  \includegraphics[width=\columnwidth]{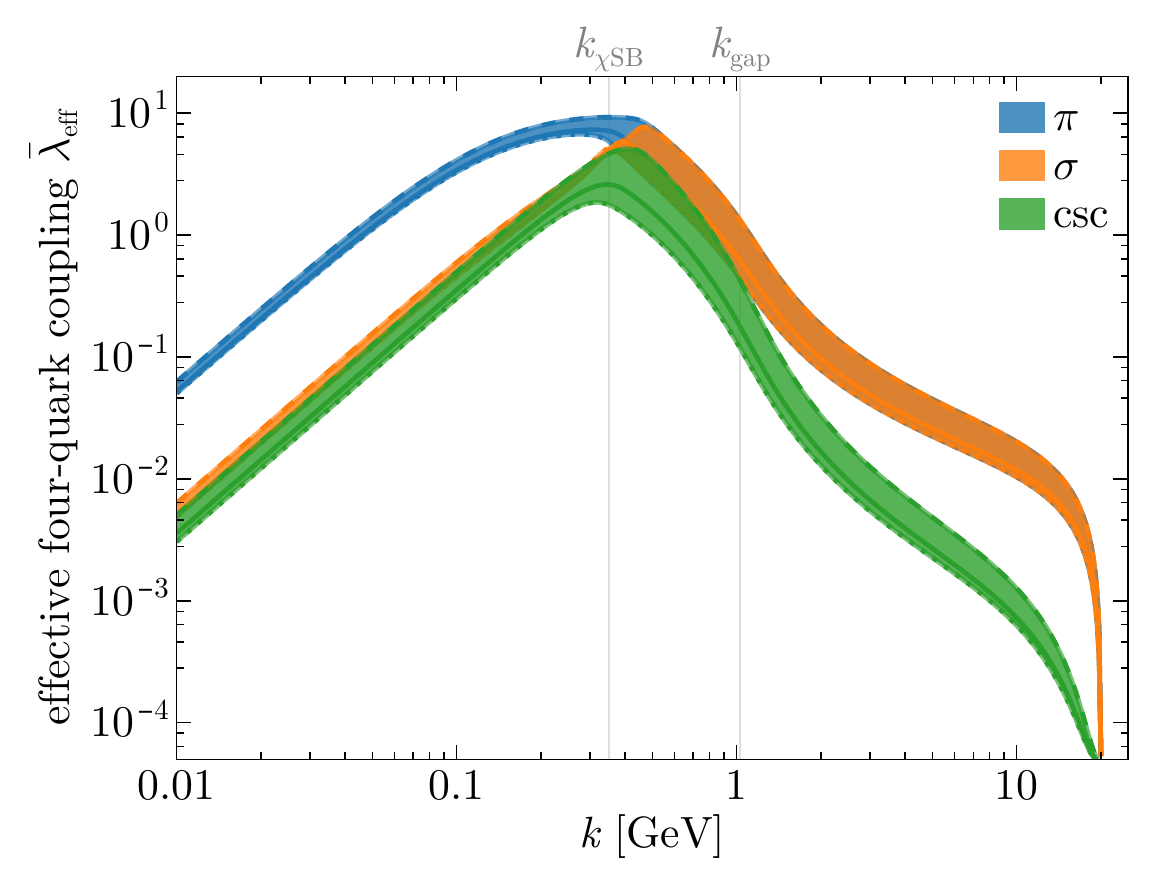}
  \caption{Dimensionless effective four-quark scalar-pseudoscalar $\bar{\lambda}_{\sigma,\pi}$ and color superconducting $\bar{\lambda}_{\text{csc}}$
    channels as functions of the \gls{RG}-scale $k$.}
  \label{fig:effective_four_quark}
\end{figure}

\begin{figure}[t]
  \centering
  \includegraphics[width=\columnwidth]{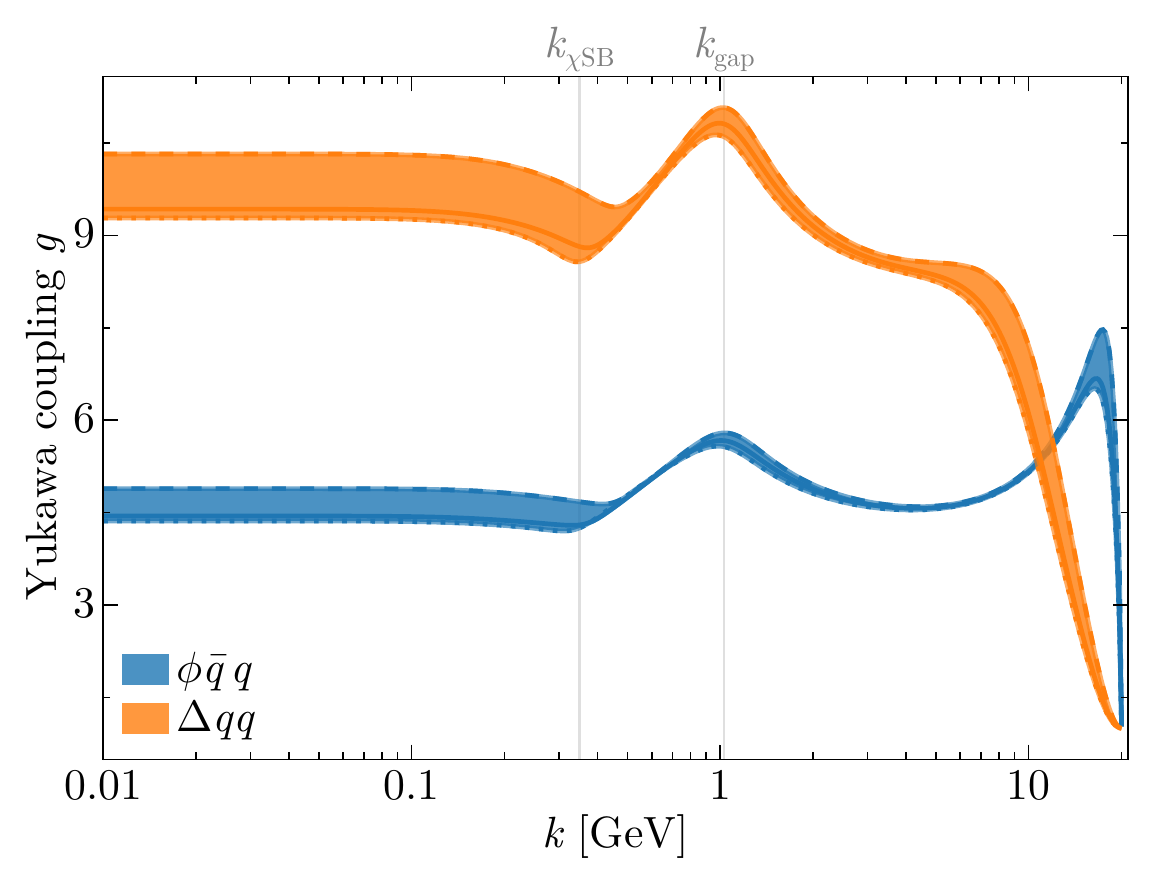}
  \caption{Quark-meson $g_{\phi\bar{q}q}$ and quark-diquark
  $g_{\Delta qq}$ Yukawa couplings as a function of the \gls{RG}-scale $k$.}
  \label{fig:yukawa_couplings}
\end{figure}

In \cref{fig:effective_four_quark}, we show as a function of the \gls{RG}-scale the dimensionless effective
four-quark interactions
\begin{equation} \label{eq:effective_4quarks}
  \bar\lambda_{\sigma/\pi,\mathrm{eff}} =
  \frac{1}{2}
  \frac{g_{\phi\bar{q}q}^2}{1+\bar{m}^2_{\sigma/\pi}} \; ,
  \quad 
  \bar\lambda_{\text{csc},\mathrm{eff}} =
  \frac{1}{2}
  \frac{g_{\Delta q q}^2}{1+\bar{m}^2_{\Delta}} \; ,
\end{equation}
see \cref{eq:lambdaeff}.
In the deep \gls{UV}, the running of the four-quark interactions is governed by the weakly interacting fixed point discussed in \cref{sec:key_aspect_qcd,sec:dynamical_hadronization}. At lower scales, four-quark interactions are generated with a clear
hierarchy: the \gls{SPS} interactions dominates over the
color superconducting channel by roughly an order of magnitude, and
this hierarchy persists for all $k$. This confirms the
role of the \gls{SPS} channel as the dominant vacuum
four-quark interactions. This is expected, as the soft modes associated with  \gls{DCSB} emerge from this channel; see also \cref{fig:masses_flow}. Consequently, below the chiral symmetry breaking
scale $k_{\chi\text{SB}}$, the Goldstone channel $\lambda_{\pi}$ starts to dominate over $\lambda_\sigma$.
Furthermore, each effective four-quark
interactions follows the canonical $k^{2}$-running of $1/\bar m^2$ once the \gls{RG}-scale reaches the mass of the associated composite. 

\begin{figure}[t]
  \centering
  \includegraphics[width=\columnwidth]{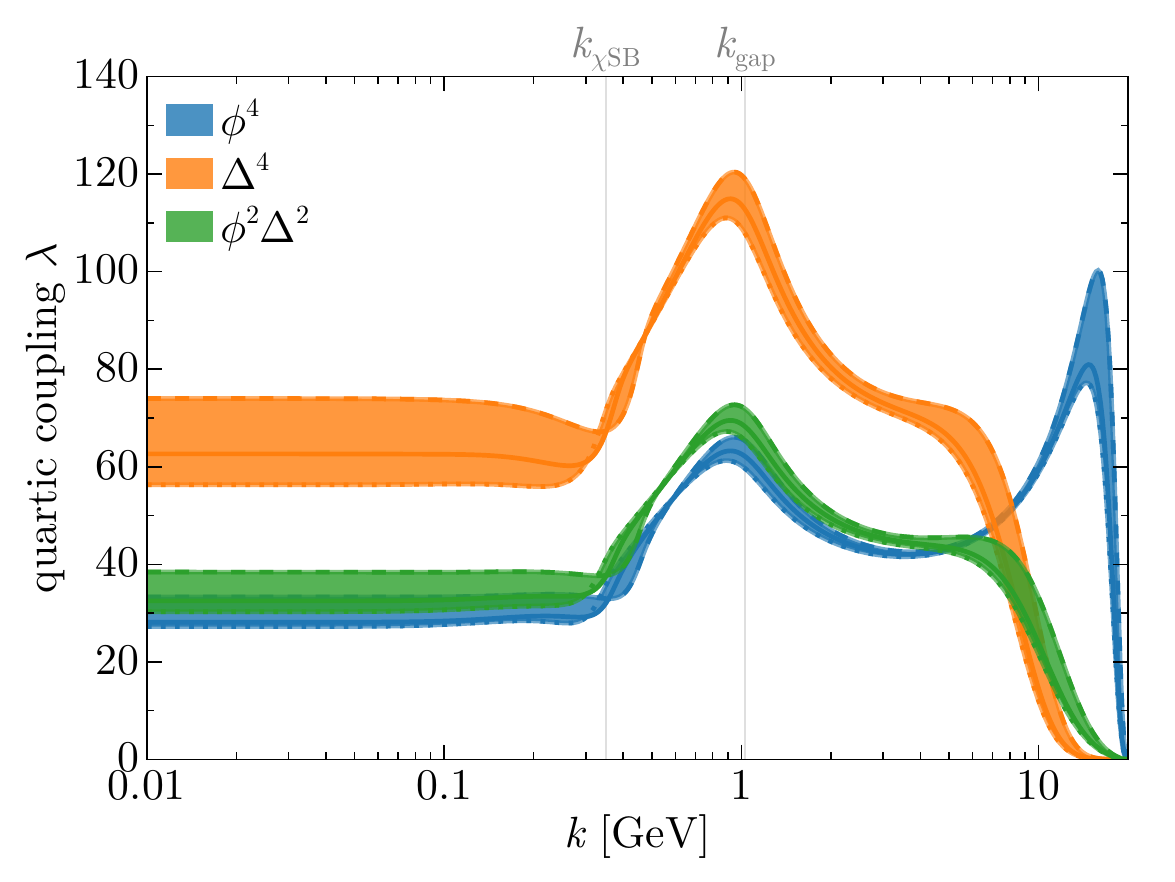}
  \caption{Quartic meson-meson $\lambda_{2,0}$, diquark-diquark
    $\lambda_{0,2}$ and meson-diquark $\lambda_{1,1}$ couplings as a
  function of the \gls{RG}-scale $k$.}
  \label{fig:quartic_couplings}
\end{figure}

In \cref{fig:yukawa_couplings} we show the quark-meson and
quark-diquark Yukawa couplings. The running of both couplings is qualitatively very similar, as the large-$k$ behavior is determined by the intermediate scale fixed point and all fields decouple in the \gls{IR}, leading to frozen flows. The differences in the deep-\gls{UV} behavior are due to the different anomalous dimensions, see \cref{fig:boson_anomalous} and the associated
discussion. Furthermore the
quark-diquark Yukawa coupling is considerably stronger than the quark-meson Yukawa coupling in the \gls{IR}. As we
will discuss in more details in \cref{sec:composite-two-point}, this relatively
high value of the Yukawa coupling plays an important role in the determination
of the diquark pole mass, and combined with the diquark curvature mass $m_\Delta
\sim 1 \GeV$, yields a pole mass below twice the quark mass $2 m_q \simeq 700\MeV$
which is consistent with the quark-diquark picture of the nucleon.

\begin{table}[t]
  \centering
  \bgroup
  \def\arraystretch{2.2}
\begin{tabular}{c @{\hspace{1em}} c c c c c @{\hspace{1em}} c}
    \multirow{2}{*}{\makecell{\textbf{Low-energy} \\ \textbf{Constants}}}
    & \multicolumn{3}{c}{\textbf{\gls{QCD} Value}} 
    & \multirow{2}{*}{$[\MeV^n]$}
    \\
    & $r_{\text{flat}}$ 
    & $r_{\text{exp},2}$ 
    & $r_{\text{exp},1}$
    &
    \\ \hline\hline
    $\langle\sigma\rangle$
    & $71.55$ 
    & $78.69$ 
    & $80.30$ 
    & $n=1$
    \\
    $g_{\phi\bar{q}q}$
    & $4.89$ 
    & $4.44$ 
    & $4.36$
    & $n=0$ 
    \\
    $g_{\Delta q q}$
    & $10.32$ 
    & $9.42$ 
    & $9.27$ 
    & $n=0$
    \\
    $\partial_\sigma^2\,\Omega$
    & $\big( 435.29 \big)^2$ 
    & $\big( 439.24 \big)^2$ 
    & $\big( 440.81 \big)^2$ 
    & $n=2$
    \\
    $\frac{1}{2} \partial_\Delta^2\,\Omega$
    & $\big( 1029.72 \big)^2$
    & $\big( 1112.67 \big)^2$
    & $\big( 1187.00 \big)^2$ 
    & $n=2$
    \\
    $\frac{1}{6} \partial_\sigma^4\,\Omega$
    & $6.56$
    & $2.47$
    & $8.70$
    & $n=0$
    \\
    $\frac{1}{2} \partial_\sigma^2\,\partial_\Delta^2\,\Omega$
    & $41.43$
    & $32.36$ 
    & $23.01$ 
    & $n=0$
    \\
    $\frac{1}{24} \partial_\Delta^4\,\Omega$
    & $36.99$
    & $31.30$ 
    & $28.14$ 
    & $n=0$
  \end{tabular}
  \egroup
  \caption{Definition of selected low-energy constants and their
    extracted values from our \gls{QCD} computation.
      All quantities are
    evaluated in the infrared. The effective potential is
    defined as $\Omega = U_{k=0} - h\sigma$. All quantities given in 
    an \gls{RG}-invariant manner,
    i.e., rescaled by the appropriate powers of the infrared meson
    and/or diquark wave-function renormalizations. As a minimal
    estimate of the systematic uncertainty of the computation, we
    quote the variation of the results obtained with different
    regulators.}
  \label{tab:qcd_parameters}
\end{table}

Finally, in \cref{fig:quartic_couplings} we show the flow of the meson-meson,
diquark-meson and diquark-diquark quartic couplings. We find the
diquark-diquark coupling values are higher than the meson-meson and meson-diquark
ones, supporting a stiffer effective potential in the $\rho_\Delta$ direction
compared to the $\rho_\phi$ direction. This result is consistent with the sub-leading role
played by the diquark sector in vacuum \gls{QCD}. And also here, the approximate plateaus at large $k$ relate to the underlying weak-coupling fixed point. We will discuss this in more detail in the next section, this is key for the predictive power of our approach.

In addition to the Yukawa
couplings in \cref{fig:yukawa_couplings} and the curvature mass in
\cref{fig:masses_flow}, these are key quantities needed to fix parameters
of commonly used \gls{LEM} of \gls{CSC} quark matter, and forms one of
the main results of our study. For convenience, we report in \cref{tab:qcd_parameters} the quark-meson coupling, quark-diquark coupling and the derivatives of the effective potential up to the fourth order for the three regulator employed in this work.

\subsection{Independence on the UV parameters}
\label{sec:indep_params_uv}

\begin{figure}[t]
  \centering
  \includegraphics[width=\onefig]{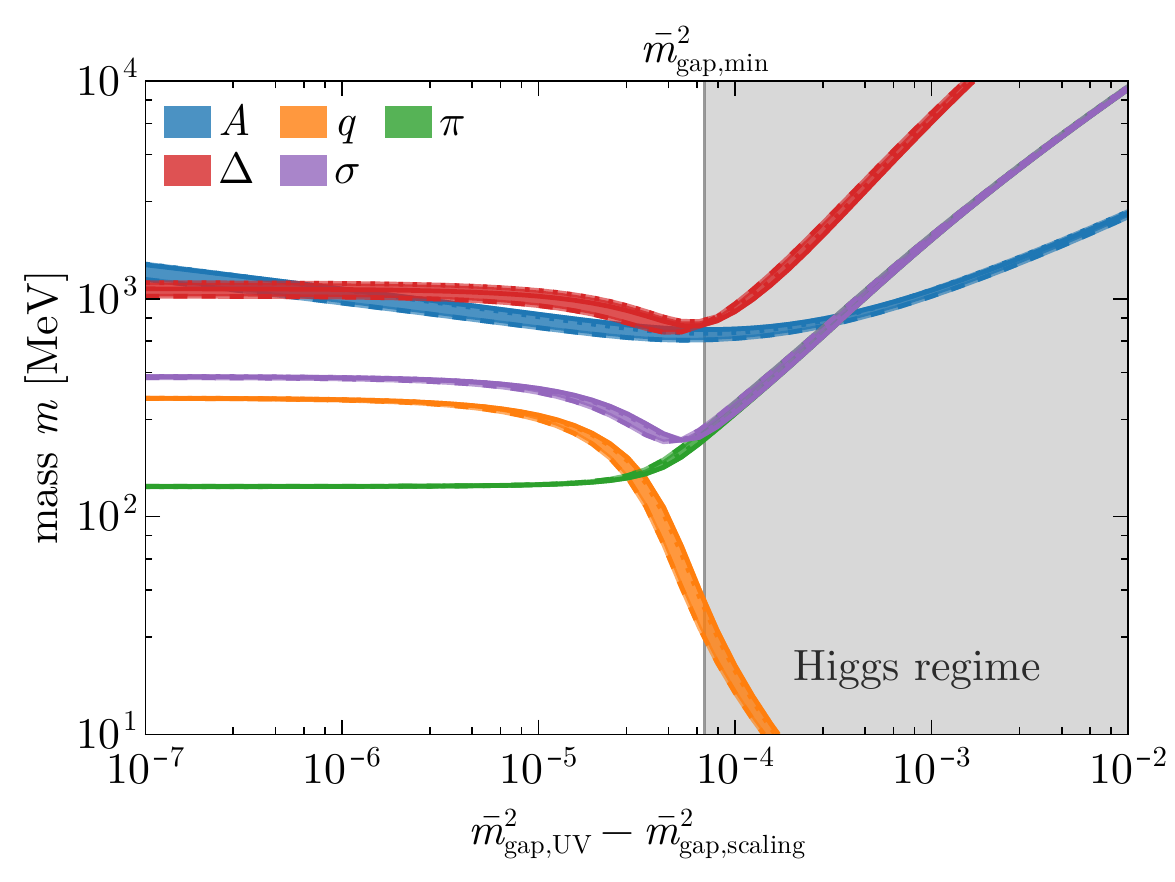}
  \caption{
  Mass of the different fields considered in this work
    against the difference between the dimensionless mass gap at the
    \gls{UV} scale, $\bar{m}^2_{\text{gap},\text{UV}}$, and
    the \gls{UV} dimensionless mass gap yielding the scaling solution,
    $\bar{m}^2_{\text{gap},\text{scaling}}$. The gluon mass parameter is given by $m_{\text{gap}}$. The minimum of
    $m_{\text{gap}}$ is indicated by a vertical line, and yields the
  delimitation between the confining (left) and Higgs branch (right).}
  \label{fig:masses_vs_uv_gap}
\end{figure}

In this section we show the predictive power of our approach by
demonstrating the independence of the infrared bound-state properties
on the choice of the initial \gls{UV} values for the gluon mass gap
and the couplings in the composite sector. Hence, as appropriate for
\gls{QCD}, all quantities, including emergent ones, are fully
determined by the initial values for the strong coupling and the
current quark masses.  For the gluon mass gap, this entails a
discussion of the scaling and decoupling solutions, and the
independence of our results on the choice of a specific solution. For
the properties of the composite fields, we explicitly confirm that the
weak-coupling fixed point at intermediate scales in conjunction with
dynamical hadronization leads to unique predictions.

\begin{figure}[t]
  \centering
  \includegraphics[width=\twofigs]{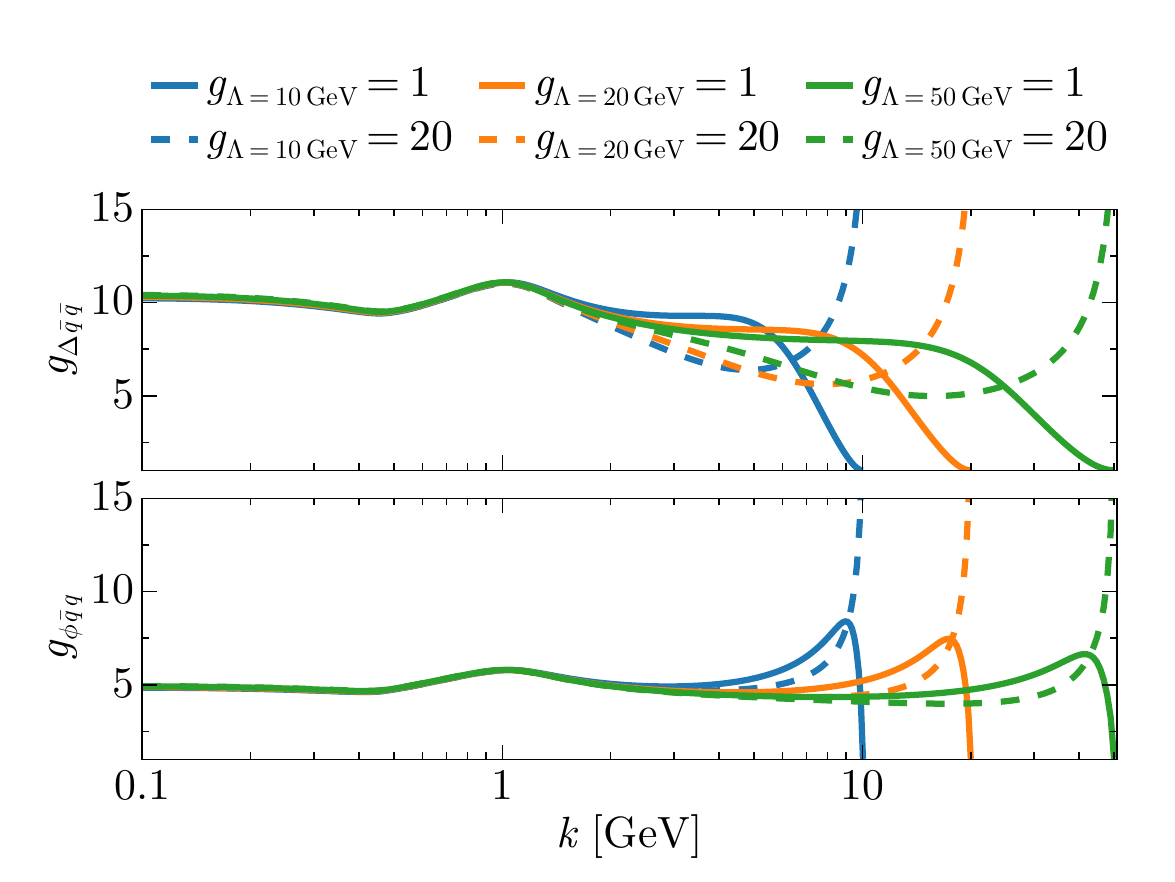}
  \includegraphics[width=\twofigs]{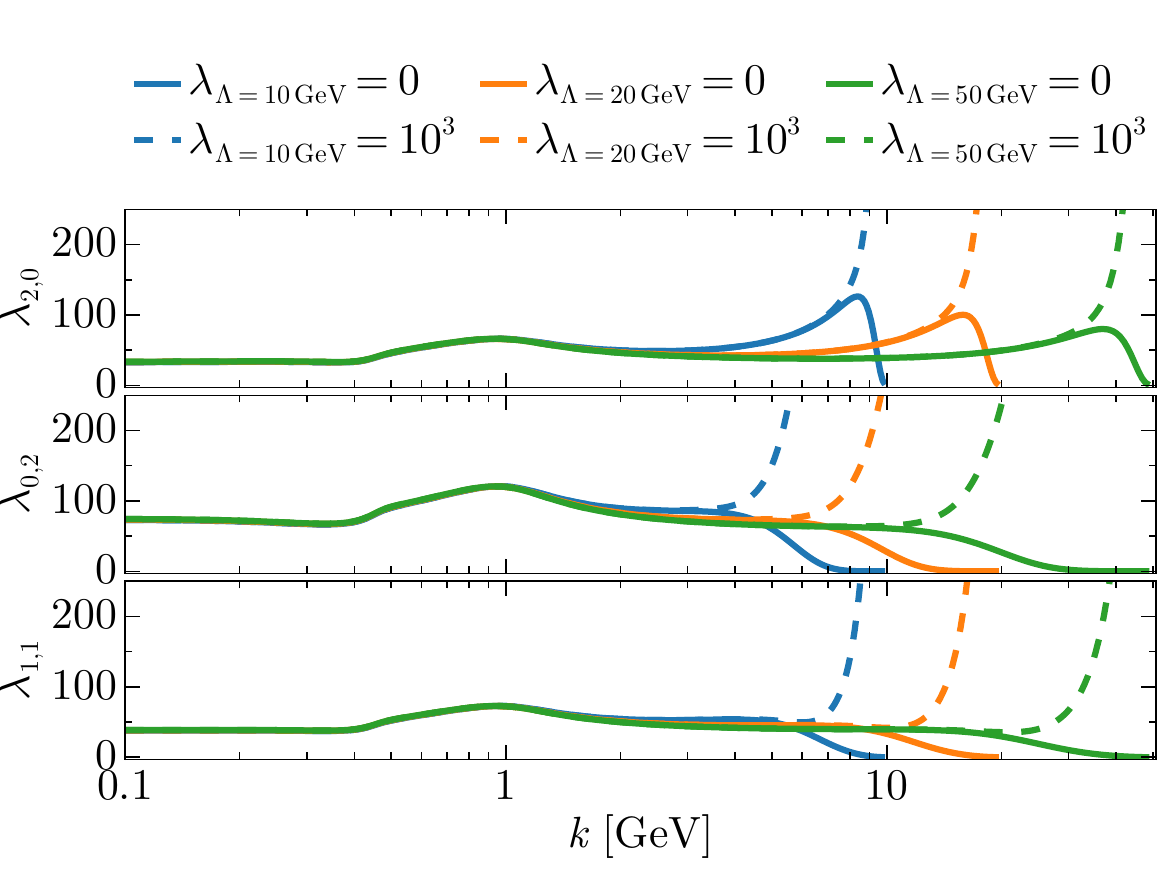}
  \caption{ Quark-meson $g_{\phi\bar{q}q}$ and quark-diquark
    $g_{\Delta\bar{q}\bar{q}}$ Yukawa couplings (top) and quartic
    meson-meson $\lambda_{2,0}$, diquark-diquark $\lambda_{0,2}$ and
    meson-diquark $\lambda_{1,1}$ couplings (bottom) as a function of
    the \gls{RG}-scale $k$ for different initial \gls{UV} conditions
    $\Lambda$ with
    $g_{\phi\bar{q}q}=g_{\Delta\bar{q}\bar{q}}=g_\Lambda$ and
    $\lambda_{2,0}=\lambda_{0,2}=\lambda_{1,1}=\lambda_{\Lambda}$. For
    clarity, we present only results obtained with the flat regulator
    shape function, but the same conclusion holds for any regulator.}
  \label{fig:couplings_diffuv}
\end{figure}

In \cref{fig:masses_vs_uv_gap}, we show the curvature mass of all
fields against the difference between the \gls{UV} mass gap,
$\bar{m}^2_{\text{gap},\text{UV}}$, and the \gls{UV} gap corresponding
to the scaling solution, $\bar{m}^2_{\text{gap},\text{scaling}}$. We
find that the mass gap parameter $m_\text{gap}$ displays a minimum for
a specific \gls{UV} value
$\bar{m}^2_{\text{gap},\text{UV}}=\bar{m}^2_{\text{gap},\text{min}}$. This
point corresponds to the delimitation between the Higgs regime, where
the gluon propagator is that of a massive vector particle, and the
confining regime, where the gluon propagator peaks at nonzero
momentum, see, e.g., \ccites{Cyrol:2016tym,Goertz:2024dnz} for
details. In the Higgs regime, we find no sign of \gls{DCSB}: the quark
mass tends towards the current quark mass and the pion and
$\sigma$-meson masses are degenerate. Furthermore, the mass of all
composite fields are large as the strong interaction is suppressed by
the large effective gluon mass.

For
$\bar{m}^2_{\text{gap},\text{UV}} \lesssim
\bar{m}^2_{\text{gap},\text{min}}$ we enter the confining
regime. However, close to $m^2_{\text{gap},\text{min}}$ we do not have
\gls{DCSB} as the pion and $\sigma$-meson masses are still
degenerate. For smaller $\bar{m}^2_{\text{gap},\text{UV}}$, we find
\gls{DCSB} with a non-negligible quark mass and non-degenerate pions
and $\sigma$-mesons. We emphasize that this behavior indicates a
direct relation between confinement and \gls{DCSB}.  In general,
\gls{DCSB} requires strong four-quark interactions which are only
obtained in the confining regime, see also \ccite{Goertz:2024dnz}.

For $\bar{m}^2_{\text{gap},\text{UV}}$ well below
$\bar{m}^2_{\text{gap},\text{min}}$, we find that all the masses, with
the exception of $m_\text{gap}$ itself, are independent of the value
of the \gls{UV} mass gap parameter. All solutions in this region
correspond to decoupling, and they are identical to those
corresponding to scaling. Thus, the precise choice of a confining
solution has no effect on physical observables. This confirms that
choosing the scaling solution to fine-tune
$\bar{m}^2_{\text{gap},\text{UV}}$ is a well-defined and, crucially,
unique procedure. Because $m_\text{gap}$ corresponds to the transverse
part of the gluon propagator at vanishing momenta, it is
\gls{RG}-variant and gauge dependent. It therefore depends on
$\bar{m}^2_{\text{gap},\text{UV}}$ even within the confining regime.

In \cref{fig:couplings_diffuv}, we show the flow of the quark-meson
and quark-diquark Yukawa couplings (upper panels), as well as the
three composite quartic couplings (lower panels), for different
initial \gls{UV} scales, $\Lambda = 10 \GeV$, $20 \GeV$ and $50 \GeV$,
and for initial coupling values differing by several orders of
magnitude.  In all cases, we follow the parameter-fixing procedure
detailed in \cref{sec:parameter-fixing}, ensuring that all results
reproduce the same \gls{IR} pion and quark masses. For all initial
conditions, the flows at large $k$ are attracted to the
intermediate-scale fixed point, such that for $k \lesssim 1 \GeV$ they
agree quantitatively; see also
\ccites{Gies:2006wv,Braun:2014ata,Rennecke:2015lur}.  Note that the
relevant \gls{UV} parameters of the composite sector are determined
through the effective four-quark couplings in
\cref{eq:lambdaeff}. Hence, \cref{eq:lambdaeff} implies that varying
the \gls{UV} Yukawa couplings is equivalent to varying the \gls{UV}
masses of the composite fields.  All in all, these results explicitly
demonstrate that the emergent properties of \gls{QCD} are independent
of the choice of initial conditions. We emphasize that, although its
precise value is unimportant for the resulting \gls{IR} physics, the
initial scale $\Lambda$ must nevertheless lie sufficiently deep in the
perturbative regime.

\section{Real-Time Correlation Functions}
\label{sec:composite-two-point}

A striking result from our \gls{QCD} computation is the large value for the diquark
curvature mass, $m_\Delta \sim 1 \GeV$. As discussed in \cref{sec:introduction}, this seems surprising because a value around twice the constituent quark mass, $\lesssim 700$\,MeV, appears to be natural. In this section, we clarify this discrepancy. To this end, we will compute the diquark two-point function in real time. For completeness, we will also do the same for the mesons.   

Within our approach, this pose two difficulties. First, our
computation is performed within a zero-momentum expansion. As
such we do not have access to correlation functions at finite
external momenta. Second, our approach is based on Euclidean time and
getting access to real time quantities (such as pole masses) requires
analytic continuation and additional care regarding the choice of regulator. Because of these two difficulties,
we focus on providing qualitative estimates here, while the quantitative focus remains on correlation functions at vanishing momenta. As explained previously, such quantities are readily accessible by \gls{LEM}.

\subsection{Curvature vs pole masses}
\label{sec:curv-vs-pol}

 A simple way to estimate the mass of the diquark field is based on the quark-diquark
picture of the nucleons. Within this picture, one
naturally expect the mass of the diquark to be
around, or slightly below twice the constituent quark mass ($350\MeV$ in this work), such that the quark and diquark form a bound state with enough
binding energy to reproduce the correct nucleon mass $m_N \sim
940\MeV$. Previous estimates for the scalar diquark mass support this
picture and predict a value around twice the quark mass \citeDiquarkMass.
However, as pointed out in \cref{sec:introduction}, such estimates apply to the diquark \textit{pole mass}, defined in \cref{eq:mpole}, and it is not obvious how it relates to the \textit{curvature mass}, see \cref{eq:mcurv},
computed here.

In order to understand the origin of possible differences between pole and curvature mass, it is instructive to consider the Euclidean meson or diquark propagator $G(p_0)$ in the vicinity of the relevant frequency $p_0$. We can set the spatial momenta to zero for our argument.
The propagator can in general be expressed as
\begin{align}
G(p_0) = \frac{1}{Z^\parallel(p_0) p_0^2 +m^2+\Delta\Pi(p_0)}\,,
\end{align} 
where $m$ is the bare mass and we split the self-energy into a temporal wave function renormalization correction, $Z^\parallel$, and all the rest, $\Delta\Pi$. Expanding these corrections to leading order around a frequency $p_{\rm exp}$ yields
\begin{align}
G(p_0) \approx \frac{1}{Z^\parallel(p_{\rm exp})p_0^2 + m^2 + \Delta\Pi(p_{\rm exp})}\,.
\end{align}
To access the pole and curvature masses defined in \cref{eq:mpole,eq:mcurv} from the Euclidean propagator, we set $p_{\rm exp} = -i m_{pole}$ and $p_{\rm exp} = 0$, respectively. We then readily read off the \gls{RG}-invariant relations
\begin{align}
\begin{split}
m_{\rm pole} &= \frac{m^2+\Delta\Pi(-i m_{\rm pole})}{Z^\parallel(-i m_{\rm pole})}\,,\\
m_{\rm curv} &= \frac{m^2+\Delta\Pi(0)}{Z^\parallel(0)}\,.
\end{split}
\end{align}
The difference between pole and curvature mass clearly lies in the momentum dependence of the self-energy corrections. If this dependence is weak and/or the pole mass is close to zero, pole and curvature mass can be very similar. This is, in fact, the case for pions, where the pole is indeed close to the origin compared to the rest of the low-energy spectrum \cite{Helmboldt:2014iya, Fu:2024rto}.

In turn, for heavy states it seems more natural to expect a considerable difference between pole and curvature masses. As we demonstrate in the following, this turns out to be true for the scalar diquark.

\begin{figure*}[t]
  \centering
  \includegraphics[width=0.89\linewidth]{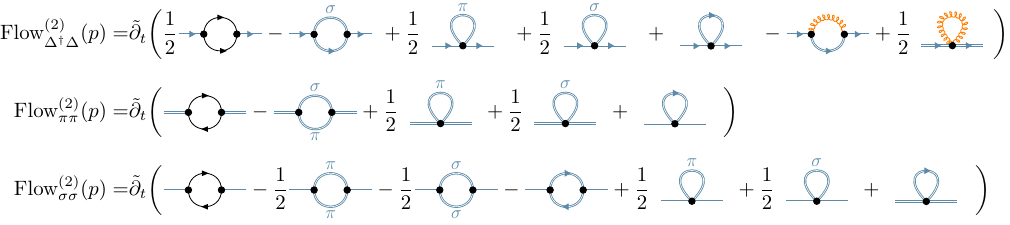}
  \caption{
    Diagrammatic
    representation of the flows of the momentum-dependent
    two-point functions of the diquark, pion, and $\sigma$-meson
    fields. The conventions are the same as in \cref{fig:two_point_flows}.}
  \label{fig:two_point_flows_finite_p}
\end{figure*}

\subsection{Flows of momentum-dependent two-point functions}

In order to analytically continue the Euclidean two-point functions to Minkowski space, we need access to their full momentum dependence. Within our zero-momentum
expansion, we therefore solve three additional flow equations for the pion, sigma
and diquark two-point functions at finite external momenta
\begin{align}
  \label{eq:Gamma2EDelta}
  \partial_t \Gamma^{(2),E}_{\Delta^\dagger\Delta}(p) & =
  \eta_\Delta \Gamma^{(2),E}_{\Delta^\dagger\Delta}(p)
  + \text{Flow}_{\Delta^\dagger\Delta}^{(2)}(p) \c \\[2ex]
  \label{eq:Gamma2Epi}
  \partial_t \Gamma^{(2),E}_{\pi\pi}(p) & =
  \eta_\phi \Gamma^{(2),E}_{\pi\pi}(p)
  + \text{Flow}_{\pi\pi}^{(2)}(p) \c \\[2ex]
  \label{eq:Gamma2Esigma}
  \partial_t \Gamma^{(2),E}_{\sigma\sigma}(p) & =
  \eta_\phi \Gamma^{(2),E}_{\sigma\sigma}(p)
  + \text{Flow}_{\sigma\sigma}^{(2)}(p) \d
\end{align}
The superscript ${}^E$ denotes the Euclidean quantities. The diagrammatic representation of these flows is shown in \cref{fig:two_point_flows_finite_p}. We use the procedure put forward in \cite{Strodthoff:2011tz, Kamikado:2013sia} and use our solutions for the $k$-dependent vertices and propagators obtained in the fully coupled system at zero momentum as input for  the diagrams shown in \cref{fig:two_point_flows_finite_p}. This will give us the full momentum dependence of the two-point function based on the zero-momentum solutions. This procedure is not self-consistent as the zero-momentum masses and couplings are used instead of fully momentum dependent dressing functions. However, it still provides a good approximation to the momentum dependence of the full correlation functions \cite{Jung:2021ipc}.

In contrast to the rest of the flow equations, we employ here a three
dimensional regulator which leaves the temporal component of
the four-momentum unaffected,
\begin{align}
  R_{\bar{q}q}^{3d}(p) & = i\slashed{\vec{p}} \left( -1 + \sqrt{1 +
  r_q(x_{3d})} \right) \; , \\[2ex]
  \Big[ R_{\phi\phi}^{3d}(p) \Big]_{ij} & = \sqvec{p} r_\phi(x_{3d})
  \delta_{ij} \; , \\[2ex]
  \Big[ R_{\Delta^\dagger\Delta}^{3d}(p) \Big]_{\da\db} & =
  \sqvec{p}r_\Delta(x_{3d})\delta_{\da\db} \c
\end{align}
with $x_{3d} = \sqvec{p}/k^2$. The flow equation resulting from this
choice can be found in \cref{app:flow_two_point}. 
A three dimensional
regulator ensure that the pole structure of the two-point functions
in the complex $p_0$-plane is not affected by potential fictitious
poles introduced by four dimensional regulators
\cite{Pawlowski:2015mia}. As in the zero-momentum \gls{QCD} expansion, we
choose the same shape function for all fields. To estimate the systematic error, we use different shape functions also here, see \cref{app:regulators}. 

For the diquark two-point function we ignore all gluonic contributions to the flow as we expect them to be subleading, consistent with the effect of the diquark-gluon interactions in the zero-momentum expansion, see \cref{sec:numerical-results-qcd}. Furthermore, the mixed gluon-diquark diagram, i.e.\ the second-to-last diagram in the first line of \cref{fig:two_point_flows_finite_p}, may seem to contain a threshold for the process $\Delta \to \Delta + g$. In the current approximation, and since we only consider correlation function at vanishing spatial momenta $\vec{p}=0$, this threshold is located at the sum of the diquark and gluon curvature masses, $m_{\Delta,\mathrm{curv}}$ + $m_{\mathrm{gap}}$. Since the mass gap parameter diverges for the scaling solution, this threshold is removed from the spectrum and can hence not affect the diquark pole mass. 

Choosing $3d$ regulators for \cref{eq:Gamma2EDelta,eq:Gamma2Epi,eq:Gamma2Esigma}, but using input obtained with $4d$ regulators can lead to inconsistencies. As a direct consequence, the curvature masses resulting from
\cref{eq:Gamma2EDelta,eq:Gamma2Epi,eq:Gamma2Esigma} and the
zero-momentum \gls{QCD} expansion differ, although they should match by
definition. For the diquark two-point functions this
inconsistency results in a small variation of the curvature mass of around
$100\MeV$, while for the pion and $\sigma$-meson the difference is more
drastic, as both curvature masses turn negative.

The curvature masses are given by the curvature of the effective potential in the corresponding field direction, evaluated at its minimum, see \cref{eq:Uk_definition,eq:curv_mass_pion,eq:curv_mass_sigma}. The inconsistency in regularization schemes in \cref{eq:Gamma2EDelta,eq:Gamma2Epi,eq:Gamma2Esigma} apparently leads to a mismatch between the underlying evaluation point and the minimum of the potential. A negative meson curvature mass implies that the potential is evaluated at a field value below its minimum in $\rho_\phi$-direction. Hence, this mismatch can in principle be resolved by adjusting the location of the minimum through the explicit symmetry breaking parameter $h$. Since the effective potential is the zero-momentum part of the effective action, this leads to constant shifts in the two-point functions of mesons and diquarks, while leaving the momentum-dependence unaffected.

In practice, we can therefore directly apply constant shifts to the two-point functions so that the curvature masses obtained with the $3d$ regulators match the ones with $4d$ regulators,
\begin{align}
  \label{eq:DeltaGamma2EDelta}
  \Delta\Gamma^{(2),E}_{\Delta^\dagger\Delta}(p) & =
  \Gamma^{(2),E}_{\Delta^\dagger\Delta}(p)
  - \Gamma^{(2),E}_{\Delta^\dagger\Delta}(0)
  + m^2_\Delta \; , \\[2ex]
  \label{eq:DeltaGamma2Epi}
  \Delta\Gamma^{(2),E}_{\pi\pi}(p) & =
  \Gamma^{(2),E}_{\pi\pi}(p)
  - \Gamma^{(2),E}_{\pi\pi}(0)
  + m^2_\pi \; , \\[2ex]
  \label{eq:DeltaGamma2Esigma}
  \Delta\Gamma^{(2),E}_{\sigma\sigma}(p) & =
  \Gamma^{(2),E}_{\sigma\sigma}(p)
  - \Gamma^{(2),E}_{\sigma\sigma}(0)
  + m^2_\sigma \; .
\end{align}
With this prescription, we find very good agreement between the
pion pole and curvature mass, as expected from other studies
\cite{Helmboldt:2014iya, Fu:2024rto}. As mentioned above, the modification of the diquark two-point function is only minor, and hence irrelevant for our estimate of the diquark pole mass.

It may be possible to directly find the pole mass from
\cref{eq:DeltaGamma2EDelta,eq:DeltaGamma2Epi,eq:DeltaGamma2Esigma} at
imaginary Euclidean frequency,
\begin{equation}
  \Delta \Gamma^{(2),E}_{\Phi_i\Phi_j}
  (p_0 = -i m_{\text{pole}}, \vec{p}=0)
  = 0 \; .
\end{equation}
This works for single particle poles, but fails for resonances where proper analytic continuation is necessary to account for the branch cuts from multi-particle thresholds. 
In the present approximation, the location of these thresholds is determined by the curvature masses of the different fields. Clearly, in a self-consistent treatment of the momentum-dependence, it would be the pole masses. For example, the cut for the decay into two quarks starts at
$2m_q \simeq 700 \MeV$ and for two pions at $2m_\pi \simeq 274\MeV$.

To access the full spectrum, we follow the direct analytic continuation of flow equations first introduced in \ccites{Kamikado:2013sia,Tripolt:2013jra}, and recently
used in \ccite{Fu:2024rto} in a setting similar to ours. This procedure exploits the simple one loop structure of the flow equations
to provide the correct $i\epsilon$ prescription for the analytic continuation of Euclidean to retarded two-point functions,
\begin{equation} \label{eq:retarded_2p_def}
  \Gamma^{(2),R}_{\Phi_i\Phi_j}(\omega) =
  \lim_{\epsilon \to 0^+}
  \Gamma^{(2),E}_{\Phi_i\Phi_j}\big(p_0 = -i(\omega + i\epsilon),
  \vec{p} = 0\big) \; .
\end{equation}
We keep $\epsilon=0.1\MeV$, and focus on vanishing spatial momentum as this is sufficient for our purposes. It is convenient to compute the spectral function from the retarded propagator,
\begin{equation}
  \rho = - \frac{1}{\pi} \text{Im} \, G^R \; .
\end{equation}
This leads to
\begin{align}
  \label{eq:spectral_def}
  \rho_{\Phi_i\Phi_j} & = - \frac{1}{\pi} \frac{\text{Im} \,
  \Gamma^{(2),R}_{\Phi_i\Phi_j}}{
    \left( \text{Re} \, \Gamma^{(2),R}_{\Phi_i\Phi_j} \right)^2 +
    \left( \text{Im} \, \Gamma^{(2),R}_{\Phi_i\Phi_j} \right)^2
  } \; .
\end{align}
From the spectral function $\rho_{\Phi_i\Phi_j}$, we can directly read off all information regarding single-particle excitations, resonances and decay channels of the field $\Phi$.

For simple truncations, such as the local potential approximation, this procedure provides a powerful way to compute spectral functions. However, difficulties can arise for more complex approximations. These difficulties are a direct consequence of the lack of self-consistency of the procedure: some of the finite momentum behavior is extracted from the \gls{RG}-scale dependence of zero-momentum correlation functions, and not from the genuine momentum structure. 

In the present work, we encounter these problems in two distinct forms. First, for the exponential shape function, \cref{eq:exp_shape_function}, the $k$-dependent quark energy $\epsilon_q = \sqrt{k^2 + m_q^2}$ has a local minimum at nonzero $k$, which in turns leads to a singularity reminiscent of a Van-Hove singularity \cite{PhysRev.89.1189}. This singularity is an approximation artifact, as self-consistent computation of the momentum dependent quark-mass indicate a monotonous quark energy, see \ccites{Bowman:2005vx,Chang:2021vvx,Cyrol:2017ewj,Fu:2025hcm,Ihssen:2024miv}. 
A similar behavior was already observed in \ccite{Jung:2016yxl}. The second problem is encountered for all shape functions, and is caused by the non-monotonic behavior of the quark-meson Yukawa coupling, see \cref{fig:yukawa_couplings}, which induces spurious oscillation in the retarded two-point functions for $p \gtrsim 800 \MeV$. 

\begin{figure*}[t]
  \centering
  \includegraphics[width=\columnwidth]{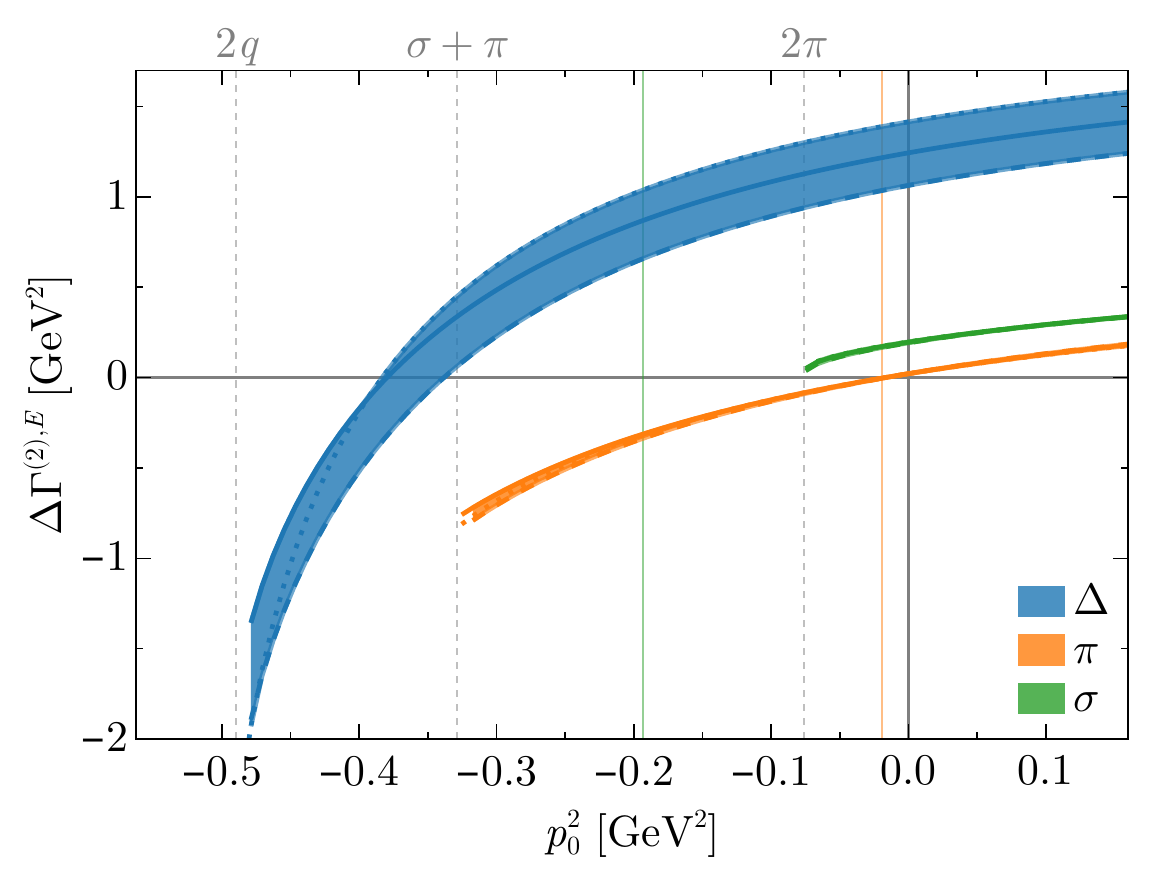}
  \includegraphics[width=\twofigs]{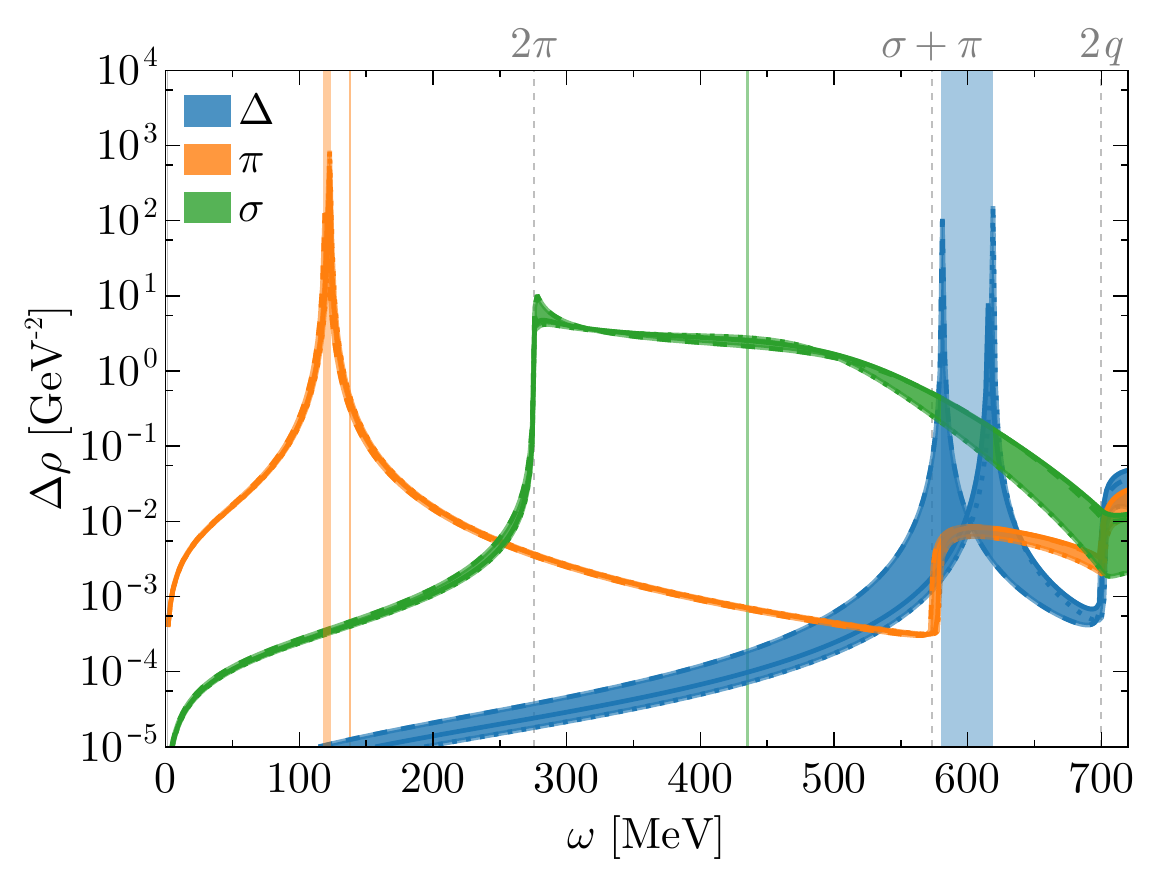}
  \caption{ Euclidean two-point functions $\Delta\Gamma^{(2),E}$ as
    functions of the squared Euclidean frequency $p_0^2$ (left) and
    spectral functions $\Delta\rho$ as functions of the real-time
    frequency $\omega$ (right) for the diquark, pion, and sigma-meson,
    obtained from the $3d$-regulated flow.  Gray dashed lines indicate
    the thresholds of different multi-particle processes, while the
    colored solid lines denote the corresponding curvature masses.}
  \label{fig:two_point_spectral}
\end{figure*}

Based on this two observations, we restrict our results to external momenta $p < 800 \MeV$. Furthermore, we compare the analytically continued results to the Euclidean results to ensure that analytical continuation artifacts are not present in the range of interest, see \cref{fig:two_point_spectral}.

\subsection{Numerical results and discussion}\label{sec:numerical_spec}

We now present the numerical results we obtain for the analytically
continued two-point functions. In \cref{fig:two_point_spectral} (left) we show the shifted euclidean two-point functions $\Delta \Gamma^{(2),E}$ of the composite fields defined in \cref{eq:DeltaGamma2EDelta,eq:DeltaGamma2Epi,eq:DeltaGamma2Esigma} against the squared temporal Euclidean momenta $p_0^2$. Negative values of $p_0^2$ corresponds to imaginary temporal Euclidean momenta, i.e., Minkowski time. As explained above, this procedure is only valid until the first branch cut on the real frequency axis, which is located at the smallest threshold for the decay of the associated field. These thresholds are indicated by gray dashed vertical lines in the figure. By definition of the shifted two-point functions, the value of the two-point function at vanishing momenta $p_0=0$ corresponds to squared curvature mass of the associated field.

A zero crossing of $\Delta \Gamma^{(2),E}$ for negative $p_0^2$ indicates the pole mass of the associated field. Asexpected for \gls{QCD}, we find a small pole mass for the pion field, and no isolated pole for the $\sigma$-meson, confirming that it is a resonance. As mentioned above, we also confirm that the pion pole mass is close to its curvature mass (indicated by the vertical solid orange line). Note that these observations justify a posteriori the shifting prescription defined in \cref{eq:DeltaGamma2EDelta,eq:DeltaGamma2Epi,eq:DeltaGamma2Esigma}. Furthermore, we find a relatively small regulator dependence for the Euclidean two-point functions of the pion and the $\sigma$-meson, supporting the idea that the current truncation can be used for quantitative predictions, at least in the \gls{SPS} sector.

We now turn to the analytically continued diquark two-point function,
which constitutes the central  result of this section. For all regulators, we
find a diquark pole mass below the two-quark threshold,
\begin{equation}
m_{\Delta,\text{pole}} = 580\text{--}619\MeV \c 
\end{equation}
consistent with the diquark as a two-quark bound state. In contrast to
the pion, the diquark curvature mass differs significantly from its
pole mass, see \cref{tab:pole_masses}.
Following our discussion in \cref{sec:curv-vs-pol}, this
indicates that the diquark self-energy corrections have a strong
frequency dependence. Furthermore, consistent with the results
obtained for the diquark curvature mass, we also find a non-negligible
regulator dependence of the corresponding pole mass. Importantly, we
predict a diquark bound state for all regulator choices considered. We
have explicitly verified that the shifting prescription in
\cref{eq:DeltaGamma2EDelta} has a negligible impact on the diquark
pole mass.

\begin{table}[t]
  \centering
  \bgroup
  \def\arraystretch{1.5}
    \begin{tabular}{c @{\hspace{2em}} c @{\hspace{1em}} c @{\hspace{1em}} c @{\hspace{1em}}}
    $[\mathrm{MeV}]$
    & $r_{\text{flat}}$ 
    & $r_{\text{exp},2}$ 
    & $r_{\text{exp},1}$
    \\ \hline\hline
    $m_{\pi,\mathrm{pole}}$
    & $120$ 
    & $122$ 
    & $123$ 
    \\
    $m_{\pi,\mathrm{curv}}$
    & $137$ 
    & $137$ 
    & $137$ 
    \\ \hline
    $m_{\Delta,\mathrm{pole}}$
    & $580$ 
    & $616$ 
    & $619$
    \\
    $m_{\Delta,\mathrm{curv}}$
    & $1030$ 
    & $1113$ 
    & $1187$
  \end{tabular}
  \egroup
  \caption{Pion and diquark pole masses for different regulators,
    extracted from the composite two-point functions shown in
    \cref{fig:two_point_spectral}. The corresponding curvature masses are also shown for comparison.}
  \label{tab:pole_masses}
\end{table}

The diquark pole mass constrains the phase structure of cold
  \gls{QCD} matter through the Silver-Blaze property
  \cite{Cohen:2003kd}, which predicts the onset of diquark
  condensation at $\mu_B = 2 m_{\Delta,\mathrm{pole}} / 3$, provided
  that no first-order transition occurs at lower chemical potential;
  see, e.g., \ccite{Mire:2026auc}.  Since nuclear matter sets in at
  $\mu_B = m_N - E_B = 923 \MeV$ (using the measured nucleon mass and
  binding energy), a diquark pole mass below this value would imply
  diquark condensation prior to the onset of nuclear matter and would
  therefore be phenomenologically excluded.  Our predicted pole mass
  corresponds to a condensation threshold at
  $\mu_B = 870\text{--}929\MeV$, consistent with the onset of nuclear
  matter within our estimated systematic uncertainty.  Moreover,
  beyond the nuclear onset the Silver-Blaze property no longer
  determines the true ground state, so a diquark pole mass slightly
  above $923\MeV$ does not necessarily imply early diquark
  condensation.

By analytically continuing our results according to
\cref{eq:retarded_2p_def}, we obtain access to real-time frequencies
also in the presence of multi-particle branch cuts. In the right panel
of \cref{fig:two_point_spectral} we show the spectral functions of the
composite fields. Since we keep a finite value of $\epsilon=0.1\MeV$
for numerical convenience, the pion and diquark single-particles poles
appear as narrow peaks rather than Dirac delta peaks. Moreover, the
pole masses extracted from the spectral functions agree with those
obtained from the Euclidean two-point functions, corroborating our
results.

We also confirm that the $\sigma$-meson is a very broad resonance. Any
putative single-particle peak in its spectral function appears to be
swallowed by the low-lying $\sigma \rightarrow \pi \pi$ contribution.

The existence of a single-particle pole in the scalar diquark channel
suggests the presence of a stable excitation in the \gls{QCD}
vacuum. Superficially, this appears to be in tension with confinement,
since the diquark field carries a net color charge.  A similar issue
arises in studies of the real-time properties of quark correlators.
However, the connection between the pole structure of colored correlation
functions and confinement remains an open question, particularly in
gauge-fixed settings
\cite{Alkofer:2000wg,Krein:1990sf,Pawlowski:2024kxc}. Furthermore,
similar poles are well-known from Bethe-Salpeter equation studies of
the hadron spectrum \cite{Eichmann:2016yit}.

To summarize, we confirm that the pion is a stable single-particle
excitation with very similar pole and curvature masses, while the
$\sigma$-meson appears as a broad resonance. Most importantly, we find
that the scalar diquark forms a two-quark bound state with a binding
energy of approximately $100\MeV$, consistent with the quark-diquark
picture of the nucleon. Our analysis demonstrates that the low-energy
constants reported in \cref{tab:qcd_parameters}, including the
large diquark curvature mass, are fully consistent with
the expected \gls{QCD} spectrum.

\section{Conclusion and Outlook}
\label{sec:conclusion}

In this work, we present a self-consistent, first-principles framework
for vacuum \gls{QCD} based on the \gls{FRG}. It allows us to smoothly
interpolate between the high-energy quark and gluon degrees of
freedom and a low-energy description in terms of dynamically generated
composite operators. In this way, it establishes a direct
first-principles connection between \gls{QCD} and low-energy
models. 
Key non-perturbative phenomena, including \gls{DCSB} and confinement,
signaled by the emergence of a gluon mass gap, arise naturally within
the framework.
This facilitates the direct computation of the \gls{LECs} required as
input in \gls{LEM}s of \gls{QCD}.

While some model parameters can be constrained by well-established
hadronic properties in vacuum, others remain largely unconstrained. A
prominent example are diquark parameters, which are essential for
describing \gls{CSC} quark matter and therefore play a key role in our
understanding of the interior neutron stars and their mergers.
As a first application of our framework, we determine for the first
time the vacuum properties of the scalar diquark directly from \gls{QCD}.
In particular, we compute its curvature and pole masses, as well as
the corresponding point-like quark-diquark, meson-diquark, and
diquark-diquark couplings.

A key result of this work is the substantial difference between the
diquark curvature and pole masses. While the curvature mass is found
to be $m_{\Delta,\text{curv}} = 1030\text{--}1187 \MeV$, the pole mass lies
in the range $m_{\Delta,\text{pole}} = 580\text{--}619 \MeV$. Since
the latter is significantly below the constituent quark-antiquark
decay threshold, it provides strong support for the quark-diquark
picture of the nucleon in terms of a diquark bound state. Beyond its
conceptual significance, the pronounced discrepancy between the
curvature and pole masses has important implications
for the parameter-fixing procedure of \gls{LEM}s, where theses
quantities are often identified,
at least implicitly.

The natural next step is to apply the derived \gls{LECs} to constrain
\gls{LEM}s of \gls{CSC} quark matter. This significantly enhances
their predictive power by replacing previously free parameters with
quantities determined directly from \gls{QCD}.
In a forthcoming companion study, we use our results to constrain the
renormalized \gls{QMD} model, \ccite{Gholami:2025afm} and the
\gls{RG}-consistent \gls{NJL} model, \ccite{Gholami:2024diy}, enabling
parameter-free predictions for dense \gls{CSC} quark matter.

The numerical efficiency and versatility of our self-consistent
\gls{QCD} framework open several avenues for further improvements and
extensions.
While regulator variations provide a first estimate of the
associated systematic uncertainties, a more comprehensive error
analysis requires also a detailed assessment of truncation effects
\cite{Balog:2019rrg, Ihssen:2024miv}.
Furthermore, it will be
important to
investigate the impact of additional interaction channels, tensor
structures, and the momentum dependence of correlation
functions. Significant progress in this direction has already been
achieved for the gauge and scalar-pseudoscalar sectors
\ccite{Cyrol:2017ewj}. It will be interesting to explore whether
similar improvements extend to other composite degrees of freedom,
such as scalar diquarks.  Finally, we are currently extending our
setup to finite temperature and density along the lines of
\cite{Cyrol:2017qkl, Fu:2019hdw} in order to gain access to the
\gls{QCD} phase diagram and the in-medium properties of hadrons and
other bound states.

\section*{Acknowledgment}

We are indebted to \'Alvaro Pastor-Guti\'errez and Chuang Huang for fruitful discussions and helpful guidance.
We thank Michael Buballa, Christian S.\ Fischer, Wei-jie Fu, Markus Q.\ Huber, Jan M.\ Pawlowski, Ralf Rapp, Lorenz von Smekal, Jonas
Wessely and Jonathan Y.\ Yigzaw for discussions. The authors gratefully
acknowledge support from the Helmholtz Graduate School for Hadron and
Ion Research (HGS-HIRe) for FAIR, the GSI Helmholtzzentrum f\"ur
Schwerionenforschung, and the Deutsche Forschungsgemeinschaft (DFG,
German Research Foundation) through the CRC-TR211 'Strong-interaction
matter under extreme conditions', project number 315477589 – TRR 211.

\section*{Data Availability}

The numerical data presented in all figures in this work are openly
available in the ancillary files of the corresponding arXiv
submission.

\appendix

\section{Regulator choice}
\label{app:regulators}

In this appendix we give more details regarding our regulators.
For clarity, we define the elements of the propagator in field space as
\begin{equation}
  R_{\Phi_i\Phi_j}(p) = \Big( R_k(p) \Big)_{ij} \; ,
\end{equation}
where the regulator matrix is defined in \cref{eq:DeltaSk_def}.
We make the following choices for the regulators,
\begin{align}
  \Big[ R_{AA}(p) \Big]^{ab}_{\mu\nu} & = p^2 r_A(x) \delta^{ab}
  \Pi^{\text{T}}_{\mu\nu}(p) \; , \\[2ex]
  \Big[ R_{\bar{c}c}(p) \Big]^{ab} & = p^2 r_c(x) \delta^{ab} \; , \\[2ex]
  R_{\bar{q}q}(p) & = i\slashed{p} \left( -1 + \sqrt{1 + r_q(x)} \right) \; , \\[2ex]
  \Big[ R_{\phi\phi}(p) \Big]_{ij} & = p^2 r_\phi(x) \delta_{ij} \; , \\[2ex]
  \Big[ R_{\Delta^\dagger\Delta}(p) \Big]_{\da\db} & = p^2
  r_\Delta(x)\delta_{\da\db}\; ,
\end{align}
while all other combinations vanish. With these, all momentum-dependent terms in the propagators are replaced by their
regularized counterparts,
\begin{equation} \label{eq:regularized_momenta_def}
  p^2_{\Phi_i} = p^2 \left( 1 + r_{\Phi_i}(x) \right)  \; ,
\end{equation}
where $r_{\Phi_i}(x)$ is a dimensionless shape function with
$x=p^2 / k^2$ for each field $\Phi_i$.

\begin{figure}
  \centering
  \includegraphics[width=\onefig]{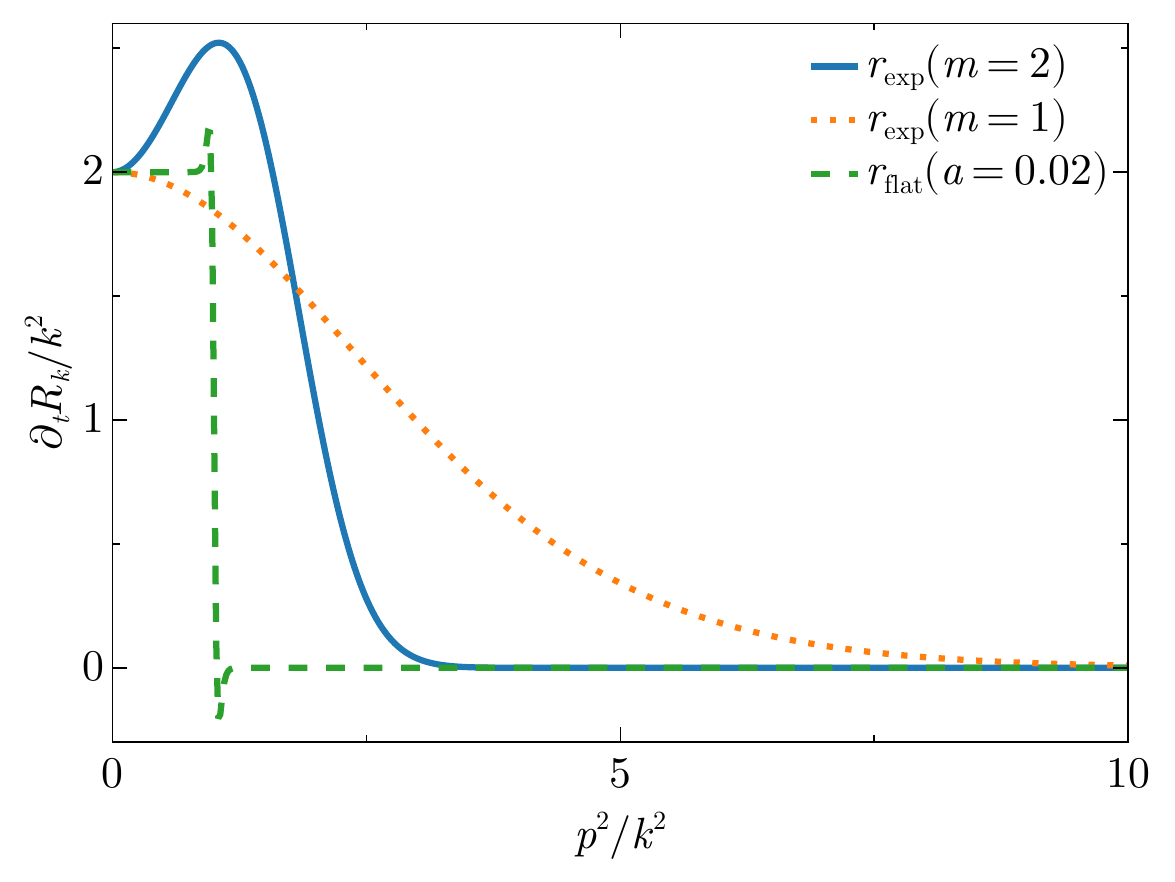}
  \caption{$\partial_t R_k/k^2$ for the bosonic regulator
      $R_k = p^2 r$ as a function of $x=p^2/k^2$ for the two regulator
      shape functions $r=r_{\text{flat}}$ and $r=r_{\text{exp}}$ used
      in this work. This figure illustrates how the different
      regulators probe
      different momentum regions in the flow, essentially determining how sharp the momentum shells in the \gls{RG} flow are. The line style associated with
      each regulator (solid, dotted, or dashed) is used consistently
      throughout most figures in this work.}
  \label{fig:shape-functions}
\end{figure}

In all expression that follow we keep our choice of regulator general, such that different fields can in principle get different shape functions. This may prove useful in choosing different scales for the bosonic and fermionic fields of the theory, see for example
\ccite{Ihssen:2023xlp}. In practice, we use the same shape function for all fields in this work, though, i.e.,
\begin{equation}
  r_A=r_c=r_q=r_\phi=r_\Delta=r \; ,
\end{equation}
where $r$ is a common shape function for every field. In order to
obtain an estimate of our systematic error, we assess the sensitivity of our results to the choice of regulator by considering three
different shape functions. We either employ the smooth flat shape function
\begin{equation} \label{eq:flat_shape_function}
  r_{\text{flat}}(x) =
  \left( \frac{1}{x} - 1 \right)
  \frac{1}{1 + e^{\frac{x-1}{a}}}
  \; ,
\end{equation}
with $a=0.02$, or the exponential shape function
\begin{equation} \label{eq:exp_shape_function}
  r_{\text{exp},m}(x) = \frac{x^{m-1} e^{-x^m}}{1 - e^{-x^m}} \; ,
\end{equation}
with $m=1$ or $m=2$. The behavior of the $t$-derivative of the
regulator associated with the two different shape functions is shown
in \cref{fig:shape-functions}.

\section{Diagrammatics}
\label{app:diagrammatic}

In this appendix, we list all the building blocks needed for the
computation of the flow equations, i.e.\ the propagators and vertices. For all $n$-point functions we choose all momenta to be incoming, which is equivalent to a similar convention for the Fourier transformation of all fields
\begin{equation}
    \Phi(x) = \int \frac{d^4p}{(2\pi)^4} e^{ipx} \Phi(p) \d
\end{equation}

\subsection{Propagators}
\label{app:propagators}

In this section we list the two-point functions and the associated
regularized propagators. We express the
propagators of the different fields through
\begin{equation}
  G_{\Phi_{i}\Phi_{j}}(p) = \Big( G_k(p) \Big)_{ij} \; ,
\end{equation}
where $G_k(p)$ is defined in \cref{eq:propagator_def}.
Furthermore, we show all expressions on a
non-vanishing diquark background, as this is necessary for the
computation of the effective potential, see the discussion in
\cref{sec:eff_pot}. Note than in the following we use the notation defined in \cref{eq:regularized_momenta_def} to denote the regularized momenta.

The gluon propagator is given by
\begin{align} \label{eq:gluon_propagator}
  & \Big[G_{AA}(p)\Big]^{ab}_{\mu\nu} =
  \bigg(\frac{1}{p^2_A + m^2_{\mathrm{gap}} +
  m^2_{\mathrm{Higgs}}}\bigg)^{ab} \Pi^\mathrm{T}_{\mu\nu}(p) \; ,
\end{align}
where we already enforced Landau gauge $\xi=0$ and hence only the
transverse parts remains. In presence of a non-vanishing diquark
condensate, the Higgs mechanism leads to the generation of a mass
term for the gluon field, which for the \gls{2SC} phase (where
$\langle\Delta_\da\rangle = \delta_{\da3}\sqrt{\rho_\Delta}$) reads
\begin{align}
  \left( m_\mathrm{Higgs}^2 \right)^{ab}
  & =
  \lambda_{A^2\Delta^\dagger\Delta}
  \rho_\Delta
  \left( \big\{T^a,T^b\big\} \right)_{33}
  \\[2ex] \nonumber
  & =
  \lambda_{A^2\Delta^\dagger\Delta}
  \rho_\Delta
  \left(
    \diag\Big(
      0, 0, 0,
      \frac{1}{2}, \frac{1}{2}, \frac{1}{2}, \frac{1}{2},
      \frac{2}{3}
    \Big)
  \right)^{ab}
  \; .
\end{align}
The ghost propagator reads
\begin{equation}
  \Big[ G_{c\bar{c}}(p) \Big]^{ab} =
  \frac{1}{p_c^2} \delta^{ab} \; .
\end{equation}

The quark propagator is given by
\begin{equation}
  G_{\bar{q}q} =
  - \frac{m_q - i \slashed{p}_q}{p^2_q + m_q^2 + \Delta_{\text{gap}}^2} P_{rg}
  - \frac{m_q - i \slashed{p}_q}{p^2_q + m_q^2} P_{b}    \; ,
\end{equation}
with the quark mass $m_q=g_{\phi\bar{q}q}\sqrt{2\rho_\phi}$ and the
diquark gap $\Delta_{\text{gap}}=g_\Delta\sqrt{\rho_\Delta}$.
Furthermore, we introduced the red-green projector $P_{rg} =
\diag(1,1,0)$ and the blue projector $P_b = \diag(0,0,1)$ in color space.

The pion propagators is given by
\begin{equation}
  \Big[ G_{\pi\pi}(p) \Big]_{ij} =
  \frac{\delta_{ij}}{p^2_\phi + m^2_\pi} \; ,
\end{equation}
with the pion curvature mass $m^2_\pi = \partial_{\rho_\phi} U_k$.

In presence of diquark condensate, there is a mixing between the
two-point function of sigma meson and the diquark field. The
propagator for the correct physical excitation is found by diagonalization, see
\ccites{Strodthoff:2011tz,Khan:2015puu,Fukushima:2021ctq} for a similar discussion in the \gls{QMD}
model. For simplicity, and since this procedure is only needed for the flow of the
effective potential, we only show the propagators in the absence of
mixing. For the $\sigma$-meson we find
\begin{align}
  G_{\sigma\sigma}(p) =
  \frac{1}{p^2_\phi + m^2_\sigma}
  \; ,
\end{align}
with the sigma curvature mass $m^2_\sigma=\partial_{\rho_\phi} U_k + 2
\rho_\phi \partial_{\rho_\phi}^2 U_k$, while for the diquark field we have
\begin{equation}
  \Big[ G_{\Delta^\dagger\Delta}(p) \Big]_{\da\db} =
  \frac{\delta_{\da\db}}{p^2_\Delta + m^2_\Delta} \; ,
\end{equation}
where the diquark curvature mass is $m^2_\Delta = \partial_{\rho_\Delta} U_k$.

\subsection{Vertices}
\label{sec:diagramatic-vertices}

\subsubsection{Strong coupling avatars}

In the pure glue sector we only consider the classical tensor structure
which are given by
\begin{align}
  \Big[\mathcal{T}_{c\bar{c}A}^{(1)}(p,q)\Big]_{\mu}^{abc}
  = & i f^{abc} q_\mu \; , \\[2ex]
  \Big[\mathcal{T}_{A^3}^{(1)}(p,q)\Big]_{\mu\nu\rho}^{abc}
  = & i f^{abc} \Big[
    (q - p)_\rho \delta_{\mu\nu} \\[2ex] \nonumber
    & - (p + 2q)_\mu \delta_{\rho\nu}
  + (2p + q)_\nu \delta_{\mu\rho} \Big] \; , \\[2ex] \nonumber
  \Big[\mathcal{T}_{A^4}^{(1)}\Big]_{\mu\nu\rho\sigma}^{abcd}
  = & f^{eab} f^{ecd} \Big( \delta_{\mu\rho} \delta_{\nu\sigma} -
  \delta_{\mu\sigma} \delta_{\nu\rho} \Big) \\[2ex] \nonumber & +
  f^{eac} f^{ebd} \Big( \delta_{\mu\nu} \delta_{\rho\sigma} -
  \delta_{\mu\sigma} \delta_{\nu\rho} \Big) \\[2ex] & +
  f^{ead} f^{ebc} \Big( \delta_{\mu\nu} \delta_{\rho\sigma} -
  \delta_{\mu\rho} \delta_{\nu\sigma} \Big) \d
\end{align}
For the quark-gluon vertex, we consider the classical tensor structure
\begin{align}
  \left[ \mathcal{T}_{A\bar{q}q}^{(1)} \right]^{a}_{\mu}
  = & i \gamma_\mu T^a \c 
\end{align}
and the non-classical tensor structure
\begin{align}
  \left[ \mathcal{T}_{A\bar{q}q}^{(4)}(p,q) \right]^{a}_{\mu}
  = & - \slashed{p} \gamma_\mu T^a \d
\end{align}

\subsubsection{Four-quark vertices}

The four-quark tensor structures are given by
\begin{align}
  \nonumber
  \left[ \mathcal{T}^{(\text{sps})}_{\qbqbqq}
  \right]_{\alpha\beta\gamma\delta} & =
  2 \Big[
    \delta_{\beta\gamma} \delta_{\alpha\delta}
    - \left(\vec{\tau}\gamma_5\right)_{\beta\gamma}
    \left(\vec{\tau}\gamma_5\right)_{\alpha\delta} \\[2ex] & \quad
    - \delta_{\beta\delta} \delta_{\alpha\gamma}
    + \left(\vec{\tau}\gamma_5\right)_{\beta\delta}
    \left(\vec{\tau}\gamma_5\right)_{\alpha\gamma}
  \Big] \; , \\[2ex]
  \left[ \mathcal{T}^{(\text{csc})}_{\qbqbqq}
  \right]_{\alpha\beta\gamma\delta} & =
  2  \big( C \gamma_5 i \epsilon_a \tau_2 \big)_{\alpha\beta}
  \big(C \gamma_5 i \epsilon_a \tau_2 \big)_{\gamma\delta} \; .
\end{align}

\subsubsection{Yukawa vertices}

The quark-meson and quark-diquark vertices are given by
\begin{align}
  \left[ \Gamma^{(3)}_{\sigma\bar{q}q} \right]_{\alpha\beta} & =
  - g_{\phi \bar{q} q} \delta_{\alpha\beta} \; ,
  \\[2ex]
  \left[ \Gamma^{(3)}_{\pi\bar{q}q} \right]_{i\alpha\beta} & =
  - i g_{\phi \bar{q} q} \Big( \gamma_5 \tau_i \Big)_{\alpha\beta} \; ,
  \\[2ex]
  \left[ \Gamma^{(3)}_{\Delta\bar{q}\bar{q}} \right]_{\da\alpha\beta} & =
  - g_{\Delta q q} \Big( C\gamma_5 \tau_2 i \epsilon_\da
  \Big)_{\alpha\beta} \; ,
  \\[2ex]
  \left[ \Gamma^{(3)}_{\Delta^\dagger q q} \right]_{\da\alpha\beta} & =
  g_{\Delta q q} \Big( C\gamma_5 \tau_2 i \epsilon_\da \Big)_{\alpha\beta} \; .
\end{align}

\subsubsection{Meson and diquark self-interactions}

The effective potential $U_k(\rho_\phi,\rho_\Delta)$ generates all
the zero-momentum meson-meson, diquark-diquark and diquark-meson
interactions allowed by symmetry. As discussed in \cref{sec:eff_pot}, we expand this potential up to tenth order in the meson and diquark fields.
For the
meson-meson interactions we find
\begin{align}
  \left[ \Gamma^{(3)}_{\pi\pi\sigma} \right]_{ij} & =
  \delta_{ij} \sqrt{2\rho_\phi} U_k^{(2,0)} \; , \\[2ex]
  \left[ \Gamma^{(4)}_{\pi\pi\sigma\sigma} \right]_{ij} & =
  \delta_{ij} \left( U_k^{(2,0)} + 2 \rho_\phi U_k^{(3,0)} \right) \; , \\[2ex]
  \left[ \Gamma^{(4)}_{\pi\pi\pi\pi} \right]_{ijkl} & =
  \Big( \delta_{ij}\delta_{kl} + \delta_{ik} \delta_{jl} +
  \delta_{il} \delta_{jk} \Big) U_k^{(2,0)} \; , \\[2ex]
  \Gamma^{(3)}_{\sigma\sigma\sigma} & =
  \sqrt{2\rho_\phi} \left( U_k^{(2,0)} + 2 \rho_\phi U_k^{(3,0)} \right) \; , \\[2ex]
  \Gamma^{(4)}_{\sigma\sigma\sigma\sigma} & =
  3 U_k^{(2,0)} + 12\rho_\phi U_k^{(3,0)} + 4\rho_\phi^2 U_k^{(4,0)} \; .
\end{align}
For the diquark-diquark and diquark-meson interaction, because of the vanishing diquark condensate in vacuum, the $U(1)_B$ symmetry
constrains every interaction to include both $\Delta$ and
$\Delta^\dagger$. This yields only one possible diquark-diquark interaction
\begin{equation}
  \left[ \Gamma^{(4)}_{\Delta^\dagger\Delta^\dagger\Delta\Delta}
  \right]_{\da\db\dc\dd} =
  \Big( \delta_{\da\dc} \delta_{\db\dd} + \delta_{\da\dd}
  \delta_{\db\dc} \Big) U_k^{(0,2)} \; ,
\end{equation}
while for the diquark-meson interactions we find
\begin{align}
  \left[ \Gamma^{(4)}_{\Delta^\dagger\Delta\pi\pi} \right]_{\da\db ij} & =
  \delta_{\da\db} \delta_{ij} U_k^{(1,1)} \; , \\[2ex]
  \left[ \Gamma^{(3)}_{\Delta^\dagger\Delta\sigma} \right]_{\da\db} & =
  \delta_{\da\db} \sqrt{2\rho_\phi} U_k^{(1,1)} \; , \\[2ex]
  \left[ \Gamma^{(4)}_{\Delta^\dagger\Delta\sigma\sigma} \right]_{\da\db} & =
  \delta_{\da\db} \left( U_k^{(1,1)} + 2\rho_\phi U_k^{(2,1)} \right) \; .
\end{align}

\subsubsection{Diquark-gluon vertices}

Finally, the tensor structure of the diquark-gluon interaction are given by
\begin{align}
  \left[ \Gamma^{(3)}_{A\Delta^\dagger\Delta}(p,q) \right]_{\mu\da\db}^a & =
  - \lambda_{A\Delta^\dagger\Delta} (p + 2q)_\mu (T^a)_{\db\da} \; , \\[2ex]
  \left[ \Gamma^{(4)}_{A^2\Delta^\dagger\Delta} \right]^{ab}_{\mu\nu\da\db} & =
  \lambda_{A^2\Delta^\dagger\Delta} \delta_{\mu\nu} \left(
  \{T^a,T^b\} \right)_{\db\da} \; .
\end{align}
The reversed diquark index ordering on each sides follows directly from the color antitriplet nature of the diquark field $\Delta$.

\section{Numerical implementation}

The analytic expressions of the flow equations were derived
using \texttt{DoFun} \cite{Huber:2011qr,Huber:2019dkb} and traced numerically with
\texttt{FormTracer} \cite{Cyrol:2016zqb}. The flow equations are solved numerically in \texttt{Julia} \cite{Julia-2017} 
making use of the \texttt{SciML} ecosystem
\cite{rackauckas2017differentialequations}. Numerical integration over loop momenta is performed using Gauss–Legendre quadrature with $N=192$ for the flat regulator and $N=64$ for the exponential regulators. We checked explicitly that the results do not change upon doubling the number of points. The time stepping of the system of flow equations is performed using the \texttt{Tsit5} \cite{tsitouras2011runge} algorithm. In such a setting, solving our system of flow equations numerically takes around ten seconds on a standard laptop, highlighting the efficiency of our approach. All figures are produced using \texttt{Makie.jl} \cite{DanischKrumbiegel2021} and \texttt{TikZ-Feynman} \cite{Ellis:2016jkw}. 

\onecolumngrid

\section{Flow equations}

For completeness, we list the analytic expressions of all flow
equations used in this work.  Details of the projection procedure used
to derive these equations can be found in
\cref{sec:correlation_function}. A \texttt{Mathematica} notebook with all these equations can be found in the supplemental material.

\subsection{Two-point functions}
\label{app:flow_two_point}

In the following, we collect the explicit expressions entering the
computation of the \gls{QCD} two-point functions discussed in
\cref{sec:composite-two-point}. Throughout, all quantities are given
in Euclidean space and evaluated at vanishing external spatial
momentum, $p=(p_0, \vec{0} \,)$. Furthermore, we employ a
three-dimensional spatial regulator. The dimensionless momentum is
defined as $x=\sqvec{p} / k^2$ and the regulated dispersion relation
of the field $\Phi_i$ is given by
\begin{equation}
    \epsilon_{\Phi_i} = \sqrt{\sqvec{p}(1+r^{3d}_{\Phi_i}) + m^2_{\Phi_i}} \; .
\end{equation}
The flow equation for the dimensionless diquark two-point function
reads
\begin{align}
    \nonumber
    & \partial_t \bar\Gamma^{(2),E}_{\Delta^\dagger\Delta}(p) = 
    \big( -2 + \eta_\Delta \big) \bar\Gamma^{(2),E}_{\Delta^\dagger\Delta}(p)
    + \frac{1}{4\pi^2} \int_0^\infty dx \, x^{3/2} \bigg\{
    \\[1ex] & \qquad \nonumber
        \lambda_{0,2} (1 + N_c) (\eta_\Delta r_\Delta + 2x r_\Delta') \frac{1}{4\bar\epsilon_\Delta^3}
        + \frac{\lambda_{1,1}}{2} (\eta_\phi r_\phi + 2x r_\phi') \left(
        \frac{3}{4\bar\epsilon_\pi^3}
        + \frac{1}{4\bar\epsilon_\sigma^3}
        \right)
        \\[1ex] & \qquad \nonumber
        - 2 \bar \rho_{\phi,0} \lambda_{1,1}^2 \left[
            \frac{(\eta_\phi r_\phi + 2x r_\phi')\big((\bar\epsilon_\sigma + \bar\epsilon_\Delta)^2 (2\bar\epsilon_\sigma + \bar\epsilon_\Delta) + p_0^2 \bar\epsilon_\Delta \big)}{4\bar\epsilon_\sigma^3 \bar\epsilon_\Delta \big( (\bar\epsilon_\sigma + \bar\epsilon_\Delta)^2 + p_0^2 \big)^2}
            +
            \frac{(\eta_\Delta r_\Delta + 2x r_\Delta')\big((\bar\epsilon_\Delta + \bar\epsilon_\sigma)^2 (2\bar\epsilon_\Delta + \bar\epsilon_\sigma) + p_0^2 \bar\epsilon_\sigma \big)}{4\bar\epsilon_\Delta^3 \bar\epsilon_\sigma \big( (\bar\epsilon_\Delta + \bar\epsilon_\sigma)^2 + p_0^2 \big)^2}
        \right]
        \\[1ex] & \qquad
        - g_{\Delta\bar{q}\bar{q}}^2 2N_f (r_q\eta_q + 2x r_q')(1 + r_q) 
        \frac{1}{\bar\epsilon_q} \left[
            \frac{2}{(2 \bar\epsilon_q - i p_0)^2}
            + \frac{2}{(2 \bar\epsilon_q + i p_0)^2}
        \right]
    \bigg\} \; .
\end{align}
The corresponding flow equations for the dimensionless pion and sigma
two-point functions are given by
\begin{align}
    \nonumber
    & \partial_t \bar\Gamma^{(2),E}_{\pi\pi}(p) = 
    \big( -2 + \eta_\phi \big) \bar\Gamma^{(2),E}_{\pi\pi}(p)
    + \frac{1}{4\pi^2} \int_0^\infty dx \, x^{3/2} \bigg\{
    \\[1ex] & \qquad \nonumber
    \lambda_{1,1} N_c (\eta_\Delta r_\Delta + 2 x r_\Delta') \frac{1}{4\bar\epsilon_\Delta^3}
    + \frac{1}{2} (\eta_\phi r_\phi + 2 x r_\phi') \left(
        \frac{3\lambda_{2,0}}{4\bar\epsilon_\pi^3}
        + \frac{\lambda_{2,0}+2\bar{\rho}_0 \bar\lambda_{3,0}}{4\bar\epsilon_\sigma^3}
    \right)
    \\[1ex] & \qquad \nonumber
    - 2 \bar \rho_{\phi,0} \lambda_{2,0}^2 (\eta_\phi r_\phi + 2x r_\phi') \left[
        \frac{(\bar\epsilon_\sigma + \bar\epsilon_\pi)^2 (2\bar\epsilon_\sigma + \bar\epsilon_\pi) + p_0^2 \bar\epsilon_\pi}{4\bar\epsilon_\sigma^3 \bar\epsilon_\pi \big( (\bar\epsilon_\sigma + \bar\epsilon_\pi)^2 + p_0^2 \big)^2}
        +
        \frac{(\bar\epsilon_\pi + \bar\epsilon_\sigma)^2 (2\bar\epsilon_\pi + \bar\epsilon_\sigma) + p_0^2 \bar\epsilon_\sigma}{4\bar\epsilon_\pi^3 \bar\epsilon_\sigma \big( (\bar\epsilon_\pi + \bar\epsilon_\sigma)^2 + p_0^2 \big)^2}
    \right]
    \\[1ex] & \qquad
    - g_{\phi\bar{q}q}^2 2N_f N_c (\eta_q r_q + 2x r_q')(1 + r_q) 
        \frac{1}{\bar\epsilon_q} \left[
            \frac{2}{(2 \bar\epsilon_q - i p_0)^2}
            + \frac{2}{(2 \bar\epsilon_q + i p_0)^2}
        \right]
    \bigg\} \; ,
\end{align}
and
\begin{align}
    \nonumber
    & \partial_t \bar\Gamma^{(2),E}_{\sigma\sigma}(p) = 
    \big( -2 + \eta_\phi \big) \bar\Gamma^{(2),E}_{\sigma\sigma}(p)
    + \frac{1}{4\pi^2} \int_0^\infty dx \, x^{3/2} \bigg\{
    - 2 N_c (\eta_\Delta r_\Delta + 2 x r_\Delta') \frac{1}{4\bar\epsilon_\Delta^3} \frac{p_0^2 + 12 \bar\epsilon_\Delta^2}{(p_0^2 + 4\bar\epsilon_\Delta^2)^2}
    \\[1ex] & \qquad \nonumber
    - (\eta_\phi r_\phi + 2x r_\phi') 2\bar{\rho}_0 \left(
        3 \lambda_{2,0}^2 \frac{1}{4\bar\epsilon_\pi^3} \frac{p_0^2 + 12\bar\epsilon_\pi^2}{(p_0^2 + 4\bar\epsilon_\pi^2)^2}
        + (\lambda_{2,0} + 2\bar\rho_{\phi,0} \bar\lambda_{3,0})^2  \frac{1}{4\bar\epsilon_\sigma^3} \frac{p_0^2 + 12\bar\epsilon_\sigma^2}{(p_0^2 + 4\bar\epsilon_\sigma^2)^2}
    \right)
    \\[1ex] & \qquad \nonumber
    + \frac{1}{2} (\eta_\phi r_\phi + 2 x r_\phi')
    \left(
        3\frac{\lambda_{2,0} + 2 \bar\rho_{\phi,0} \bar\lambda_{3,0}}{4\bar\epsilon_\pi^3}
        + \frac{3\lambda_{2,0} + 12 \bar\rho_{\phi,0} \bar\lambda_{3,0} + 4\bar\rho_{\phi,0}^2 \bar\lambda_{4,0}}{4\bar\epsilon_\sigma^3}
    \right)
    + N_c (\eta_\Delta r_\Delta + 2 x r_\Delta')
    \frac{\lambda_{1,1} + 2\bar\rho_{\phi,0} \bar\lambda_{2,1}}{4\bar\epsilon_\Delta^3}
    \\[1ex] & \qquad
    + 8N_f N_c g_{\phi\bar{q}q}^2 (\eta_q r_q + 2 x r_q') (1 + r_q) \frac{1}{\bar\epsilon_q^3}
    \frac{8 \bar\epsilon_q^4 + 2 \left( p_0^2 - 6x(1+r_q)^2 \right) \bar\epsilon_q^2 - p_0^2 x (1+r_q)^2}{(p_0^2 + 4 \bar\epsilon_q^2)^2}
    \bigg\} \; .
\end{align}

\subsection{Zero-momentum expansion}
\label{app:zero-momentum-flows}

\newcommand{\autoLineBreakStart}[2]{
\begin{equation*}
\parbox[b]{0.9\textwidth}{\raggedright\hangafter=1\hangindent=1em$\displaystyle
    #1 \frac{1}{16\pi^2} \int_0^\infty dx \, x^3 \bigg\{ #2
  $}
\end{equation*}
}

\newcommand{\autoLineBreakContinue}[1]{
    \begin{equation*}
    \parbox[b]{0.9\textwidth}{\raggedright\hangafter=1\hangindent=2.2em$\displaystyle
      \qquad\; #1
      $}
    \end{equation*}
}

\newcommand{\autoLineBreakEnd}[1]{
    \begin{equation}
    \parbox[b]{0.9\textwidth}{\raggedright\hangafter=1\hangindent=2.2em$\displaystyle
      \qquad\; #1 \bigg\} \; .
      $}
    \end{equation}
}

\newcommand{\autoLineBreak}[2]{
    \begin{equation}
    \parbox[b]{0.9\textwidth}{\raggedright\hangafter=1\hangindent=1em$\displaystyle
      #1 \frac{1}{16\pi^2} \int_0^\infty dx \, x^3 \bigg\{ #2 \bigg\} \; .
      $}
    \end{equation}
}

\subsubsection{Effective potential}

The flow of the effective potential is given by
\autoLineBreak{
\partial_t \bar{U}_k =  - 4 \bar{U}_k
  + (2 + \eta_\phi) \bar\rho_\phi \partial_{\bar\rho_\phi} \bar{U}_k
  + (2 + \eta_\Delta) \bar\rho_\Delta \partial_{\bar\rho_\Delta} \bar{U}_k
  +
}{
\frac{8 \big(r_c \eta _c+2 x r_c'\big)}{x \big(1+r_c\big)}+8 \big(1+r_q\big) \big(\frac{1}{x \big(1+r_q\big){}^2+\bar{m}_q^2}+\frac{2}{x \big(1+r_q\big){}^2+\bar{m}_q^2+g_{\Delta  \bar{q} \bar{q}}^2 \bar{\rho }_{\Delta }}\big) \big(r_q \eta _q+2 x r_q'\big)-\frac{3}{2} \big(\frac{3}{x+x r_A+\bar{m}_{\text{gap}}^2}+\frac{4}{x+x r_A+\bar{m}_{\text{gap}}^2+\frac{1}{2} \lambda _{A^2 \Delta ^{\dagger } \Delta } \bar{\rho }_{\Delta }}+\frac{1}{x+x r_A+\bar{m}_{\text{gap}}^2+\frac{2}{3} \lambda _{A^2 \Delta ^{\dagger } \Delta } \bar{\rho }_{\Delta }}\big) \big(r_A \eta _A+2 x r_A'\big)+\big(-\frac{2}{x+x r_{\Delta }+\bar{m}_{\Delta }^2}-\frac{\big(x+x r_{\phi }+\bar{m}_{\sigma }^2\big) \big(x+x r_{\Delta }+\bar{m}_{\Delta _3}^2\big)-\bar\chi}{\big(x+x r_{\Delta }+\bar{m}_{\Delta _3}^2\big) \big(\big(x+x r_{\phi }+\bar{m}_{\sigma }^2\big) \big(x+x r_{\Delta }+\bar{m}_{\Delta _3}^2\big)-2\bar\chi\big)}\big) \big(r_{\Delta } \eta _{\Delta }+2 x r_{\Delta }'\big)+\frac{1}{2} \big(-\frac{3}{x+x r_{\phi }+\bar{m}_{\pi }^2}-\frac{x+x r_{\Delta }+\bar{m}_{\Delta _3}^2}{\big(x+x r_{\phi }+\bar{m}_{\sigma }^2\big) \big(x+x r_{\Delta }+\bar{m}_{\Delta _3}^2\big)-2\bar{\chi}}\big) \big(r_{\phi } \eta _{\phi }+2 x r_{\phi }'\big)
}
Note that, in this subsection only, we allow for  a non-vanishing diquark
field configuration $\rho_\Delta$; see \cref{sec:eff_pot} for
details. Furthermore, we introduce the mass of the third diquark in
the absence of $\sigma$-$\Delta$ mixing,
$m_{\Delta_3}^2 = \lambda_{0,1} + \rho_\Delta \lambda_{0,2}$ and the
mixing parameter $\chi = 2\rho_\phi \rho_\Delta \lambda_{1,1}^2$.

\subsubsection{Anomalous dimensions}

\autoLineBreak{
\eta_c =
}{
\lambda _{A \bar{c} c}^2 \big(\frac{9 \big(r_c \eta _c+2 x r_c'\big)}{4 x^2 \big(1+r_c\big){}^2 \big(x+x r_A+\bar{m}_{\text{gap}}^2\big)}+\frac{9 \big(r_A \eta _A+2 x r_A'\big)}{4 x \big(1+r_c\big) \big(x+x r_A+\bar{m}_{\text{gap}}^2\big){}^2}\big)
}

\autoLineBreak{\eta_A = }{
\frac{24 \big(1+r_q\big){}^2 \bar{m}_q \lambda _{A \bar{q} q}^{\text{(1)}} \lambda _{A \bar{q} q}^{\text{(4)}} \big(r_q \eta _q+2 x r_q'\big)}{\big(x+2 x r_q+x r_q^2+\bar{m}_q^2\big){}^3}-\frac{16 \big(1+r_q\big) \bar{m}_q^2 \big(\lambda _{A \bar{q} q}^{\text{(4)}}\big){}^2 \big(r_q \eta _q+2 x r_q'\big)}{\big(x+2 x r_q+x r_q^2+\bar{m}_q^2\big){}^3}+\frac{\lambda _{A \bar{c} c}^2 \big(r_c \eta _c+2 x r_c'\big) \big(1+r_c^2+x r_c'-2 x^2 \big(r_c'\big){}^2+x^2 r_c''+r_c \big(2+x r_c'+x^2 r_c''\big)\big)}{2 x^3 \big(1+r_c\big){}^5}
-
\frac{1}{3 \big(x+2 x r_q+x r_q^2+\bar{m}_q^2\big){}^5} 8 \big(\lambda _{A \bar{q} q}^{\text{(1)}}\big){}^2 \big(r_q \eta _q+2 x r_q'\big) \big(\big(1+r_q\big){}^3 \big(x^2+4 x^2 r_q^3+x^2 r_q^4+2 x \bar{m}_q^2+3 \bar{m}_q^4+4 x r_q \big(x+\bar{m}_q^2\big)+2 x r_q^2 \big(3 x+\bar{m}_q^2\big)\big)+\big(x^3+6 x^3 r_q^5+x^3 r_q^6+5 x^2 \bar{m}_q^2+15 x \bar{m}_q^4+3 \bar{m}_q^6+20 x^2 r_q^3 \big(x+\bar{m}_q^2\big)+15 x r_q^2 \big(x+\bar{m}_q^2\big){}^2+5 x^2 r_q^4 \big(3 x+\bar{m}_q^2\big)+2 x r_q \big(3 x^2+10 x \bar{m}_q^2+15 \bar{m}_q^4\big)\big) r_q'-2 x^3 \big(1+r_q\big){}^3 \big(x+2 x r_q+x r_q^2+5 \bar{m}_q^2\big) \big(r_q'\big){}^2+x \big(x^3+6 x^3 r_q^5+x^3 r_q^6+5 x^2 \bar{m}_q^2+5 x \bar{m}_q^4+\bar{m}_q^6+20 x^2 r_q^3 \big(x+\bar{m}_q^2\big)+5 x^2 r_q^4 \big(3 x+\bar{m}_q^2\big)+5 x r_q^2 \big(3 x^2+6 x \bar{m}_q^2+\bar{m}_q^4\big)+2 x r_q \big(3 x^2+10 x \bar{m}_q^2+5 \bar{m}_q^4\big)\big) r_q''\big)
+
\frac{1}{2 \big(x+x r_A+\bar{m}_{\text{gap}}^2\big){}^5}\lambda _{A^3}^2 \big(r_A \eta _A+2 x r_A'\big) \big(25 x^2+25 x^2 r_A^2+44 x \bar{m}_{\text{gap}}^2+31 \bar{m}_{\text{gap}}^4-6 x^2 \big(x+5 \bar{m}_{\text{gap}}^2\big) r_A'+12 x^4 \big(r_A'\big){}^2-6 x^4 r_A''-6 x^3 \bar{m}_{\text{gap}}^2 r_A''+2 x r_A \big(25 x+22 \bar{m}_{\text{gap}}^2-3 x^2 r_A'-3 x^3 r_A''\big)\big)
-
\frac{1}{3 \big(x+x r_{\Delta }+\bar{m}_{\Delta }^2\big){}^5}x \lambda _{A \Delta ^{\dagger } \Delta }^2 \big(r_{\Delta } \eta _{\Delta }+2 x r_{\Delta }'\big) \big(x+x r_{\Delta }^2+3 \bar{m}_{\Delta }^2+x \big(x+5 \bar{m}_{\Delta }^2\big) r_{\Delta }'-2 x^3 \big(r_{\Delta }'\big){}^2+x^3 r_{\Delta }''+x^2 \bar{m}_{\Delta }^2 r_{\Delta }''+r_{\Delta } \big(2 x+3 \bar{m}_{\Delta }^2+x^2 r_{\Delta }'+x^3 r_{\Delta }''\big)\big)
}

\autoLineBreak{
\eta_q =
}{
\big(\lambda _{A \bar{q} q}^{\text{(1)}}\big){}^2 \bigg[\frac{\big(-12 \big(1+r_q\big) \bar{m}_q^2+12 x \big(x+2 x r_q+x r_q^2-\bar{m}_q^2\big) r_q'\big) \big(r_A \eta _A+2 x r_A'\big)}{6 \big(x+x r_A+\bar{m}_{\text{gap}}^2\big){}^2 \big(x+2 x r_q+x r_q^2+\bar{m}_q^2\big){}^2}+\frac{2 \big(x+2 x r_q+x r_q^2-\bar{m}_q^2\big) \big(r_q \eta _q+2 x r_q'\big) \big(\bar{m}_{\text{gap}}^2-x^2 r_A'\big)}{x \big(x+x r_A+\bar{m}_{\text{gap}}^2\big){}^2 \big(x+2 x r_q+x r_q^2+\bar{m}_q^2\big){}^2}\bigg]
+
\lambda _{A \bar{q} q}^{\text{(1)}} \lambda _{A \bar{q} q}^{\text{(4)}} 
\bigg[-\frac{4 x \big(1+r_q\big) \bar{m}_q \big(1+r_q+2 x r_q'\big) \big(r_A \eta _A+2 x r_A'\big)}{\big(x+x r_A+\bar{m}_{\text{gap}}^2\big){}^2 \big(x+2 x r_q+x r_q^2+\bar{m}_q^2\big){}^2}-\frac{8 \big(1+r_q\big) \bar{m}_q \big(r_q \eta _q+2 x r_q'\big) \big(x+x r_A+2 \bar{m}_{\text{gap}}^2-x^2 r_A'\big)}{\big(x+x r_A+\bar{m}_{\text{gap}}^2\big){}^2 \big(x+2 x r_q+x r_q^2+\bar{m}_q^2\big){}^2}\bigg]
+
\big(\lambda _{A \bar{q} q}^{\text{(4)}}\big){}^2 \bigg[-\frac{2 x^2 \big(\big(1+r_q\big){}^3+\big(x+2 x r_q+x r_q^2-\bar{m}_q^2\big) r_q'\big) \big(r_A \eta _A+2 x r_A'\big)}{\big(x+x r_A+\bar{m}_{\text{gap}}^2\big){}^2 \big(x+2 x r_q+x r_q^2+\bar{m}_q^2\big){}^2}-\frac{2 \big(x+2 x r_q+x r_q^2-\bar{m}_q^2\big) \big(r_q \eta _q+2 x r_q'\big) \big(2 x+2 x r_A+3 \bar{m}_{\text{gap}}^2-x^2 r_A'\big)}{\big(x+x r_A+\bar{m}_{\text{gap}}^2\big){}^2 \big(x+2 x r_q+x r_q^2+\bar{m}_q^2\big){}^2}\bigg]
+
g_{\Delta  \bar{q} \bar{q}}^2 \bigg[-\frac{\big(x+2 x r_q+x r_q^2-\bar{m}_q^2\big) \big(r_q \eta _q+2 x r_q'\big) \big(1+r_{\Delta }+x r_{\Delta }'\big)}{\big(x+2 x r_q+x r_q^2+\bar{m}_q^2\big){}^2 \big(x+x r_{\Delta }+\bar{m}_{\Delta }^2\big){}^2}+\frac{\big(-6 \big(1+r_q\big) \big(x+2 x r_q+x r_q^2+2 \bar{m}_q^2\big)+6 x \big(x+2 x r_q+x r_q^2-\bar{m}_q^2\big) r_q'\big) \big(r_{\Delta } \eta _{\Delta }+2 x r_{\Delta }'\big)}{6 \big(x+2 x r_q+x r_q^2+\bar{m}_q^2\big){}^2 \big(x+x r_{\Delta }+\bar{m}_{\Delta }^2\big){}^2}\bigg]
+ g_{\phi  \bar{q} q}^2 \bigg[
-\frac{1}{2 \big(x+x r_{\phi }+\bar{m}_{\pi }^2\big){}^2 \big(x+2 x r_q+x r_q^2+\bar{m}_q^2\big){}^2 \big(x+x r_{\phi }+\bar{m}_{\sigma }^2\big){}^2}\big(x+2 x r_q+x r_q^2-\bar{m}_q^2\big) \big(4 x^2+4 x^2 r_{\phi }^2+2 x \bar{m}_{\pi }^2+\bar{m}_{\pi }^4+6 x \bar{m}_{\sigma }^2+3 \bar{m}_{\sigma }^4+2 x r_{\phi } \big(4 x+\bar{m}_{\pi }^2+3 \bar{m}_{\sigma }^2\big)\big) \big(r_q \eta _q+2 x r_q'\big) \big(1+r_{\phi }+x r_{\phi }'\big)
+
\frac{1}{2 \big(x+x r_{\phi }+\bar{m}_{\pi }^2\big){}^2 \big(x+2 x r_q+x r_q^2+\bar{m}_q^2\big){}^2 \big(x+x r_{\phi }+\bar{m}_{\sigma }^2\big){}^2}\big(4 x^2+4 x^2 r_{\phi }^2+2 x \bar{m}_{\pi }^2+\bar{m}_{\pi }^4+6 x \bar{m}_{\sigma }^2+3 \bar{m}_{\sigma }^4+2 x r_{\phi } \big(4 x+\bar{m}_{\pi }^2+3 \bar{m}_{\sigma }^2\big)\big) \big(-\big(\big(1+r_q\big) \big(x+2 x r_q+x r_q^2+2 \bar{m}_q^2\big)\big)+x \big(x+2 x r_q+x r_q^2-\bar{m}_q^2\big) r_q'\big) \big(r_{\phi } \eta _{\phi }+2 x r_{\phi }'\big)
\bigg]
}

\autoLineBreak{
\eta_\phi =
}{
-\frac{1}{\big(x+2 x r_q+x r_q^2+\bar{m}_q^2\big){}^3}24 g_{\phi  \bar{q} q}^2 \big(r_q \eta _q+2 x r_q'\big) \big(\big(1+r_q\big){}^3+\big(x+2 x r_q+x r_q^2+3 \bar{m}_q^2\big) r_q'-2 x^2 \big(1+r_q\big) \big(r_q'\big){}^2+x \big(x+2 x r_q+x r_q^2+\bar{m}_q^2\big) r_q''\big)
-
\frac{1}{\big(x+x r_{\phi }+\bar{m}_{\pi }^2\big){}^3 \big(x+x r_{\phi }+\bar{m}_{\sigma }^2\big){}^3}\bar{\sigma }_0^2 \lambda _{2,0}^2 \big(r_{\phi } \eta _{\phi }+2 x r_{\phi }'\big) \big(x \bar{m}_{\pi }^2+x \bar{m}_{\sigma }^2+2 \bar{m}_{\pi }^2 \bar{m}_{\sigma }^2+2 x \big(x \bar{m}_{\sigma }^2+\bar{m}_{\pi }^2 \big(x+2 \bar{m}_{\sigma }^2\big)\big) r_{\phi }'-x^3 \big(2 x+\bar{m}_{\pi }^2+\bar{m}_{\sigma }^2\big) \big(r_{\phi }'\big){}^2+x^4 r_{\phi }''+x^3 \bar{m}_{\pi }^2 r_{\phi }''+x^3 \bar{m}_{\sigma }^2 r_{\phi }''+x^2 \bar{m}_{\pi }^2 \bar{m}_{\sigma }^2 r_{\phi }''+x r_{\phi }^2 \big(\bar{m}_{\pi }^2+\bar{m}_{\sigma }^2+x^3 r_{\phi }''\big)+r_{\phi } \big(2 x \bar{m}_{\sigma }^2+2 \bar{m}_{\pi }^2 \big(x+\bar{m}_{\sigma }^2\big)+2 x^2 \big(\bar{m}_{\pi }^2+\bar{m}_{\sigma }^2\big) r_{\phi }'-2 x^4 \big(r_{\phi }'\big){}^2+x^3 \big(2 x+\bar{m}_{\pi }^2+\bar{m}_{\sigma }^2\big) r_{\phi }''\big)\big)
}

\autoLineBreak{
\eta_\Delta =
}{
\lambda _{A \Delta ^{\dagger } \Delta }^2 \bigg[
\frac{4 \big(r_A \eta _A+2 x r_A'\big)}{\big(x+x r_A+\bar{m}_{\text{gap}}^2\big){}^2 \big(x+x r_{\Delta }+\bar{m}_{\Delta }^2\big)}+\frac{4 \big(r_{\Delta } \eta _{\Delta }+2 x r_{\Delta }'\big)}{\big(x+x r_A+\bar{m}_{\text{gap}}^2\big) \big(x+x r_{\Delta }+\bar{m}_{\Delta }^2\big){}^2}
\bigg]
-
\frac{8 g_{\Delta  \bar{q} \bar{q}}^2 \big(r_q \eta _q+2 x r_q'\big) \big(\big(1+r_q\big){}^3+\big(x+2 x r_q+x r_q^2+3 \bar{m}_q^2\big) r_q'-2 x^2 \big(1+r_q\big) \big(r_q'\big){}^2+x \big(x+2 x r_q+x r_q^2+\bar{m}_q^2\big) r_q''\big)}{\big(x+2 x r_q+x r_q^2+\bar{m}_q^2\big){}^3}
-
\bar{\sigma }_0^2 \lambda _{1,1}^2 \bigg[
\frac{\big(r_{\phi } \eta _{\phi }+2 x r_{\phi }'\big) \big(2 \bar{m}_{\Delta }^2+4 x \bar{m}_{\Delta }^2 r_{\Delta }'-2 x^3 \big(r_{\Delta }'\big){}^2+x^3 r_{\Delta }''+x^2 \bar{m}_{\Delta }^2 r_{\Delta }''+r_{\Delta } \big(2 \bar{m}_{\Delta }^2+x^3 r_{\Delta }''\big)\big)}{2 \big(x+x r_{\Delta }+\bar{m}_{\Delta }^2\big){}^3 \big(x+x r_{\phi }+\bar{m}_{\sigma }^2\big){}^2}
+ \frac{\big(r_{\Delta } \eta _{\Delta }+2 x r_{\Delta }'\big) \big(2 \bar{m}_{\sigma }^2+4 x \bar{m}_{\sigma }^2 r_{\phi }'-2 x^3 \big(r_{\phi }'\big){}^2+x^3 r_{\phi }''+x^2 \bar{m}_{\sigma }^2 r_{\phi }''+r_{\phi } \big(2 \bar{m}_{\sigma }^2+x^3 r_{\phi }''\big)\big)}{2 \big(x+x r_{\Delta }+\bar{m}_{\Delta }^2\big){}^2 \big(x+x r_{\phi }+\bar{m}_{\sigma }^2\big){}^3}\bigg]
}

\subsubsection{Strong coupling avatars and gluon mass gap}

\autoLineBreak{
\partial_t \bar{m}^2_{\text{gap}} = (-2 + \eta_A) \bar{m}^2_\text{gap} +
}{
\frac{3 \lambda _{A \bar{c} c}^2 \big(r_c \eta _c+2 x r_c'\big)}{2 x^2 \big(1+r_c\big){}^3}-\frac{4 \big(1+r_q\big) \big(x+2 x r_q+x r_q^2+3 \bar{m}_q^2\big) \big(\lambda _{A \bar{q} q}^{\text{(1)}}\big){}^2 \big(r_q \eta _q+2 x r_q'\big)}{\big(x+2 x r_q+x r_q^2+\bar{m}_q^2\big){}^3}-\frac{9 x \lambda _{A^3}^2 \big(r_A \eta _A+2 x r_A'\big)}{\big(x+x r_A+\bar{m}_{\text{gap}}^2\big){}^3}+\frac{27 \lambda _{A^4} \big(r_A \eta _A+2 x r_A'\big)}{4 \big(x+x r_A+\bar{m}_{\text{gap}}^2\big){}^2}-\frac{x \lambda _{A \Delta ^{\dagger } \Delta }^2 \big(r_{\Delta } \eta _{\Delta }+2 x r_{\Delta }'\big)}{\big(x+x r_{\Delta }+\bar{m}_{\Delta }^2\big){}^3}+\frac{\lambda _{A^2 \Delta ^{\dagger } \Delta } \big(r_{\Delta } \eta _{\Delta }+2 x r_{\Delta }'\big)}{\big(x+x r_{\Delta }+\bar{m}_{\Delta }^2\big){}^2}
}

\autoLineBreakStart{
\partial_t \lambda_{A^3} = \frac{3}{2} \eta_A \lambda_{A^3} +
}{
-\frac{3 \lambda _{A \bar{c} c}^3 \big(r_c \eta _c+2 x r_c'\big)}{8 x^3 \big(1+r_c\big){}^4}
+
\frac{4 \big(1+r_q\big){}^2 \big(x+2 x r_q+x r_q^2+2 \bar{m}_q^2\big) \big(\lambda _{A \bar{q} q}^{\text{(1)}}\big){}^3 \big(r_q \eta _q+2 x r_q'\big)}{\big(x+2 x r_q+x r_q^2+\bar{m}_q^2\big){}^4}
-
\frac{16 \big(1+r_q\big) \bar{m}_q \big(\lambda _{A \bar{q} q}^{\text{(1)}}\big){}^2 \lambda _{A \bar{q} q}^{\text{(4)}} \big(r_q \eta _q+2 x r_q'\big)}{\big(x+2 x r_q+x r_q^2+\bar{m}_q^2\big){}^3}
+
}
\autoLineBreakEnd{
\frac{9 x \lambda _{A^3}^3 \big(r_A \eta _A+2 x r_A'\big)}{2 \big(x+x r_A+\bar{m}_{\text{gap}}^2\big){}^4}-\frac{45 \lambda _{A^3} \lambda _{A^4} \big(r_A \eta _A+2 x r_A'\big)}{4 \big(x+x r_A+\bar{m}_{\text{gap}}^2\big){}^3}+\frac{x \lambda _{A \Delta ^{\dagger } \Delta }^3 \big(r_{\Delta } \eta _{\Delta }+2 x r_{\Delta }'\big)}{2 \big(x+x r_{\Delta }+\bar{m}_{\Delta }^2\big){}^4}
}

\autoLineBreak{
\partial_t \lambda_{A^4} = 2\eta_A \lambda_{A^4} +
}{
\frac{8 \big(1+r_q\big) \big(2 x+4 x r_q+2 x r_q^2+5 \bar{m}_q^2\big) \big(\lambda _{A \bar{q} q}^{\text{(1)}}\big){}^4 \big(r_q \eta _q+2 x r_q'\big)}{3 \big(x+2 x r_q+x r_q^2+\bar{m}_q^2\big){}^4}+\frac{6 x \lambda _{A^3}^2 \lambda _{A^4} \big(r_A \eta _A+2 x r_A'\big)}{\big(x+x r_A+\bar{m}_{\text{gap}}^2\big){}^4}-\frac{8 \lambda _{A^4}^2 \big(r_A \eta _A+2 x r_A'\big)}{\big(x+x r_A+\bar{m}_{\text{gap}}^2\big){}^3}+\frac{x \lambda _{A \Delta ^{\dagger } \Delta }^2 \lambda _{A^2 \Delta ^{\dagger } \Delta } \big(r_{\Delta } \eta _{\Delta }+2 x r_{\Delta }'\big)}{\big(x+x r_{\Delta }+\bar{m}_{\Delta }^2\big){}^4}-\frac{\lambda _{A^2 \Delta ^{\dagger } \Delta }^2 \big(r_{\Delta } \eta _{\Delta }+2 x r_{\Delta }'\big)}{3 \big(x+x r_{\Delta }+\bar{m}_{\Delta }^2\big){}^3}
}

\autoLineBreakStart{
\partial_t \lambda_{A\bar{q}q}^{(1)} = (\eta_q + \frac{1}{2} \eta_A) \lambda_{A\bar{q}q}^{(1)} +
}{
\lambda _{A \bar{q} q}^{\text{(1)}} \big(\lambda _{A \bar{q} q}^{\text{(4)}}\big){}^2 \big(-\frac{x \big(1+r_q\big) \big(x+2 x r_q+x r_q^2-\bar{m}_q^2\big) \big(r_q \eta _q+2 x r_q'\big)}{2 \big(x+x r_A+\bar{m}_{\text{gap}}^2\big) \big(x+2 x r_q+x r_q^2+\bar{m}_q^2\big){}^3}-\frac{x^2 \big(1+r_q\big){}^2 \big(r_A \eta _A+2 x r_A'\big)}{4 \big(x+x r_A+\bar{m}_{\text{gap}}^2\big){}^2 \big(x+2 x r_q+x r_q^2+\bar{m}_q^2\big){}^2}\big)
+
\big(\lambda _{A \bar{q} q}^{\text{(1)}}\big){}^2 \lambda _{A \bar{q} q}^{\text{(4)}} \big(\frac{\bar{m}_q \big(-3 x-6 x r_q-3 x r_q^2+\bar{m}_q^2\big) \big(r_q \eta _q+2 x r_q'\big)}{2 \big(x+x r_A+\bar{m}_{\text{gap}}^2\big) \big(x+2 x r_q+x r_q^2+\bar{m}_q^2\big){}^3}-\frac{x \big(1+r_q\big) \bar{m}_q \big(r_A \eta _A+2 x r_A'\big)}{2 \big(x+x r_A+\bar{m}_{\text{gap}}^2\big){}^2 \big(x+2 x r_q+x r_q^2+\bar{m}_q^2\big){}^2}\big)
+
\big(\lambda _{A \bar{q} q}^{\text{(1)}}\big){}^3 \big(-\frac{\big(1+r_q\big) \bar{m}_q^2 \big(r_q \eta _q+2 x r_q'\big)}{\big(x+x r_A+\bar{m}_{\text{gap}}^2\big) \big(x+2 x r_q+x r_q^2+\bar{m}_q^2\big){}^3}-\frac{\bar{m}_q^2 \big(r_A \eta _A+2 x r_A'\big)}{4 \big(x+x r_A+\bar{m}_{\text{gap}}^2\big){}^2 \big(x+2 x r_q+x r_q^2+\bar{m}_q^2\big){}^2}\big)
+
}
\autoLineBreakEnd{
\lambda _{A^3} \big(\lambda _{A \bar{q} q}^{\text{(1)}}\big){}^2 \big(\frac{9 \big(x+2 x r_q+x r_q^2-\bar{m}_q^2\big) \big(r_q \eta _q+2 x r_q'\big)}{4 \big(x+x r_A+\bar{m}_{\text{gap}}^2\big){}^2 \big(x+2 x r_q+x r_q^2+\bar{m}_q^2\big){}^2}+\frac{9 x \big(1+r_q\big) \big(r_A \eta _A+2 x r_A'\big)}{2 \big(x+x r_A+\bar{m}_{\text{gap}}^2\big){}^3 \big(x+2 x r_q+x r_q^2+\bar{m}_q^2\big)}\big)+\lambda _{A^3} \big(\lambda _{A \bar{q} q}^{\text{(4)}}\big){}^2 \big(-\frac{9 x \big(x+2 x r_q+x r_q^2-\bar{m}_q^2\big) \big(r_q \eta _q+2 x r_q'\big)}{4 \big(x+x r_A+\bar{m}_{\text{gap}}^2\big){}^2 \big(x+2 x r_q+x r_q^2+\bar{m}_q^2\big){}^2}-\frac{9 x^2 \big(1+r_q\big) \big(r_A \eta _A+2 x r_A'\big)}{2 \big(x+x r_A+\bar{m}_{\text{gap}}^2\big){}^3 \big(x+2 x r_q+x r_q^2+\bar{m}_q^2\big)}\big)
+
\lambda _{A^3} \lambda _{A \bar{q} q}^{\text{(1)}} \lambda _{A \bar{q} q}^{\text{(4)}} \big(-\frac{9 x \big(1+r_q\big) \bar{m}_q \big(r_q \eta _q+2 x r_q'\big)}{\big(x+x r_A+\bar{m}_{\text{gap}}^2\big){}^2 \big(x+2 x r_q+x r_q^2+\bar{m}_q^2\big){}^2}-\frac{9 x \bar{m}_q \big(r_A \eta _A+2 x r_A'\big)}{\big(x+x r_A+\bar{m}_{\text{gap}}^2\big){}^3 \big(x+2 x r_q+x r_q^2+\bar{m}_q^2\big)}\big)
+
g_{\Delta  \bar{q} \bar{q}}^2 \lambda _{A \Delta ^{\dagger } \Delta } \big(\frac{\big(x+2 x r_q+x r_q^2-\bar{m}_q^2\big) \big(r_q \eta _q+2 x r_q'\big)}{2 \big(x+2 x r_q+x r_q^2+\bar{m}_q^2\big){}^2 \big(x+x r_{\Delta }+\bar{m}_{\Delta }^2\big){}^2}+\frac{x \big(1+r_q\big) \big(r_{\Delta } \eta _{\Delta }+2 x r_{\Delta }'\big)}{\big(x+2 x r_q+x r_q^2+\bar{m}_q^2\big) \big(x+x r_{\Delta }+\bar{m}_{\Delta }^2\big){}^3}\big)
+
g_{\Delta  \bar{q} \bar{q}}^2 \lambda _{A \bar{q} q}^{\text{(1)}} \big(\frac{\big(1+r_q\big) \big(x+2 x r_q+x r_q^2+3 \bar{m}_q^2\big) \big(r_q \eta _q+2 x r_q'\big)}{\big(x+2 x r_q+x r_q^2+\bar{m}_q^2\big){}^3 \big(x+x r_{\Delta }+\bar{m}_{\Delta }^2\big)}+\frac{\big(x+2 x r_q+x r_q^2+2 \bar{m}_q^2\big) \big(r_{\Delta } \eta _{\Delta }+2 x r_{\Delta }'\big)}{2 \big(x+2 x r_q+x r_q^2+\bar{m}_q^2\big){}^2 \big(x+x r_{\Delta }+\bar{m}_{\Delta }^2\big){}^2}\big)
+
g_{\phi  \bar{q} q}^2 \lambda _{A \bar{q} q}^{\text{(1)}} \big(\frac{\big(1+r_q\big) \big(x+2 x r_q+x r_q^2+3 \bar{m}_q^2\big) \big(4 x+4 x r_{\phi }+\bar{m}_{\pi }^2+3 \bar{m}_{\sigma }^2\big) \big(r_q \eta _q+2 x r_q'\big)}{\big(x+x r_{\phi }+\bar{m}_{\pi }^2\big) \big(x+2 x r_q+x r_q^2+\bar{m}_q^2\big){}^3 \big(x+x r_{\phi }+\bar{m}_{\sigma }^2\big)}+\frac{1}{2 \big(x+x r_{\phi }+\bar{m}_{\pi }^2\big){}^2 \big(x+2 x r_q+x r_q^2+\bar{m}_q^2\big){}^2 \big(x+x r_{\phi }+\bar{m}_{\sigma }^2\big){}^2}\big(x+2 x r_q+x r_q^2+2 \bar{m}_q^2\big) \big(4 x^2+4 x^2 r_{\phi }^2+2 x \bar{m}_{\pi }^2+\bar{m}_{\pi }^4+6 x \bar{m}_{\sigma }^2+3 \bar{m}_{\sigma }^4+2 x r_{\phi } \big(4 x+\bar{m}_{\pi }^2+3 \bar{m}_{\sigma }^2\big)\big) \big(r_{\phi } \eta _{\phi }+2 x r_{\phi }'\big)\big)
}

\autoLineBreakStart{
\partial_t \lambda_{A\bar{q}q}^{(4)} = (1 + \eta_q + \frac{1}{2} \eta_A) \lambda_{A\bar{q}q}^{(4)} +
}{
\big(\lambda _{A \bar{q} q}^{\text{(1)}}\big){}^2 \lambda _{A \bar{q} q}^{\text{(4)}} \big(\frac{\big(1+r_q\big) \big(49 x+98 x r_q+49 x r_q^2-57 \bar{m}_q^2\big) \big(r_q \eta _q+2 x r_q'\big)}{72 \big(x+x r_A+\bar{m}_{\text{gap}}^2\big) \big(x+2 x r_q+x r_q^2+\bar{m}_q^2\big){}^3}+\frac{x \big(1+r_q\big){}^2 \big(r_A \eta _A+2 x r_A'\big)}{3 \big(x+x r_A+\bar{m}_{\text{gap}}^2\big){}^2 \big(x+2 x r_q+x r_q^2+\bar{m}_q^2\big){}^2}\big)
+
\big(\lambda _{A \bar{q} q}^{\text{(4)}}\big){}^3 \big(-\frac{x \big(1+r_q\big) \big(x+2 x r_q+x r_q^2-51 \bar{m}_q^2\big) \big(r_q \eta _q+2 x r_q'\big)}{72 \big(x+x r_A+\bar{m}_{\text{gap}}^2\big) \big(x+2 x r_q+x r_q^2+\bar{m}_q^2\big){}^3}+\frac{x \bar{m}_q^2 \big(r_A \eta _A+2 x r_A'\big)}{6 \big(x+x r_A+\bar{m}_{\text{gap}}^2\big){}^2 \big(x+2 x r_q+x r_q^2+\bar{m}_q^2\big){}^2}\big)
+
\lambda _{A^3} \lambda _{A \bar{q} q}^{\text{(1)}} \lambda _{A \bar{q} q}^{\text{(4)}} \big(\frac{33 \big(x+2 x r_q+x r_q^2-\bar{m}_q^2\big) \big(r_q \eta _q+2 x r_q'\big)}{4 \big(x+x r_A+\bar{m}_{\text{gap}}^2\big){}^2 \big(x+2 x r_q+x r_q^2+\bar{m}_q^2\big){}^2}+\frac{133 x \big(1+r_q\big) \big(r_A \eta _A+2 x r_A'\big)}{8 \big(x+x r_A+\bar{m}_{\text{gap}}^2\big){}^3 \big(x+2 x r_q+x r_q^2+\bar{m}_q^2\big)}\big)
+
\lambda _{A^3} \big(\lambda _{A \bar{q} q}^{\text{(1)}}\big){}^2 \big(\frac{15 \big(1+r_q\big) \bar{m}_q \big(r_q \eta _q+2 x r_q'\big)}{2 \big(x+x r_A+\bar{m}_{\text{gap}}^2\big){}^2 \big(x+2 x r_q+x r_q^2+\bar{m}_q^2\big){}^2}+\frac{15 \bar{m}_q \big(r_A \eta _A+2 x r_A'\big)}{2 \big(x+x r_A+\bar{m}_{\text{gap}}^2\big){}^3 \big(x+2 x r_q+x r_q^2+\bar{m}_q^2\big)}\big)
+
\lambda _{A^3} \big(\lambda _{A \bar{q} q}^{\text{(4)}}\big){}^2 \big(-\frac{9 x \big(1+r_q\big) \bar{m}_q \big(r_q \eta _q+2 x r_q'\big)}{\big(x+x r_A+\bar{m}_{\text{gap}}^2\big){}^2 \big(x+2 x r_q+x r_q^2+\bar{m}_q^2\big){}^2}-\frac{73 x \bar{m}_q \big(r_A \eta _A+2 x r_A'\big)}{8 \big(x+x r_A+\bar{m}_{\text{gap}}^2\big){}^3 \big(x+2 x r_q+x r_q^2+\bar{m}_q^2\big)}\big) 
+ 
\lambda _{A \bar{q} q}^{\text{(1)}} \big(\lambda _{A \bar{q} q}^{\text{(4)}}\big){}^2 \big(
-\frac{x \bar{m}_q \big(6+6 r_q+x r_q'\big) \big(r_A \eta _A+2 x r_A'\big)}{12 \big(x+x r_A+\bar{m}_{\text{gap}}^2\big){}^2 \big(x+2 x r_q+x r_q^2+\bar{m}_q^2\big){}^2}
-
\frac{1}{36 \big(x+x r_A+\bar{m}_{\text{gap}}^2\big){}^2 \big(x+2 x r_q+x r_q^2+\bar{m}_q^2\big){}^3}\bar{m}_q \big(r_q \eta _q+2 x r_q'\big) \big(58 x^2+116 x^2 r_q+58 x^2 r_q^2+61 x \bar{m}_{\text{gap}}^2+122 x r_q \bar{m}_{\text{gap}}^2+61 x r_q^2 \bar{m}_{\text{gap}}^2-13 x \bar{m}_q^2-10 \bar{m}_{\text{gap}}^2 \bar{m}_q^2+x r_A \big(58 x+116 x r_q+58 x r_q^2-13 \bar{m}_q^2\big)-3 x^2 \big(x+2 x r_q+x r_q^2+\bar{m}_q^2\big) r_A'\big)
\big)
+
}
\autoLineBreakEnd{
\big(\lambda _{A \bar{q} q}^{\text{(1)}}\big){}^3 \big(-\frac{x \bar{m}_q r_q' \big(r_A \eta _A+2 x r_A'\big)}{12 \big(x+x r_A+\bar{m}_{\text{gap}}^2\big){}^2 \big(x+2 x r_q+x r_q^2+\bar{m}_q^2\big){}^2}
+
\frac{1}{36 x \big(x+x r_A+\bar{m}_{\text{gap}}^2\big){}^2 \big(x+2 x r_q+x r_q^2+\bar{m}_q^2\big){}^3}\bar{m}_q \big(r_q \eta _q+2 x r_q'\big) \big(5 x^2+10 x^2 r_q+5 x^2 r_q^2+2 x \bar{m}_{\text{gap}}^2+4 x r_q \bar{m}_{\text{gap}}^2+2 x r_q^2 \bar{m}_{\text{gap}}^2-3 x \bar{m}_q^2-6 \bar{m}_{\text{gap}}^2 \bar{m}_q^2+x r_A \big(5 x+10 x r_q+5 x r_q^2-3 \bar{m}_q^2\big)+3 x^2 \big(x+2 x r_q+x r_q^2+\bar{m}_q^2\big) r_A'\big)
\big)
+g_{\Delta  \bar{q} \bar{q}}^2 \lambda _{A \bar{q} q}^{\text{(4)}} \big(-\frac{4 \big(1+r_q\big) \bar{m}_q^2 \big(r_q \eta _q+2 x r_q'\big)}{\big(x+2 x r_q+x r_q^2+\bar{m}_q^2\big){}^3 \big(x+x r_{\Delta }+\bar{m}_{\Delta }^2\big)}-\frac{\bar{m}_q^2 \big(r_{\Delta } \eta _{\Delta }+2 x r_{\Delta }'\big)}{\big(x+2 x r_q+x r_q^2+\bar{m}_q^2\big){}^2 \big(x+x r_{\Delta }+\bar{m}_{\Delta }^2\big){}^2}\big)
+g_{\Delta  \bar{q} \bar{q}}^2 \lambda _{A \bar{q} q}^{\text{(1)}} 
\big(
\frac{\bar{m}_q \big(2+2 r_q+x r_q'\big) \big(r_{\Delta } \eta _{\Delta }+2 x r_{\Delta }'\big)}{2 \big(x+2 x r_q+x r_q^2+\bar{m}_q^2\big){}^2 \big(x+x r_{\Delta }
+\bar{m}_{\Delta }^2\big){}^2}
-
\frac{1}{2 \big(x+2 x r_q+x r_q^2+\bar{m}_q^2\big){}^3 \big(x+x r_{\Delta }+\bar{m}_{\Delta }^2\big){}^2}\bar{m}_q \big(r_q \eta _q+2 x r_q'\big) \big(-5 x-10 x r_q-5 x r_q^2+\bar{m}_q^2+r_{\Delta } \big(-5 x-10 x r_q-5 x r_q^2+\bar{m}_q^2\big)-6 \bar{m}_{\Delta }^2-12 r_q \bar{m}_{\Delta }^2-6 r_q^2 \bar{m}_{\Delta }^2+x \big(x+2 x r_q+x r_q^2+\bar{m}_q^2\big) r_{\Delta }'\big)
\big)
+ g_{\phi  \bar{q} q}^2 \lambda _{A \bar{q} q}^{\text{(4)}} 
\big(\frac{4 \big(1+r_q\big) \bar{m}_q^2 \big(-2 x-2 x r_{\phi }+\bar{m}_{\pi }^2-3 \bar{m}_{\sigma }^2\big) \big(r_q \eta _q+2 x r_q'\big)}{\big(x+x r_{\phi }+\bar{m}_{\pi }^2\big) \big(x+2 x r_q+x r_q^2+\bar{m}_q^2\big){}^3 \big(x+x r_{\phi }+\bar{m}_{\sigma }^2\big)}+\frac{\bar{m}_q^2 \big(-2 x^2-2 x^2 r_{\phi }^2+2 x \bar{m}_{\pi }^2+\bar{m}_{\pi }^4-6 x \bar{m}_{\sigma }^2-3 \bar{m}_{\sigma }^4+2 x r_{\phi } \big(-2 x+\bar{m}_{\pi }^2-3 \bar{m}_{\sigma }^2\big)\big) \big(r_{\phi } \eta _{\phi }+2 x r_{\phi }'\big)}{\big(x+x r_{\phi }+\bar{m}_{\pi }^2\big){}^2 \big(x+2 x r_q+x r_q^2+\bar{m}_q^2\big){}^2 \big(x+x r_{\phi }+\bar{m}_{\sigma }^2\big){}^2}\big)
+g_{\phi  \bar{q} q}^2 \lambda _{A \bar{q} q}^{\text{(1)}} \big(
-\frac{\bar{m}_q \big(-2 x^2-2 x^2 r_{\phi }^2+2 x \bar{m}_{\pi }^2+\bar{m}_{\pi }^4-6 x \bar{m}_{\sigma }^2-3 \bar{m}_{\sigma }^4+2 x r_{\phi } \big(-2 x+\bar{m}_{\pi }^2-3 \bar{m}_{\sigma }^2\big)\big) \big(2+2 r_q+x r_q'\big) \big(r_{\phi } \eta _{\phi }+2 x r_{\phi }'\big)}{2 \big(x+x r_{\phi }+\bar{m}_{\pi }^2\big){}^2 \big(x+2 x r_q+x r_q^2+\bar{m}_q^2\big){}^2 \big(x+x r_{\phi }+\bar{m}_{\sigma }^2\big){}^2}
+
\frac{1}{{2 \big(x+x r_{\phi }+\bar{m}_{\pi }^2\big){}^2 \big(x+2 x r_q+x r_q^2+\bar{m}_q^2\big){}^3 \big(x+x r_{\phi }+\bar{m}_{\sigma }^2\big){}^2}}
\bar{m}_q \big(r_q \eta _q+2 x r_q'\big) \big(10 x^3+20 x^3 r_q+10 x^3 r_q^2+8 x^2 \bar{m}_{\pi }^2+16 x^2 r_q \bar{m}_{\pi }^2+8 x^2 r_q^2 \bar{m}_{\pi }^2-5 x \bar{m}_{\pi }^4-10 x r_q \bar{m}_{\pi }^4-5 x r_q^2 \bar{m}_{\pi }^4-2 x^2 \bar{m}_q^2+2 x \bar{m}_{\pi }^2 \bar{m}_q^2+\bar{m}_{\pi }^4 \bar{m}_q^2+2 x^2 r_{\phi }^3 \big(5 x+10 x r_q+5 x r_q^2-\bar{m}_q^2\big)+24 x^2 \bar{m}_{\sigma }^2+48 x^2 r_q \bar{m}_{\sigma }^2+24 x^2 r_q^2 \bar{m}_{\sigma }^2+24 x \bar{m}_{\pi }^2 \bar{m}_{\sigma }^2+48 x r_q \bar{m}_{\pi }^2 \bar{m}_{\sigma }^2+24 x r_q^2 \bar{m}_{\pi }^2 \bar{m}_{\sigma }^2-6 \bar{m}_{\pi }^4 \bar{m}_{\sigma }^2-12 r_q \bar{m}_{\pi }^4 \bar{m}_{\sigma }^2-6 r_q^2 \bar{m}_{\pi }^4 \bar{m}_{\sigma }^2-6 x \bar{m}_q^2 \bar{m}_{\sigma }^2+15 x \bar{m}_{\sigma }^4+30 x r_q \bar{m}_{\sigma }^4+15 x r_q^2 \bar{m}_{\sigma }^4+18 \bar{m}_{\pi }^2 \bar{m}_{\sigma }^4+36 r_q \bar{m}_{\pi }^2 \bar{m}_{\sigma }^4+18 r_q^2 \bar{m}_{\pi }^2 \bar{m}_{\sigma }^4-3 \bar{m}_q^2 \bar{m}_{\sigma }^4+x \big(x+2 x r_q+x r_q^2+\bar{m}_q^2\big) \big(-2 x^2+2 x \bar{m}_{\pi }^2+\bar{m}_{\pi }^4-6 x \bar{m}_{\sigma }^2-3 \bar{m}_{\sigma }^4\big) r_{\phi }'+2 x r_{\phi }^2 \big(15 x^2+4 x \bar{m}_{\pi }^2-3 x \bar{m}_q^2+\bar{m}_{\pi }^2 \bar{m}_q^2+12 x \bar{m}_{\sigma }^2-3 \bar{m}_q^2 \bar{m}_{\sigma }^2+2 x r_q \big(15 x+4 \bar{m}_{\pi }^2+12 \bar{m}_{\sigma }^2\big)+x r_q^2 \big(15 x+4 \bar{m}_{\pi }^2+12 \bar{m}_{\sigma }^2\big)-x^2 \big(x+2 x r_q+x r_q^2+\bar{m}_q^2\big) r_{\phi }'\big)+r_{\phi } \big(30 x^3+16 x^2 \bar{m}_{\pi }^2-5 x \bar{m}_{\pi }^4-6 x^2 \bar{m}_q^2+4 x \bar{m}_{\pi }^2 \bar{m}_q^2+\bar{m}_{\pi }^4 \bar{m}_q^2+48 x^2 \bar{m}_{\sigma }^2+24 x \bar{m}_{\pi }^2 \bar{m}_{\sigma }^2-12 x \bar{m}_q^2 \bar{m}_{\sigma }^2+15 x \bar{m}_{\sigma }^4-3 \bar{m}_q^2 \bar{m}_{\sigma }^4+2 x r_q \big(30 x^2+16 x \bar{m}_{\pi }^2-5 \bar{m}_{\pi }^4+48 x \bar{m}_{\sigma }^2+24 \bar{m}_{\pi }^2 \bar{m}_{\sigma }^2+15 \bar{m}_{\sigma }^4\big)+x r_q^2 \big(30 x^2+16 x \bar{m}_{\pi }^2-5 \bar{m}_{\pi }^4+48 x \bar{m}_{\sigma }^2+24 \bar{m}_{\pi }^2 \bar{m}_{\sigma }^2+15 \bar{m}_{\sigma }^4\big)+2 x^2 \big(x+2 x r_q+x r_q^2+\bar{m}_q^2\big) \big(-2 x+\bar{m}_{\pi }^2-3 \bar{m}_{\sigma }^2\big) r_{\phi }'\big)\big)
\big)
}

\autoLineBreak{
\partial_t \lambda_{A\Delta^\dagger\Delta} = 
(\eta_\Delta + \frac{1}{2} \eta_A) \lambda_{A\Delta^\dagger\Delta} +
}{
\frac{12 g_{\Delta  \bar{q} \bar{q}}^2 \big(1+r_q\big){}^2 \lambda _{A \bar{q} q}^{\text{(1)}} \big(r_q \eta _q+2 x r_q'\big)}{\big(x+2 x r_q+x r_q^2+\bar{m}_q^2\big){}^3}
+
\lambda _{A \Delta ^{\dagger } \Delta } \lambda _{A^2 \Delta ^{\dagger } \Delta } \big(-\frac{7 \big(r_A \eta _A+2 x r_A'\big)}{4 \big(x+x r_A+\bar{m}_{\text{gap}}^2\big){}^2 \big(x+x r_{\Delta }+\bar{m}_{\Delta }^2\big)}-\frac{7 \big(r_{\Delta } \eta _{\Delta }+2 x r_{\Delta }'\big)}{4 \big(x+x r_A+\bar{m}_{\text{gap}}^2\big) \big(x+x r_{\Delta }+\bar{m}_{\Delta }^2\big){}^2}\big)
+
\lambda _{A \Delta ^{\dagger } \Delta } \lambda _{1,1}^2 \big(\frac{x \bar{\sigma }_0^2 \big(r_{\Delta } \eta _{\Delta }+2 x r_{\Delta }'\big) \big(1+r_{\phi }+x r_{\phi }'\big)}{\big(x+x r_{\Delta }+\bar{m}_{\Delta }^2\big){}^3 \big(x+x r_{\phi }+\bar{m}_{\sigma }^2\big){}^2}+\frac{\bar{\sigma }_0^2 \big(\bar{m}_{\Delta }^2-x^2 r_{\Delta }'\big) \big(r_{\phi } \eta _{\phi }+2 x r_{\phi }'\big)}{\big(x+x r_{\Delta }+\bar{m}_{\Delta }^2\big){}^3 \big(x+x r_{\phi }+\bar{m}_{\sigma }^2\big){}^2}\big)
}

\autoLineBreakStart{
\partial_t \lambda_{A^2\Delta^\dagger\Delta} = 
(\eta_\Delta + \eta_A) \lambda_{A^2\Delta^\dagger\Delta} +
}{
\frac{16 g_{\Delta  \bar{q} \bar{q}}^2 \big(1+r_q\big) \big(x+2 x r_q+x r_q^2+4 \bar{m}_q^2\big) \big(\lambda _{A \bar{q} q}^{\text{(1)}}\big){}^2 \big(r_q \eta _q+2 x r_q'\big)}{\big(x+2 x r_q+x r_q^2+\bar{m}_q^2\big){}^4}
+
\frac{27 x \lambda _{A^3}^2 \lambda _{A^2 \Delta ^{\dagger } \Delta } \big(r_A \eta _A+2 x r_A'\big)}{\big(x+x r_A+\bar{m}_{\text{gap}}^2\big){}^4}
-
\frac{27 \lambda _{A^4} \lambda _{A^2 \Delta ^{\dagger } \Delta } \big(r_A \eta _A+2 x r_A'\big)}{2 \big(x+x r_A+\bar{m}_{\text{gap}}^2\big){}^3}
+
\frac{12 x \lambda _{A \Delta ^{\dagger } \Delta }^2 \lambda _{0,2} \big(r_{\Delta } \eta _{\Delta }+2 x r_{\Delta }'\big)}{\big(x+x r_{\Delta }+\bar{m}_{\Delta }^2\big){}^4}
-
\frac{8 \lambda _{A^2 \Delta ^{\dagger } \Delta } \lambda _{0,2} \big(r_{\Delta } \eta _{\Delta }+2 x r_{\Delta }'\big)}{\big(x+x r_{\Delta }+\bar{m}_{\Delta }^2\big){}^3}
+
\lambda _{A^2 \Delta ^{\dagger } \Delta }^2 \big(-\frac{7 \big(r_A \eta _A+2 x r_A'\big)}{4 \big(x+x r_A+\bar{m}_{\text{gap}}^2\big){}^2 \big(x+x r_{\Delta }+\bar{m}_{\Delta }^2\big)}-\frac{7 \big(r_{\Delta } \eta _{\Delta }+2 x r_{\Delta }'\big)}{4 \big(x+x r_A+\bar{m}_{\text{gap}}^2\big) \big(x+x r_{\Delta }+\bar{m}_{\Delta }^2\big){}^2}\big)
+
\lambda _{A \Delta ^{\dagger } \Delta }^2 \lambda _{1,1}^2 \big(-\frac{3 x \bar{\sigma }_0^2 \big(r_{\Delta } \eta _{\Delta }+2 x r_{\Delta }'\big)}{\big(x+x r_{\Delta }+\bar{m}_{\Delta }^2\big){}^4 \big(x+x r_{\phi }+\bar{m}_{\sigma }^2\big)}-\frac{x \bar{\sigma }_0^2 \big(r_{\phi } \eta _{\phi }+2 x r_{\phi }'\big)}{\big(x+x r_{\Delta }+\bar{m}_{\Delta }^2\big){}^3 \big(x+x r_{\phi }+\bar{m}_{\sigma }^2\big){}^2}\big)
+
}
\autoLineBreakEnd{
\lambda _{A^2 \Delta ^{\dagger } \Delta } \lambda _{1,1}^2 \big(\frac{2 \bar{\sigma }_0^2 \big(r_{\Delta } \eta _{\Delta }+2 x r_{\Delta }'\big)}{\big(x+x r_{\Delta }+\bar{m}_{\Delta }^2\big){}^3 \big(x+x r_{\phi }+\bar{m}_{\sigma }^2\big)}+\frac{\bar{\sigma }_0^2 \big(r_{\phi } \eta _{\phi }+2 x r_{\phi }'\big)}{\big(x+x r_{\Delta }+\bar{m}_{\Delta }^2\big){}^2 \big(x+x r_{\phi }+\bar{m}_{\sigma }^2\big){}^2}\big)
}

\subsubsection{Yukawa couplings}

\autoLineBreak{
\partial_t g_{\phi\bar{q}q} = (\eta_q + \frac{1}{2} \eta_\phi) g_{\phi\bar{q}q} +
}{
\lambda _{A \bar{q} q}^{\text{(1)}} \lambda _{A \bar{q} q}^{\text{(4)}} \big(\frac{8 \big(x+2 x r_q+x r_q^2-\bar{m}_q^2\big) \big(r_q \eta _q+2 x r_q'\big)}{\big(x+x r_A+\bar{m}_{\text{gap}}^2\big) \big(x+2 x r_q+x r_q^2+\bar{m}_q^2\big){}^2 \bar{\sigma }_0}+\frac{8 x \big(1+r_q\big) \big(r_A \eta _A+2 x r_A'\big)}{\big(x+x r_A+\bar{m}_{\text{gap}}^2\big){}^2 \big(x+2 x r_q+x r_q^2+\bar{m}_q^2\big) \bar{\sigma }_0}\big)+\big(\lambda _{A \bar{q} q}^{\text{(1)}}\big){}^2 \big(\frac{8 \big(1+r_q\big) \bar{m}_q \big(r_q \eta _q+2 x r_q'\big)}{\big(x+x r_A+\bar{m}_{\text{gap}}^2\big) \big(x+2 x r_q+x r_q^2+\bar{m}_q^2\big){}^2 \bar{\sigma }_0}+\frac{4 \bar{m}_q \big(r_A \eta _A+2 x r_A'\big)}{\big(x+x r_A+\bar{m}_{\text{gap}}^2\big){}^2 \big(x+2 x r_q+x r_q^2+\bar{m}_q^2\big) \bar{\sigma }_0}\big)+\big(\lambda _{A \bar{q} q}^{\text{(4)}}\big){}^2 \big(-\frac{8 x \big(1+r_q\big) \bar{m}_q \big(r_q \eta _q+2 x r_q'\big)}{\big(x+x r_A+\bar{m}_{\text{gap}}^2\big) \big(x+2 x r_q+x r_q^2+\bar{m}_q^2\big){}^2 \bar{\sigma }_0}-\frac{4 x \bar{m}_q \big(r_A \eta _A+2 x r_A'\big)}{\big(x+x r_A+\bar{m}_{\text{gap}}^2\big){}^2 \big(x+2 x r_q+x r_q^2+\bar{m}_q^2\big) \bar{\sigma }_0}\big)+g_{\Delta  \bar{q} \bar{q}}^2 \big(\frac{4 \big(1+r_q\big) \bar{m}_q \big(r_q \eta _q+2 x r_q'\big)}{\big(x+2 x r_q+x r_q^2+\bar{m}_q^2\big){}^2 \big(x+x r_{\Delta }+\bar{m}_{\Delta }^2\big) \bar{\sigma }_0}+\frac{2 \bar{m}_q \big(r_{\Delta } \eta _{\Delta }+2 x r_{\Delta }'\big)}{\big(x+2 x r_q+x r_q^2+\bar{m}_q^2\big) \big(x+x r_{\Delta }+\bar{m}_{\Delta }^2\big){}^2 \bar{\sigma }_0}\big)+g_{\phi  \bar{q} q}^2 \big(-\frac{2 \big(1+r_q\big) \bar{m}_q \big(-2 x-2 x r_{\phi }+\bar{m}_{\pi }^2-3 \bar{m}_{\sigma }^2\big) \big(r_q \eta _q+2 x r_q'\big)}{\big(x+x r_{\phi }+\bar{m}_{\pi }^2\big) \big(x+2 x r_q+x r_q^2+\bar{m}_q^2\big){}^2 \big(x+x r_{\phi }+\bar{m}_{\sigma }^2\big) \bar{\sigma }_0}-\frac{\bar{m}_q \big(-2 x^2-2 x^2 r_{\phi }^2+2 x \bar{m}_{\pi }^2+\bar{m}_{\pi }^4-6 x \bar{m}_{\sigma }^2-3 \bar{m}_{\sigma }^4+2 x r_{\phi } \big(-2 x+\bar{m}_{\pi }^2-3 \bar{m}_{\sigma }^2\big)\big) \big(r_{\phi } \eta _{\phi }+2 x r_{\phi }'\big)}{\big(x+x r_{\phi }+\bar{m}_{\pi }^2\big){}^2 \big(x+2 x r_q+x r_q^2+\bar{m}_q^2\big) \big(x+x r_{\phi }+\bar{m}_{\sigma }^2\big){}^2 \bar{\sigma }_0}\big)
}

\autoLineBreak{
\partial_t g_{\Delta\bar{q}\bar{q}} = (\eta_q + \frac{1}{2} \eta_\Delta) g_{\Delta\bar{q}\bar{q}} +
}{
g_{\Delta  \bar{q} \bar{q}} \big(\lambda _{A \bar{q} q}^{\text{(1)}}\big){}^2 \big(\frac{4 \big(1+r_q\big) \big(r_q \eta _q+2 x r_q'\big)}{\big(x+x r_A+\bar{m}_{\text{gap}}^2\big) \big(x+2 x r_q+x r_q^2+\bar{m}_q^2\big){}^2}+\frac{2 \big(r_A \eta _A+2 x r_A'\big)}{\big(x+x r_A+\bar{m}_{\text{gap}}^2\big){}^2 \big(x+2 x r_q+x r_q^2+\bar{m}_q^2\big)}\big)
+
g_{\Delta  \bar{q} \bar{q}} \big(\lambda _{A \bar{q} q}^{\text{(4)}}\big){}^2 \big(\frac{4 x \big(1+r_q\big) \big(r_q \eta _q+2 x r_q'\big)}{\big(x+x r_A+\bar{m}_{\text{gap}}^2\big) \big(x+2 x r_q+x r_q^2+\bar{m}_q^2\big){}^2}+\frac{2 x \big(r_A \eta _A+2 x r_A'\big)}{\big(x+x r_A+\bar{m}_{\text{gap}}^2\big){}^2 \big(x+2 x r_q+x r_q^2+\bar{m}_q^2\big)}\big)
+
g_{\phi  \bar{q} q}^2 g_{\Delta  \bar{q} \bar{q}} \big(\frac{2 \big(1+r_q\big) \big(4 x+4 x r_{\phi }+\bar{m}_{\pi }^2+3 \bar{m}_{\sigma }^2\big) \big(r_q \eta _q+2 x r_q'\big)}{\big(x+x r_{\phi }+\bar{m}_{\pi }^2\big) \big(x+2 x r_q+x r_q^2+\bar{m}_q^2\big){}^2 \big(x+x r_{\phi }+\bar{m}_{\sigma }^2\big)}+\frac{\big(4 x^2+4 x^2 r_{\phi }^2+2 x \bar{m}_{\pi }^2+\bar{m}_{\pi }^4+6 x \bar{m}_{\sigma }^2+3 \bar{m}_{\sigma }^4+2 x r_{\phi } \big(4 x+\bar{m}_{\pi }^2+3 \bar{m}_{\sigma }^2\big)\big) \big(r_{\phi } \eta _{\phi }+2 x r_{\phi }'\big)}{\big(x+x r_{\phi }+\bar{m}_{\pi }^2\big){}^2 \big(x+2 x r_q+x r_q^2+\bar{m}_q^2\big) \big(x+x r_{\phi }+\bar{m}_{\sigma }^2\big){}^2}\big)
+
g_{\phi  \bar{q} q} g_{\Delta  \bar{q} \bar{q}} \lambda _{1,1} \big(\frac{4 \big(1+r_q\big) \bar{m}_q \bar{\sigma }_0 \big(r_q \eta _q+2 x r_q'\big)}{\big(x+2 x r_q+x r_q^2+\bar{m}_q^2\big){}^2 \big(x+x r_{\Delta }+\bar{m}_{\Delta }^2\big) \big(x+x r_{\phi }+\bar{m}_{\sigma }^2\big)}+\frac{2 \bar{m}_q \bar{\sigma }_0 \big(r_{\Delta } \eta _{\Delta }+2 x r_{\Delta }'\big)}{\big(x+2 x r_q+x r_q^2+\bar{m}_q^2\big) \big(x+x r_{\Delta }+\bar{m}_{\Delta }^2\big){}^2 \big(x+x r_{\phi }+\bar{m}_{\sigma }^2\big)}+\frac{2 \bar{m}_q \bar{\sigma }_0 \big(r_{\phi } \eta _{\phi }+2 x r_{\phi }'\big)}{\big(x+2 x r_q+x r_q^2+\bar{m}_q^2\big) \big(x+x r_{\Delta }+\bar{m}_{\Delta }^2\big) \big(x+x r_{\phi }+\bar{m}_{\sigma }^2\big){}^2}\big)
}

\subsubsection{Four-quark flows}

\autoLineBreakStart{
\left[ \mathcal{P}^{(\text{sps})}_{\qbqbqq}
  \right]_{\alpha\beta\gamma\delta}
  \left[ \overline{\text{Flow}}_{\qbqbqq}^{(4)}
  \right]_{\alpha\beta\gamma\delta} =
}{
\big(\lambda _{A \bar{q} q}^{\text{(1)}}\big){}^3 \lambda _{A \bar{q} q}^{\text{(4)}} \big(\frac{\bar{m}_q \big(3 x+6 x r_q+3 x r_q^2-\bar{m}_q^2\big) \big(r_q \eta _q+2 x r_q'\big)}{6 \big(x+x r_A+\bar{m}_{\text{gap}}^2\big){}^2 \big(x+2 x r_q+x r_q^2+\bar{m}_q^2\big){}^3}+\frac{x \big(1+r_q\big) \bar{m}_q \big(r_A \eta _A+2 x r_A'\big)}{3 \big(x+x r_A+\bar{m}_{\text{gap}}^2\big){}^3 \big(x+2 x r_q+x r_q^2+\bar{m}_q^2\big){}^2}\big)
+
}
\autoLineBreakContinue{
\lambda _{A \bar{q} q}^{\text{(1)}} \big(\lambda _{A \bar{q} q}^{\text{(4)}}\big){}^3 \big(\frac{x \bar{m}_q \big(-3 x-6 x r_q-3 x r_q^2+\bar{m}_q^2\big) \big(r_q \eta _q+2 x r_q'\big)}{6 \big(x+x r_A+\bar{m}_{\text{gap}}^2\big){}^2 \big(x+2 x r_q+x r_q^2+\bar{m}_q^2\big){}^3}-\frac{x^2 \big(1+r_q\big) \bar{m}_q \big(r_A \eta _A+2 x r_A'\big)}{3 \big(x+x r_A+\bar{m}_{\text{gap}}^2\big){}^3 \big(x+2 x r_q+x r_q^2+\bar{m}_q^2\big){}^2}\big)
+
\big(\lambda _{A \bar{q} q}^{\text{(1)}}\big){}^4 \big(-\frac{\big(1+r_q\big) \big(57 x+114 x r_q+57 x r_q^2+41 \bar{m}_q^2\big) \big(r_q \eta _q+2 x r_q'\big)}{96 \big(x+x r_A+\bar{m}_{\text{gap}}^2\big){}^2 \big(x+2 x r_q+x r_q^2+\bar{m}_q^2\big){}^3}-\frac{\big(57 x+114 x r_q+57 x r_q^2+49 \bar{m}_q^2\big) \big(r_A \eta _A+2 x r_A'\big)}{96 \big(x+x r_A+\bar{m}_{\text{gap}}^2\big){}^3 \big(x+2 x r_q+x r_q^2+\bar{m}_q^2\big){}^2}\big)
+
\big(\lambda _{A \bar{q} q}^{\text{(4)}}\big){}^4 \big(-\frac{x^2 \big(1+r_q\big) \big(57 x+114 x r_q+57 x r_q^2+41 \bar{m}_q^2\big) \big(r_q \eta _q+2 x r_q'\big)}{96 \big(x+x r_A+\bar{m}_{\text{gap}}^2\big){}^2 \big(x+2 x r_q+x r_q^2+\bar{m}_q^2\big){}^3}-\frac{x^2 \big(57 x+114 x r_q+57 x r_q^2+49 \bar{m}_q^2\big) \big(r_A \eta _A+2 x r_A'\big)}{96 \big(x+x r_A+\bar{m}_{\text{gap}}^2\big){}^3 \big(x+2 x r_q+x r_q^2+\bar{m}_q^2\big){}^2}\big)
+
\big(\lambda _{A \bar{q} q}^{\text{(1)}}\big){}^2 \big(\lambda _{A \bar{q} q}^{\text{(4)}}\big){}^2 \big(-\frac{x \big(1+r_q\big) \big(41 x+82 x r_q+41 x r_q^2+89 \bar{m}_q^2\big) \big(r_q \eta _q+2 x r_q'\big)}{48 \big(x+x r_A+\bar{m}_{\text{gap}}^2\big){}^2 \big(x+2 x r_q+x r_q^2+\bar{m}_q^2\big){}^3}-\frac{x \big(41 x+82 x r_q+41 x r_q^2+65 \bar{m}_q^2\big) \big(r_A \eta _A+2 x r_A'\big)}{48 \big(x+x r_A+\bar{m}_{\text{gap}}^2\big){}^3 \big(x+2 x r_q+x r_q^2+\bar{m}_q^2\big){}^2}\big)
+
g_{\Delta  \bar{q} \bar{q}}^4 \big(-\frac{\big(1+r_q\big) \big(x+2 x r_q+x r_q^2+5 \bar{m}_q^2\big) \big(r_q \eta _q+2 x r_q'\big)}{8 \big(x+2 x r_q+x r_q^2+\bar{m}_q^2\big){}^3 \big(x+x r_{\Delta }+\bar{m}_{\Delta }^2\big){}^2}-\frac{\big(x+2 x r_q+x r_q^2+3 \bar{m}_q^2\big) \big(r_{\Delta } \eta _{\Delta }+2 x r_{\Delta }'\big)}{8 \big(x+2 x r_q+x r_q^2+\bar{m}_q^2\big){}^2 \big(x+x r_{\Delta }+\bar{m}_{\Delta }^2\big){}^3}\big)
+
g_{\Delta  \bar{q} \bar{q}}^2 \lambda _{A \bar{q} q}^{\text{(1)}} \lambda _{A \bar{q} q}^{\text{(4)}} \big(\frac{3 \bar{m}_q \big(3 x+6 x r_q+3 x r_q^2-\bar{m}_q^2\big) \big(r_q \eta _q+2 x r_q'\big)}{4 \big(x+x r_A+\bar{m}_{\text{gap}}^2\big) \big(x+2 x r_q+x r_q^2+\bar{m}_q^2\big){}^3 \big(x+x r_{\Delta }+\bar{m}_{\Delta }^2\big)}+\frac{3 x \big(1+r_q\big) \bar{m}_q \big(r_A \eta _A+2 x r_A'\big)}{4 \big(x+x r_A+\bar{m}_{\text{gap}}^2\big){}^2 \big(x+2 x r_q+x r_q^2+\bar{m}_q^2\big){}^2 \big(x+x r_{\Delta }+\bar{m}_{\Delta }^2\big)}+\frac{3 x \big(1+r_q\big) \bar{m}_q \big(r_{\Delta } \eta _{\Delta }+2 x r_{\Delta }'\big)}{4 \big(x+x r_A+\bar{m}_{\text{gap}}^2\big) \big(x+2 x r_q+x r_q^2+\bar{m}_q^2\big){}^2 \big(x+x r_{\Delta }+\bar{m}_{\Delta }^2\big){}^2}\big)
+
g_{\Delta  \bar{q} \bar{q}}^2 \big(\lambda _{A \bar{q} q}^{\text{(1)}}\big){}^2 \big(\frac{\big(1+r_q\big) \big(-2 x-4 x r_q-2 x r_q^2+\bar{m}_q^2\big) \big(r_q \eta _q+2 x r_q'\big)}{2 \big(x+x r_A+\bar{m}_{\text{gap}}^2\big) \big(x+2 x r_q+x r_q^2+\bar{m}_q^2\big){}^3 \big(x+x r_{\Delta }+\bar{m}_{\Delta }^2\big)}-\frac{\big(4 x+8 x r_q+4 x r_q^2+\bar{m}_q^2\big) \big(r_A \eta _A+2 x r_A'\big)}{8 \big(x+x r_A+\bar{m}_{\text{gap}}^2\big){}^2 \big(x+2 x r_q+x r_q^2+\bar{m}_q^2\big){}^2 \big(x+x r_{\Delta }+\bar{m}_{\Delta }^2\big)}-\frac{\big(4 x+8 x r_q+4 x r_q^2+\bar{m}_q^2\big) \big(r_{\Delta } \eta _{\Delta }+2 x r_{\Delta }'\big)}{8 \big(x+x r_A+\bar{m}_{\text{gap}}^2\big) \big(x+2 x r_q+x r_q^2+\bar{m}_q^2\big){}^2 \big(x+x r_{\Delta }+\bar{m}_{\Delta }^2\big){}^2}\big)
+
g_{\Delta  \bar{q} \bar{q}}^2 \big(\lambda _{A \bar{q} q}^{\text{(4)}}\big){}^2 \big(-\frac{x \big(1+r_q\big) \big(x+2 x r_q+x r_q^2+7 \bar{m}_q^2\big) \big(r_q \eta _q+2 x r_q'\big)}{4 \big(x+x r_A+\bar{m}_{\text{gap}}^2\big) \big(x+2 x r_q+x r_q^2+\bar{m}_q^2\big){}^3 \big(x+x r_{\Delta }+\bar{m}_{\Delta }^2\big)}-\frac{x \big(x+2 x r_q+x r_q^2+4 \bar{m}_q^2\big) \big(r_A \eta _A+2 x r_A'\big)}{8 \big(x+x r_A+\bar{m}_{\text{gap}}^2\big){}^2 \big(x+2 x r_q+x r_q^2+\bar{m}_q^2\big){}^2 \big(x+x r_{\Delta }+\bar{m}_{\Delta }^2\big)}-\frac{x \big(x+2 x r_q+x r_q^2+4 \bar{m}_q^2\big) \big(r_{\Delta } \eta _{\Delta }+2 x r_{\Delta }'\big)}{8 \big(x+x r_A+\bar{m}_{\text{gap}}^2\big) \big(x+2 x r_q+x r_q^2+\bar{m}_q^2\big){}^2 \big(x+x r_{\Delta }+\bar{m}_{\Delta }^2\big){}^2}\big)
+
g_{\phi  \bar{q} q}^2 \lambda _{A \bar{q} q}^{\text{(1)}} \lambda _{A \bar{q} q}^{\text{(4)}} \big(\frac{\bar{m}_q \big(-3 x-6 x r_q-3 x r_q^2+\bar{m}_q^2\big) \big(x+x r_{\phi }+7 \bar{m}_{\pi }^2-6 \bar{m}_{\sigma }^2\big) \big(r_q \eta _q+2 x r_q'\big)}{12 \big(x+x r_A+\bar{m}_{\text{gap}}^2\big) \big(x+x r_{\phi }+\bar{m}_{\pi }^2\big) \big(x+2 x r_q+x r_q^2+\bar{m}_q^2\big){}^3 \big(x+x r_{\phi }+\bar{m}_{\sigma }^2\big)}-\frac{x \big(1+r_q\big) \bar{m}_q \big(x+x r_{\phi }+7 \bar{m}_{\pi }^2-6 \bar{m}_{\sigma }^2\big) \big(r_A \eta _A+2 x r_A'\big)}{12 \big(x+x r_A+\bar{m}_{\text{gap}}^2\big){}^2 \big(x+x r_{\phi }+\bar{m}_{\pi }^2\big) \big(x+2 x r_q+x r_q^2+\bar{m}_q^2\big){}^2 \big(x+x r_{\phi }+\bar{m}_{\sigma }^2\big)}-\frac{x \big(1+r_q\big) \bar{m}_q \big(x^2+x^2 r_{\phi }^2+14 x \bar{m}_{\pi }^2+7 \bar{m}_{\pi }^4-12 x \bar{m}_{\sigma }^2-6 \bar{m}_{\sigma }^4+2 x r_{\phi } \big(x+7 \bar{m}_{\pi }^2-6 \bar{m}_{\sigma }^2\big)\big) \big(r_{\phi } \eta _{\phi }+2 x r_{\phi }'\big)}{12 \big(x+x r_A+\bar{m}_{\text{gap}}^2\big) \big(x+x r_{\phi }+\bar{m}_{\pi }^2\big){}^2 \big(x+2 x r_q+x r_q^2+\bar{m}_q^2\big){}^2 \big(x+x r_{\phi }+\bar{m}_{\sigma }^2\big){}^2}\big)
+
g_{\phi  \bar{q} q}^2 g_{\Delta  \bar{q} \bar{q}}^2 \big(-\frac{\big(1+r_q\big) \big(x+2 x r_q+x r_q^2-\bar{m}_q^2\big) \big(4 x+4 x r_{\phi }+\bar{m}_{\pi }^2+3 \bar{m}_{\sigma }^2\big) \big(r_q \eta _q+2 x r_q'\big)}{4 \big(x+x r_{\phi }+\bar{m}_{\pi }^2\big) \big(x+2 x r_q+x r_q^2+\bar{m}_q^2\big){}^3 \big(x+x r_{\Delta }+\bar{m}_{\Delta }^2\big) \big(x+x r_{\phi }+\bar{m}_{\sigma }^2\big)}-\frac{x \big(1+r_q\big){}^2 \big(4 x+4 x r_{\phi }+\bar{m}_{\pi }^2+3 \bar{m}_{\sigma }^2\big) \big(r_{\Delta } \eta _{\Delta }+2 x r_{\Delta }'\big)}{8 \big(x+x r_{\phi }+\bar{m}_{\pi }^2\big) \big(x+2 x r_q+x r_q^2+\bar{m}_q^2\big){}^2 \big(x+x r_{\Delta }+\bar{m}_{\Delta }^2\big){}^2 \big(x+x r_{\phi }+\bar{m}_{\sigma }^2\big)}-\frac{x \big(1+r_q\big){}^2 \big(4 x^2+4 x^2 r_{\phi }^2+2 x \bar{m}_{\pi }^2+\bar{m}_{\pi }^4+6 x \bar{m}_{\sigma }^2+3 \bar{m}_{\sigma }^4+2 x r_{\phi } \big(4 x+\bar{m}_{\pi }^2+3 \bar{m}_{\sigma }^2\big)\big) \big(r_{\phi } \eta _{\phi }+2 x r_{\phi }'\big)}{8 \big(x+x r_{\phi }+\bar{m}_{\pi }^2\big){}^2 \big(x+2 x r_q+x r_q^2+\bar{m}_q^2\big){}^2 \big(x+x r_{\Delta }+\bar{m}_{\Delta }^2\big) \big(x+x r_{\phi }+\bar{m}_{\sigma }^2\big){}^2}\big)
+
}
\autoLineBreakEnd{
g_{\phi  \bar{q} q}^2 \big(\lambda _{A \bar{q} q}^{\text{(4)}}\big){}^2 \big(
-
\frac{1}{12 \big(x+x r_A+\bar{m}_{\text{gap}}^2\big) \big(x+x r_{\phi }+\bar{m}_{\pi }^2\big) \big(x+2 x r_q+x r_q^2+\bar{m}_q^2\big){}^3 \big(x+x r_{\phi }+\bar{m}_{\sigma }^2\big)}x \big(1+r_q\big) \big(9 x^2+9 x \bar{m}_{\pi }^2+18 x r_q \big(x+\bar{m}_{\pi }^2\big)+9 x r_q^2 \big(x+\bar{m}_{\pi }^2\big)+7 x \bar{m}_q^2-5 \bar{m}_{\pi }^2 \bar{m}_q^2+x r_{\phi } \big(9 x+18 x r_q+9 x r_q^2+7 \bar{m}_q^2\big)+12 \bar{m}_q^2 \bar{m}_{\sigma }^2\big) \big(r_q \eta _q+2 x r_q'\big)
-
\frac{1}{24 \big(x+x r_A+\bar{m}_{\text{gap}}^2\big){}^2 \big(x+x r_{\phi }+\bar{m}_{\pi }^2\big) \big(x+2 x r_q+x r_q^2+\bar{m}_q^2\big){}^2 \big(x+x r_{\phi }+\bar{m}_{\sigma }^2\big)}x \big(9 x^2+9 x \bar{m}_{\pi }^2+18 x r_q \big(x+\bar{m}_{\pi }^2\big)+9 x r_q^2 \big(x+\bar{m}_{\pi }^2\big)+8 x \bar{m}_q^2+2 \bar{m}_{\pi }^2 \bar{m}_q^2+x r_{\phi } \big(9 x+18 x r_q+9 x r_q^2+8 \bar{m}_q^2\big)+6 \bar{m}_q^2 \bar{m}_{\sigma }^2\big) \big(r_A \eta _A+2 x r_A'\big)
-
\frac{1}{24 \big(x+x r_A+\bar{m}_{\text{gap}}^2\big) \big(x+x r_{\phi }+\bar{m}_{\pi }^2\big){}^2 \big(x+2 x r_q+x r_q^2+\bar{m}_q^2\big){}^2 \big(x+x r_{\phi }+\bar{m}_{\sigma }^2\big){}^2}x \big(9 x^3+18 x^2 \bar{m}_{\pi }^2+9 x \bar{m}_{\pi }^4+18 x r_q \big(x+\bar{m}_{\pi }^2\big){}^2+9 x r_q^2 \big(x+\bar{m}_{\pi }^2\big){}^2+8 x^2 \bar{m}_q^2+4 x \bar{m}_{\pi }^2 \bar{m}_q^2+2 \bar{m}_{\pi }^4 \bar{m}_q^2+x^2 r_{\phi }^2 \big(9 x+18 x r_q+9 x r_q^2+8 \bar{m}_q^2\big)+12 x \bar{m}_q^2 \bar{m}_{\sigma }^2+6 \bar{m}_q^2 \bar{m}_{\sigma }^4+2 x r_{\phi } \big(9 x^2+9 x \bar{m}_{\pi }^2+18 x r_q \big(x+\bar{m}_{\pi }^2\big)+9 x r_q^2 \big(x+\bar{m}_{\pi }^2\big)+8 x \bar{m}_q^2+2 \bar{m}_{\pi }^2 \bar{m}_q^2+6 \bar{m}_q^2 \bar{m}_{\sigma }^2\big)\big) \big(r_{\phi } \eta _{\phi }+2 x r_{\phi }'\big)\big)
+
g_{\phi  \bar{q} q}^4 \big(-\frac{\big(1+r_q\big) \bar{m}_q^2 \big(10 x^2+10 x^2 r_{\phi }^2+8 x \bar{m}_{\pi }^2+\bar{m}_{\pi }^4+12 x \bar{m}_{\sigma }^2+6 \bar{m}_{\pi }^2 \bar{m}_{\sigma }^2+3 \bar{m}_{\sigma }^4+4 x r_{\phi } \big(5 x+2 \bar{m}_{\pi }^2+3 \bar{m}_{\sigma }^2\big)\big) \big(r_q \eta _q+2 x r_q'\big)}{\big(x+x r_{\phi }+\bar{m}_{\pi }^2\big){}^2 \big(x+2 x r_q+x r_q^2+\bar{m}_q^2\big){}^3 \big(x+x r_{\phi }+\bar{m}_{\sigma }^2\big){}^2}
-
\frac{1}{2 \big(x+x r_{\phi }+\bar{m}_{\pi }^2\big){}^3 \big(x+2 x r_q+x r_q^2+\bar{m}_q^2\big){}^2 \big(x+x r_{\phi }+\bar{m}_{\sigma }^2\big){}^3}\bar{m}_q^2 \big(10 x^3+10 x^3 r_{\phi }^3+\bar{m}_{\pi }^6+18 x^2 \bar{m}_{\sigma }^2+12 x \bar{m}_{\sigma }^4+3 \bar{m}_{\sigma }^6+3 \bar{m}_{\pi }^4 \big(2 x+\bar{m}_{\sigma }^2\big)+3 \bar{m}_{\pi }^2 \big(2 x+\bar{m}_{\sigma }^2\big){}^2+6 x^2 r_{\phi }^2 \big(5 x+2 \bar{m}_{\pi }^2+3 \bar{m}_{\sigma }^2\big)+6 x r_{\phi } \big(5 x^2+\bar{m}_{\pi }^4+6 x \bar{m}_{\sigma }^2+2 \bar{m}_{\sigma }^4+2 \bar{m}_{\pi }^2 \big(2 x+\bar{m}_{\sigma }^2\big)\big)\big) \big(r_{\phi } \eta _{\phi }+2 x r_{\phi }'\big)\big)
+
g_{\phi  \bar{q} q}^2 \big(\lambda _{A \bar{q} q}^{\text{(1)}}\big){}^2 \big(
-
\frac{1}{6 \big(x+x r_A+\bar{m}_{\text{gap}}^2\big) \big(x+x r_{\phi }+\bar{m}_{\pi }^2\big) \big(x+2 x r_q+x r_q^2+\bar{m}_q^2\big){}^3 \big(x+x r_{\phi }+\bar{m}_{\sigma }^2\big)}\big(1+r_q\big) \big(4 x^2+x \bar{m}_{\pi }^2+5 x \bar{m}_q^2+8 \bar{m}_{\pi }^2 \bar{m}_q^2+x r_{\phi } \big(4 x+8 x r_q+4 x r_q^2+5 \bar{m}_q^2\big)+3 x \bar{m}_{\sigma }^2-3 \bar{m}_q^2 \bar{m}_{\sigma }^2+2 x r_q \big(4 x+\bar{m}_{\pi }^2+3 \bar{m}_{\sigma }^2\big)+x r_q^2 \big(4 x+\bar{m}_{\pi }^2+3 \bar{m}_{\sigma }^2\big)\big) \big(r_q \eta _q+2 x r_q'\big)
-
\frac{1}{24 \big(x+x r_A+\bar{m}_{\text{gap}}^2\big){}^2 \big(x+x r_{\phi }+\bar{m}_{\pi }^2\big) \big(x+2 x r_q+x r_q^2+\bar{m}_q^2\big){}^2 \big(x+x r_{\phi }+\bar{m}_{\sigma }^2\big)}\big(8 x^2+2 x \bar{m}_{\pi }^2+9 x \bar{m}_q^2+9 \bar{m}_{\pi }^2 \bar{m}_q^2+x r_{\phi } \big(8 x+16 x r_q+8 x r_q^2+9 \bar{m}_q^2\big)+6 x \bar{m}_{\sigma }^2+4 x r_q \big(4 x+\bar{m}_{\pi }^2+3 \bar{m}_{\sigma }^2\big)+2 x r_q^2 \big(4 x+\bar{m}_{\pi }^2+3 \bar{m}_{\sigma }^2\big)\big) \big(r_A \eta _A+2 x r_A'\big)
-
\frac{1}{24 \big(x+x r_A+\bar{m}_{\text{gap}}^2\big) \big(x+x r_{\phi }+\bar{m}_{\pi }^2\big){}^2 \big(x+2 x r_q+x r_q^2+\bar{m}_q^2\big){}^2 \big(x+x r_{\phi }+\bar{m}_{\sigma }^2\big){}^2}\big(8 x^3+4 x^2 \bar{m}_{\pi }^2+2 x \bar{m}_{\pi }^4+9 x^2 \bar{m}_q^2+18 x \bar{m}_{\pi }^2 \bar{m}_q^2+9 \bar{m}_{\pi }^4 \bar{m}_q^2+x^2 r_{\phi }^2 \big(8 x+16 x r_q+8 x r_q^2+9 \bar{m}_q^2\big)+12 x^2 \bar{m}_{\sigma }^2+6 x \bar{m}_{\sigma }^4+4 x r_q \big(4 x^2+2 x \bar{m}_{\pi }^2+\bar{m}_{\pi }^4+6 x \bar{m}_{\sigma }^2+3 \bar{m}_{\sigma }^4\big)+2 x r_q^2 \big(4 x^2+2 x \bar{m}_{\pi }^2+\bar{m}_{\pi }^4+6 x \bar{m}_{\sigma }^2+3 \bar{m}_{\sigma }^4\big)+2 x r_{\phi } \big(8 x^2+2 x \bar{m}_{\pi }^2+9 x \bar{m}_q^2+9 \bar{m}_{\pi }^2 \bar{m}_q^2+6 x \bar{m}_{\sigma }^2+4 x r_q \big(4 x+\bar{m}_{\pi }^2+3 \bar{m}_{\sigma }^2\big)+2 x r_q^2 \big(4 x+\bar{m}_{\pi }^2+3 \bar{m}_{\sigma }^2\big)\big)\big) \big(r_{\phi } \eta _{\phi }+2 x r_{\phi }'\big)\big)
}

\autoLineBreakStart{
  \left[ \mathcal{P}^{(\text{csc})}_{\qbqbqq}
  \right]_{\alpha\beta\gamma\delta}
  \left[ \overline{\text{Flow}}_{\qbqbqq}^{(4)}
  \right]_{\alpha\beta\gamma\delta} =
}{
\big(\lambda _{A \bar{q} q}^{\text{(1)}}\big){}^3 \lambda _{A \bar{q} q}^{\text{(4)}} \big(\frac{5 \bar{m}_q \big(3 x+6 x r_q+3 x r_q^2-\bar{m}_q^2\big) \big(r_q \eta _q+2 x r_q'\big)}{4 \big(x+x r_A+\bar{m}_{\text{gap}}^2\big){}^2 \big(x+2 x r_q+x r_q^2+\bar{m}_q^2\big){}^3}+\frac{5 x \big(1+r_q\big) \bar{m}_q \big(r_A \eta _A+2 x r_A'\big)}{2 \big(x+x r_A+\bar{m}_{\text{gap}}^2\big){}^3 \big(x+2 x r_q+x r_q^2+\bar{m}_q^2\big){}^2}\big)
+
\lambda _{A \bar{q} q}^{\text{(1)}} \big(\lambda _{A \bar{q} q}^{\text{(4)}}\big){}^3 \big(\frac{5 x \bar{m}_q \big(-3 x-6 x r_q-3 x r_q^2+\bar{m}_q^2\big) \big(r_q \eta _q+2 x r_q'\big)}{4 \big(x+x r_A+\bar{m}_{\text{gap}}^2\big){}^2 \big(x+2 x r_q+x r_q^2+\bar{m}_q^2\big){}^3}-\frac{5 x^2 \big(1+r_q\big) \bar{m}_q \big(r_A \eta _A+2 x r_A'\big)}{2 \big(x+x r_A+\bar{m}_{\text{gap}}^2\big){}^3 \big(x+2 x r_q+x r_q^2+\bar{m}_q^2\big){}^2}\big)
+
\big(\lambda _{A \bar{q} q}^{\text{(1)}}\big){}^4 \big(\frac{5 \big(1+r_q\big) \bar{m}_q^2 \big(r_q \eta _q+2 x r_q'\big)}{4 \big(x+x r_A+\bar{m}_{\text{gap}}^2\big){}^2 \big(x+2 x r_q+x r_q^2+\bar{m}_q^2\big){}^3}+\frac{5 \bar{m}_q^2 \big(r_A \eta _A+2 x r_A'\big)}{8 \big(x+x r_A+\bar{m}_{\text{gap}}^2\big){}^3 \big(x+2 x r_q+x r_q^2+\bar{m}_q^2\big){}^2}\big)
+
\big(\lambda _{A \bar{q} q}^{\text{(4)}}\big){}^4 \big(\frac{5 x^2 \big(1+r_q\big) \bar{m}_q^2 \big(r_q \eta _q+2 x r_q'\big)}{4 \big(x+x r_A+\bar{m}_{\text{gap}}^2\big){}^2 \big(x+2 x r_q+x r_q^2+\bar{m}_q^2\big){}^3}+\frac{5 x^2 \bar{m}_q^2 \big(r_A \eta _A+2 x r_A'\big)}{8 \big(x+x r_A+\bar{m}_{\text{gap}}^2\big){}^3 \big(x+2 x r_q+x r_q^2+\bar{m}_q^2\big){}^2}\big)
+
\big(\lambda _{A \bar{q} q}^{\text{(1)}}\big){}^2 \big(\lambda _{A \bar{q} q}^{\text{(4)}}\big){}^2 \big(\frac{5 x \big(1+r_q\big) \big(x+2 x r_q+x r_q^2-2 \bar{m}_q^2\big) \big(r_q \eta _q+2 x r_q'\big)}{2 \big(x+x r_A+\bar{m}_{\text{gap}}^2\big){}^2 \big(x+2 x r_q+x r_q^2+\bar{m}_q^2\big){}^3}+\frac{5 x \big(2 x+4 x r_q+2 x r_q^2-\bar{m}_q^2\big) \big(r_A \eta _A+2 x r_A'\big)}{4 \big(x+x r_A+\bar{m}_{\text{gap}}^2\big){}^3 \big(x+2 x r_q+x r_q^2+\bar{m}_q^2\big){}^2}\big)
+
}
\autoLineBreakEnd{
g_{\Delta  \bar{q} \bar{q}}^4 \big(-\frac{\big(1+r_q\big) \bar{m}_q^2 \big(r_q \eta _q+2 x r_q'\big)}{\big(x+2 x r_q+x r_q^2+\bar{m}_q^2\big){}^3 \big(x+x r_{\Delta }+\bar{m}_{\Delta }^2\big){}^2}-\frac{\bar{m}_q^2 \big(r_{\Delta } \eta _{\Delta }+2 x r_{\Delta }'\big)}{2 \big(x+2 x r_q+x r_q^2+\bar{m}_q^2\big){}^2 \big(x+x r_{\Delta }+\bar{m}_{\Delta }^2\big){}^3}\big)
+
g_{\Delta  \bar{q} \bar{q}}^2 \big(\lambda _{A \bar{q} q}^{\text{(1)}}\big){}^2 \big(-\frac{2 \big(1+r_q\big) \big(x+2 x r_q+x r_q^2-\bar{m}_q^2\big) \big(r_q \eta _q+2 x r_q'\big)}{\big(x+x r_A+\bar{m}_{\text{gap}}^2\big) \big(x+2 x r_q+x r_q^2+\bar{m}_q^2\big){}^3 \big(x+x r_{\Delta }+\bar{m}_{\Delta }^2\big)}-\frac{x \big(1+r_q\big){}^2 \big(r_A \eta _A+2 x r_A'\big)}{\big(x+x r_A+\bar{m}_{\text{gap}}^2\big){}^2 \big(x+2 x r_q+x r_q^2+\bar{m}_q^2\big){}^2 \big(x+x r_{\Delta }+\bar{m}_{\Delta }^2\big)}-\frac{x \big(1+r_q\big){}^2 \big(r_{\Delta } \eta _{\Delta }+2 x r_{\Delta }'\big)}{\big(x+x r_A+\bar{m}_{\text{gap}}^2\big) \big(x+2 x r_q+x r_q^2+\bar{m}_q^2\big){}^2 \big(x+x r_{\Delta }+\bar{m}_{\Delta }^2\big){}^2}\big)
+
g_{\Delta  \bar{q} \bar{q}}^2 \lambda _{A \bar{q} q}^{\text{(1)}} \lambda _{A \bar{q} q}^{\text{(4)}} \big(\frac{2 \bar{m}_q \big(3 x+6 x r_q+3 x r_q^2-\bar{m}_q^2\big) \big(r_q \eta _q+2 x r_q'\big)}{\big(x+x r_A+\bar{m}_{\text{gap}}^2\big) \big(x+2 x r_q+x r_q^2+\bar{m}_q^2\big){}^3 \big(x+x r_{\Delta }+\bar{m}_{\Delta }^2\big)}+\frac{2 x \big(1+r_q\big) \bar{m}_q \big(r_A \eta _A+2 x r_A'\big)}{\big(x+x r_A+\bar{m}_{\text{gap}}^2\big){}^2 \big(x+2 x r_q+x r_q^2+\bar{m}_q^2\big){}^2 \big(x+x r_{\Delta }+\bar{m}_{\Delta }^2\big)}+\frac{2 x \big(1+r_q\big) \bar{m}_q \big(r_{\Delta } \eta _{\Delta }+2 x r_{\Delta }'\big)}{\big(x+x r_A+\bar{m}_{\text{gap}}^2\big) \big(x+2 x r_q+x r_q^2+\bar{m}_q^2\big){}^2 \big(x+x r_{\Delta }+\bar{m}_{\Delta }^2\big){}^2}\big)
+
g_{\Delta  \bar{q} \bar{q}}^2 \big(\lambda _{A \bar{q} q}^{\text{(4)}}\big){}^2 \big(-\frac{4 x \big(1+r_q\big) \bar{m}_q^2 \big(r_q \eta _q+2 x r_q'\big)}{\big(x+x r_A+\bar{m}_{\text{gap}}^2\big) \big(x+2 x r_q+x r_q^2+\bar{m}_q^2\big){}^3 \big(x+x r_{\Delta }+\bar{m}_{\Delta }^2\big)}-\frac{x \bar{m}_q^2 \big(r_A \eta _A+2 x r_A'\big)}{\big(x+x r_A+\bar{m}_{\text{gap}}^2\big){}^2 \big(x+2 x r_q+x r_q^2+\bar{m}_q^2\big){}^2 \big(x+x r_{\Delta }+\bar{m}_{\Delta }^2\big)}-\frac{x \bar{m}_q^2 \big(r_{\Delta } \eta _{\Delta }+2 x r_{\Delta }'\big)}{\big(x+x r_A+\bar{m}_{\text{gap}}^2\big) \big(x+2 x r_q+x r_q^2+\bar{m}_q^2\big){}^2 \big(x+x r_{\Delta }+\bar{m}_{\Delta }^2\big){}^2}\big)
+
g_{\phi  \bar{q} q}^2 g_{\Delta  \bar{q} \bar{q}}^2 \big(\frac{8 \big(1+r_q\big) \bar{m}_q^2 \big(r_q \eta _q+2 x r_q'\big)}{\big(x+2 x r_q+x r_q^2+\bar{m}_q^2\big){}^3 \big(x+x r_{\Delta }+\bar{m}_{\Delta }^2\big) \big(x+x r_{\phi }+\bar{m}_{\sigma }^2\big)}+\frac{2 \bar{m}_q^2 \big(r_{\Delta } \eta _{\Delta }+2 x r_{\Delta }'\big)}{\big(x+2 x r_q+x r_q^2+\bar{m}_q^2\big){}^2 \big(x+x r_{\Delta }+\bar{m}_{\Delta }^2\big){}^2 \big(x+x r_{\phi }+\bar{m}_{\sigma }^2\big)}+\frac{2 \bar{m}_q^2 \big(r_{\phi } \eta _{\phi }+2 x r_{\phi }'\big)}{\big(x+2 x r_q+x r_q^2+\bar{m}_q^2\big){}^2 \big(x+x r_{\Delta }+\bar{m}_{\Delta }^2\big) \big(x+x r_{\phi }+\bar{m}_{\sigma }^2\big){}^2}\big)
+
g_{\phi  \bar{q} q}^2 \big(\lambda _{A \bar{q} q}^{\text{(1)}}\big){}^2 \big(\frac{5 \big(1+r_q\big) \big(x+2 x r_q+x r_q^2-\bar{m}_q^2\big) \big(4 x
+
4 x r_{\phi }+\bar{m}_{\pi }^2+3 \bar{m}_{\sigma }^2\big) \big(r_q \eta _q+2 x r_q'\big)}{3 \big(x+x r_A+\bar{m}_{\text{gap}}^2\big) \big(x+x r_{\phi }+\bar{m}_{\pi }^2\big) \big(x+2 x r_q+x r_q^2+\bar{m}_q^2\big){}^3 \big(x+x r_{\phi }+\bar{m}_{\sigma }^2\big)}
+
\frac{5 x \big(1+r_q\big){}^2 \big(4 x+4 x r_{\phi }+\bar{m}_{\pi }^2+3 \bar{m}_{\sigma }^2\big) \big(r_A \eta _A+2 x r_A'\big)}{6 \big(x+x r_A+\bar{m}_{\text{gap}}^2\big){}^2 \big(x+x r_{\phi }+\bar{m}_{\pi }^2\big) \big(x+2 x r_q+x r_q^2+\bar{m}_q^2\big){}^2 \big(x+x r_{\phi }+\bar{m}_{\sigma }^2\big)}+\frac{5 x \big(1+r_q\big){}^2 \big(4 x^2+4 x^2 r_{\phi }^2+2 x \bar{m}_{\pi }^2+\bar{m}_{\pi }^4+6 x \bar{m}_{\sigma }^2+3 \bar{m}_{\sigma }^4+2 x r_{\phi } \big(4 x+\bar{m}_{\pi }^2+3 \bar{m}_{\sigma }^2\big)\big) \big(r_{\phi } \eta _{\phi }+2 x r_{\phi }'\big)}{6 \big(x+x r_A+\bar{m}_{\text{gap}}^2\big) \big(x+x r_{\phi }+\bar{m}_{\pi }^2\big){}^2 \big(x+2 x r_q+x r_q^2+\bar{m}_q^2\big){}^2 \big(x+x r_{\phi }+\bar{m}_{\sigma }^2\big){}^2}\big)
+
g_{\phi  \bar{q} q}^2 \lambda _{A \bar{q} q}^{\text{(1)}} \lambda _{A \bar{q} q}^{\text{(4)}} \big(\frac{5 \bar{m}_q \big(-3 x-6 x r_q-3 x r_q^2+\bar{m}_q^2\big) \big(4 x+4 x r_{\phi }+\bar{m}_{\pi }^2+3 \bar{m}_{\sigma }^2\big) \big(r_q \eta _q+2 x r_q'\big)}{3 \big(x+x r_A+\bar{m}_{\text{gap}}^2\big) \big(x+x r_{\phi }+\bar{m}_{\pi }^2\big) \big(x+2 x r_q+x r_q^2+\bar{m}_q^2\big){}^3 \big(x+x r_{\phi }+\bar{m}_{\sigma }^2\big)}-\frac{5 x \big(1+r_q\big) \bar{m}_q \big(4 x+4 x r_{\phi }+\bar{m}_{\pi }^2+3 \bar{m}_{\sigma }^2\big) \big(r_A \eta _A+2 x r_A'\big)}{3 \big(x+x r_A+\bar{m}_{\text{gap}}^2\big){}^2 \big(x+x r_{\phi }+\bar{m}_{\pi }^2\big) \big(x+2 x r_q+x r_q^2+\bar{m}_q^2\big){}^2 \big(x+x r_{\phi }+\bar{m}_{\sigma }^2\big)}-\frac{5 x \big(1+r_q\big) \bar{m}_q \big(4 x^2+4 x^2 r_{\phi }^2+2 x \bar{m}_{\pi }^2+\bar{m}_{\pi }^4+6 x \bar{m}_{\sigma }^2+3 \bar{m}_{\sigma }^4+2 x r_{\phi } \big(4 x+\bar{m}_{\pi }^2+3 \bar{m}_{\sigma }^2\big)\big) \big(r_{\phi } \eta _{\phi }+2 x r_{\phi }'\big)}{3 \big(x+x r_A+\bar{m}_{\text{gap}}^2\big) \big(x+x r_{\phi }+\bar{m}_{\pi }^2\big){}^2 \big(x+2 x r_q+x r_q^2+\bar{m}_q^2\big){}^2 \big(x+x r_{\phi }+\bar{m}_{\sigma }^2\big){}^2}\big)
+
g_{\phi  \bar{q} q}^2 \big(\lambda _{A \bar{q} q}^{\text{(4)}}\big){}^2 \big(\frac{10 x \big(1+r_q\big) \bar{m}_q^2 \big(4 x+4 x r_{\phi }+\bar{m}_{\pi }^2+3 \bar{m}_{\sigma }^2\big) \big(r_q \eta _q+2 x r_q'\big)}{3 \big(x+x r_A+\bar{m}_{\text{gap}}^2\big) \big(x+x r_{\phi }+\bar{m}_{\pi }^2\big) \big(x+2 x r_q+x r_q^2+\bar{m}_q^2\big){}^3 \big(x+x r_{\phi }+\bar{m}_{\sigma }^2\big)}+\frac{5 x \bar{m}_q^2 \big(4 x+4 x r_{\phi }+\bar{m}_{\pi }^2+3 \bar{m}_{\sigma }^2\big) \big(r_A \eta _A+2 x r_A'\big)}{6 \big(x+x r_A+\bar{m}_{\text{gap}}^2\big){}^2 \big(x+x r_{\phi }+\bar{m}_{\pi }^2\big) \big(x+2 x r_q+x r_q^2+\bar{m}_q^2\big){}^2 \big(x+x r_{\phi }+\bar{m}_{\sigma }^2\big)}+\frac{5 x \bar{m}_q^2 \big(4 x^2+4 x^2 r_{\phi }^2+2 x \bar{m}_{\pi }^2+\bar{m}_{\pi }^4+6 x \bar{m}_{\sigma }^2+3 \bar{m}_{\sigma }^4+2 x r_{\phi } \big(4 x+\bar{m}_{\pi }^2+3 \bar{m}_{\sigma }^2\big)\big) \big(r_{\phi } \eta _{\phi }+2 x r_{\phi }'\big)}{6 \big(x+x r_A+\bar{m}_{\text{gap}}^2\big) \big(x+x r_{\phi }+\bar{m}_{\pi }^2\big){}^2 \big(x+2 x r_q+x r_q^2+\bar{m}_q^2\big){}^2 \big(x+x r_{\phi }+\bar{m}_{\sigma }^2\big){}^2}\big)
}

\twocolumngrid

\bibliography{literature.bib}

\end{document}